\documentclass[a4paper,11pt]{article}
\usepackage{jheppub} % for details on the use of the package, please see the JINST-author-manual
\usepackage{lineno}
\usepackage{ifthen}
\usepackage{subcaption}
\usepackage{cleveref}
\crefname{appendix}{appendix}{appendices}
\Crefname{appendix}{Appendix}{Appendices}
\usepackage{mathrsfs}
\usepackage{tikz}
\usepackage[normalem]{ulem}

\DeclareMathOperator{\Li}{Li}

\newcommand{\comment}[2][]{
 {\color{red}\texttt{[\ifthenelse{\equal{#1}{}}{}{#1: }#2]}}
}

\newcommand{\Veq}{\mathbb{V}}  % equivariant volume

\def \calb {{\cal B}}

\def \cali {{\cal I}}

\def \calm {{\cal M}}
\def \caln {{\cal N}}
\def \calo {{\cal O}}
\def \calp {{\cal P}}

\def \calv {{\cal V}}

\def \A {{\mathrm A}}
\def \v {{\mathrm v}}
\def \d {{\mathrm d}}

\title{Higgs-branch Giant Gravitons and Equivariant Volume}

\author[a,b]{Paul Luis Roehl,}
\author[c]{Canberk \c{S}anl\i}

\affiliation[a]{DAMTP, Centre for Mathematical Sciences, University of Cambridge\\ Wilberforce Road, Cambridge CB3 0WA, U.K.}
\affiliation[b]{Department of Physics, Korea Advanced Institute of Science and Technology\\291 Daehak-ro, Yuseong-gu, Daejeon 34141, Republic of Korea}
\affiliation[c]{Sidney Sussex College, University of Cambridge, Sidney Street, Cambridge CB2 3HU, U.K.}

\emailAdd{plr30@cam.ac.uk}
\emailAdd{cs2412@cam.ac.uk}

\abstract{We study the giant graviton expansion of the superconformal index for M2 branes probing the orbifold singularity $X_4=\mathbb C^2/\mathbb Z_K\times\mathbb C^2/\mathbb Z_L$ in the Higgs-branch limit, where the index reduces to the Hilbert series of the Higgs-branch moduli space of $U(K)$ instantons on $\mathbb{C}^2 / \mathbb{Z}_L$.
In this limit, the giant graviton coefficients are governed by affine fermionic forms and are expected to be related to $q$-deformations of the vacuum characters of associated vertex operator algebras.

In a Cardy-like regime of small chemical potentials, we derive trilogarithmic expressions for the giant graviton free energy and show that this contribution is governed by a resummation of the equivariant volume of $\mathbb C^2/\mathbb Z_L$.
At fixed wrapping, the leading behaviour reproduces the Cardy growth associated with the effective central charge of the corresponding vertex algebra.
At large wrapping, the leading cubic term agrees with the Higgs-branch reduction of the seven-dimensional master-volume prediction, while the equivariant-volume resummation determines the subleading terms in the leading Cardy free energy.
We show that the addition of baryonic background fluxes has the effect of resolving the cone $\mathbb{C}^2 / \mathbb{Z}_L$, with flux differences determining the exceptional-curve areas.
The asymptotics also contain an additional term quadratic in the chemical potentials of $SU(K)$, which can contribute at leading Cardy order.
}

\begin{document}
\maketitle

\section{Introduction}

Supersymmetric indices have played a central role in holography, providing a rare class of protected observables that can be computed exactly and matched across the duality. Recent advances have pushed this programme into a finite-$N$ domain of AdS/CFT, where the goal is to understand non-perturbative corrections beyond the large-$N$ limit.  %, where $N$ is both the number of wrapped branes and the rank of the corresponding worldvolume gauge theory. 
One remarkable development in this direction is the giant graviton expansion of the superconformal index \cite{Gaiotto:2021xce,Imamura:2021ytr,Arai:2020uwd}.
In this expansion, the finite-$N$ index is reorganised into a series of corrections interpreted as contributions from flux-stabilised branes in the dual geometry, localised to a discrete set of wrapping numbers of the internal space.
For $\text{AdS}_4 \times S^7$, there are four wrapping numbers\footnote{This goes ultimately back to Mikhailov \cite{mikhailov2000}. Embedding $S^7$ into $\mathbb{C}^4$ via $|z_1|^2 + \ldots + |z_4|^2 = 1$, the four types of configurations $n_I$ correspond to the divisors $S^5 \subset S^7$ with $z_I = 0$.} and the superconformal index $\mathcal{I}_N$ is of the form
\begin{equation}
  \label{eq:gge-a46}
  \mathcal{I}_N = \mathcal{I}_{\infty} \sum_{n_1, n_2, n_3, n_4 = 0}^{\infty} e^{- N \sum_{I=1}^{4} n_I \Delta_I} \mathcal{I}_{n_1, n_2, n_3, n_4}.
\end{equation}

For low configuration numbers $n_I$, the coefficients $\mathcal{I}_{n_1,n_2,n_3,n_4}$ of these brane configurations can (in principle) be recovered from a probe brane computation in the AdS background \cite{Beccaria:2023cuo}.
Alternatively, they can be identified with the superconformal indices of the worldvolume theories after suitable analytic continuation \cite{Imamura:2021ytr}.
When the configuration numbers $n_I$ become large, the branes backreact with and deform the closed-string geometry.
The near-horizon geometries of these backreacted branes have been identified in a recent proposal \cite{Chen:2024erz} with a certain off-shell family of deformations of $S^7$, which has previously appeared in the context of volume minimisation \cite{Martelli:2006yb,Gauntlett:2019roi}.
In this geometric correspondence, the asymptotics of the state count $\mathcal{I}_{n_1, n_2, n_3, n_4}$ are governed by the \emph{master volume} of the auxiliary deformed geometry.
The dominant configurations\footnote{Heuristically, for D3-branes these are configuration numbers $n_I$ of order $N$ and for M2-branes they are of order $N^{1/2}$.} backreact to form the black hole near-horizon geometry, with their master volume crucial to recovering the entropy function.

While the master volume correspondence is difficult to prove even for $S^7$, it is expected to also hold for the more general Sasaki--Einstein seven-manifolds $\operatorname{SE}_7$.
In this work, we will verify this relation explicitly for a class of theories where the Sasaki--Einstein seven-manifold $\text{SE}_7$ is the link of the Calabi--Yau cone $X_4 = \mathbb{C}^2 / \mathbb{Z}_K \times \mathbb{C}^2 / \mathbb{Z}_L$.

\begin{figure}
	\centering
	\begin{tikzpicture}[
			gauge/.style={circle, draw, thick, fill=white, minimum size=1.2cm, inner sep=0pt},
			flavor/.style={rectangle, draw, thick, fill=white, minimum size=1.0cm, inner sep=0pt},
			edge/.style={thick}
		]
		% Radii for the circular layout
		\def\R{2.5}
		\def\Rf{4.5}

		% 1. Draw the circular arcs (Gauge to Gauge)
		% Solid arcs
		\draw[edge] (150:\R) arc (150:-30:\R);
		\draw[edge] (210:\R) arc (210:150:\R);
		% Dashed arc
		\draw[edge, dashed] (-30:\R) arc (330:210:\R);

		% 2. Draw the straight edges (Gauge to Flavor)
		\draw[edge] (90:\R) -- (90:\Rf);
		\draw[edge] (30:\R) -- (30:\Rf);
		\draw[edge] (-30:\R) -- (-30:\Rf);
		\draw[edge] (150:\R) -- (150:\Rf);
		\draw[edge] (210:\R) -- (210:\Rf);

		% 3. Place the nodes on top of the lines
		% Gauge nodes
		\node[gauge] (g0) at (90:\R) {$N$};
		\node[gauge] (g1) at (30:\R) {$N$};
		\node[gauge] (g2) at (-30:\R) {$N$};
		\node[gauge] (gL1) at (150:\R) {$N$};
		\node[gauge] (gL2) at (210:\R) {$N$};

		% Flavor nodes
		\node[flavor] (f0) at (90:\Rf) {$K_0$};
		\node[flavor] (f1) at (30:\Rf) {$K_1$};
		\node[flavor] (f2) at (-30:\Rf) {$K_2$};
		\node[flavor] (fL1) at (150:\Rf) {$K_{L-1}$};
		\node[flavor] (fL2) at (210:\Rf) {$K_{L-2}$};

	\end{tikzpicture}
	\caption{Quiver diagram of the 3d $\mathcal{N} = 4$ gauge theory that describes in the infrared the dynamics of $N$ M2-branes probing a $\mathbb{C}^2 / \mathbb{Z}_K \times \mathbb{C}^2 / \mathbb{Z}_L$ orbifold singularity. While in this work we keep $K = \sum_{A=0}^{L-1} K_{A}$ for generality, we may focus on the case $K_0 = K$ and $K_{A > 0} = 0$.}
	\label{fig:quiver-diag}
\end{figure}
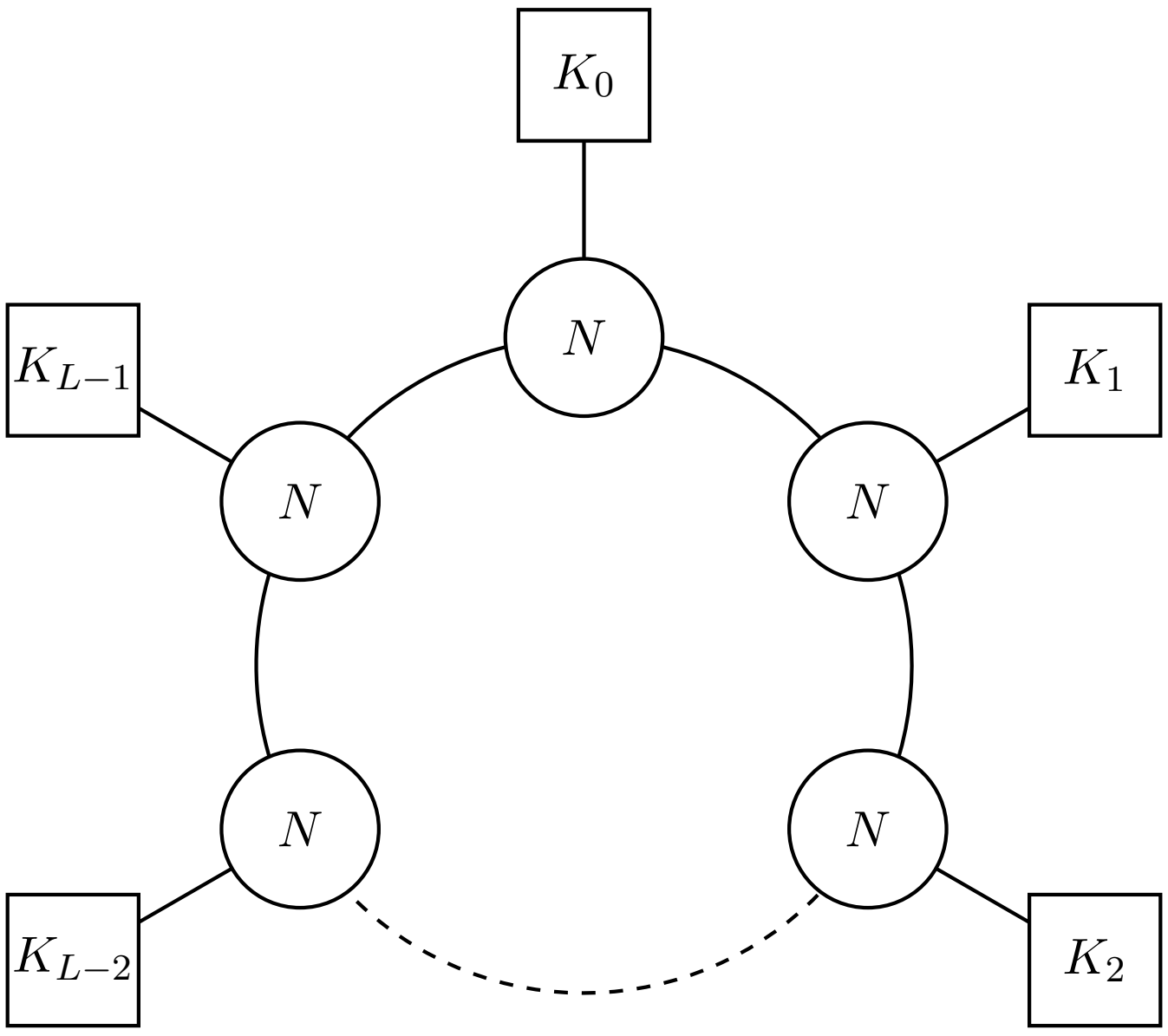

Before describing our main findings in computing the asymptotics of the giant graviton coefficients, we continue by further elaborating on the field theoretic setup.
We consider the giant graviton expansion of the superconformal index $\mathcal{I}$ of $N$ M2-branes at the tip of the Calabi--Yau cone $X_4 = \mathbb{C}^2 / \mathbb{Z}_K \times \mathbb{C}^2 / \mathbb{Z}_L$.
The worldvolume theory is a three-dimensional conformal field theory with $\mathcal{N} = 4$ supersymmetry which can be described as the infrared superconformal fixed point of the quiver diagram in figure~\ref{fig:quiver-diag}.%
\footnote{%
The UV quiver description explicitly breaks the $L \leftrightarrow K$ symmetry of the theory.
Three-dimensional mirror symmetry \cite{Intriligator:1996ex, deBoer:1996mp} asserts that both the quiver in figure~\ref{fig:quiver-diag} and its mirror quiver flow to the same superconformal fixed point.%
}
The field theory possesses a $\operatorname{SU}(2)_H \times \operatorname{SU}(2)_C$ $R$-symmetry and a $U(1)_x \times U(1)_y \times \operatorname{SU}(K) \times \operatorname{SU}(L)$ global symmetry group.
The UV quiver description naturally splits the global symmetries into flavour symmetries $U(1)_x \times \operatorname{SU}(K)$ and topological symmetries $U(1)_y \times \operatorname{SU}(L)$.\footnote{In the UV, only a $U(1)^L$ topological symmetry is manifest, which enhances in the infrared to $U(1)_y \times \operatorname{SU}(L)$.}

The theory is holographically dual to M-theory on $\text{AdS}_4 \times \operatorname{SE}_7$, where the seven-dimensional Sasaki--Einstein manifold $\operatorname{SE}_7$ is the \emph{link} of the cone $X_4$.
We denote the Cartan generators of the conformal symmetry by $E$ and $J_3$ and those of the $R$-symmetry by $H$ and $C$.
It is useful to introduce chemical potentials $\Delta_I$, $I = 1,2,3,4$, mixing the Cartan generators of the $R$-symmetry with those of $U(1)_x \times U(1)_y$ as
\begin{equation}
  \label{eq:fugacities-def}
  \begin{aligned}
    \Delta_1 &= \Delta_H + \Delta_x / L, &
    \Delta_2 &= \Delta_H - \Delta_x / L, \\
    \Delta_3 &= \Delta_C + \Delta_y / K \quad \text{and} &
    \Delta_4 &= \Delta_C - \Delta_y / K.
  \end{aligned}
\end{equation}
When $K = L = 1$, these are the generators of an enhanced $\operatorname{SO}(8)$ $R$-symmetry at the superconformal fixed point.
We parametrise the chemical potentials of the remaining generators as $\v_{\alpha}$, $\alpha = 0, \ldots, K-1$, for $\operatorname{SU}(K)$ (subject to the constraint $\sum_{\alpha = 0}^{K-1} \v_{\alpha} = \Delta_x$) and similarly $\mathrm{u}_{A}$, $A = 0, \ldots, L-1$, for $\operatorname{SU}(L)$.
In the bulk, the $\Delta_I$ describe the $U(1)^4 \subset \operatorname{SO}(8)$ isometry of $S^7$ induced from $X_4 = \mathbb{C}^4$.
This induced $U(1)^4$ isometry is preserved on $\operatorname{SE}_7$ even for $K, L \ne 1$.

The superconformal index is the supersymmetric partition function on $S^1 \times S^2$ or equivalently the trace over all states in the Hilbert space on $S^2$, subject to the BPS condition
\begin{equation}
  \Delta \equiv E - J_3 - H - C = 0.
\end{equation}
We write this as
\begin{equation}
  \label{eq:SCI-defn}
  \mathcal{I}_N(\omega, \Delta_I; \v_{\alpha}, \mathrm{u}_{A}) =
  \operatorname{Tr}_{\text{BPS}}
  \left[
    (-1)^F e^{- \omega J_3} \prod_{I=1}^{4} e^{-\Delta_I Q_I}
    \prod_{\alpha=0}^{K-1} e^{- \v_{\alpha} Q_{\alpha}}
    \prod_{A=0}^{L-1} e^{-\mathrm{u}_{A} Q_A}
  \right]. 
\end{equation}
In order for the combination of generators to anti-commute with the supercharge, we need to impose the constraint
\begin{equation}
  \label{eq:SCI-constraint}
  \omega - 2 \Delta_H - 2 \Delta_C = \omega - \sum_{I=1}^{4} \Delta_I \in 4 \pi i \mathbb{Z}.
\end{equation}
The configuration numbers $n_I$ in the giant graviton expansion of toric quiver gauge theories are determined by the underlying toric diagram \cite{Arai:2019wgv,Arai:2019aou}.
Roughly, every vertex gives rise to one type of configuration.
For the orbifold quiver gauge theories considered here, the giant graviton expansion is a sum over four types of configurations $n_I$ just as in the case $K = L = 1$ in \eqref{eq:gge-a46}, except that $\Delta_{1,2} \to L \Delta_{1,2}$ and $\Delta_{3,4} \to K \Delta_{3,4}$ in the exponent.
The nature of the orbifold presents itself more clearly in the giant graviton expansion when introducing background fluxes for the global symmetries, as we will elaborate on below.
We call background fluxes for the flavour symmetries `magnetic' and those for the topological symmetries `baryonic'.

The crucial simplification we perform to get the giant graviton expansion under control is the \emph{Higgs-branch limit} $\Delta_C \to \infty$ \cite{Razamat:2014pta, Hayashi:2024aaf, Dorey:2025qht}.
In order to keep the constraint \eqref{eq:SCI-constraint} satisfied, we simultaneously send $\omega \to \infty$ such that $\omega - 2 \Delta_C = \omega - \Delta_3 - \Delta_4 = \text{const}$.
Note that from \eqref{eq:fugacities-def} it follows that also $\Delta_3 / \Delta_4 \to 1$.
In the Higgs-branch limit, the superconformal index reduces to the Hilbert series of the Higgs-branch moduli space.
The dependence on the Coulomb branch chemical potentials $\Delta_3, \Delta_4$ and $\mathrm{u}_A$ drops out and only the sector without magnetic background fluxes remains non-zero. The remaining quantity depends on $\Delta_1, \Delta_2$, $\v_{\alpha}$ and potential baryonic background fluxes $\boldsymbol{b} \in \mathbb{Z}^L$ and will be denoted by\footnote{This construction also works analogously in the \emph{Coulomb-branch limit} $\omega, \Delta_H \to \infty$.}
\begin{equation}
  H_N^{\boldsymbol{b}}(\Delta_1, \Delta_2; \v_{\alpha}).
\end{equation}
The giant graviton expansion for the Hilbert series without baryonic flux is then
\begin{equation}
  H_N(\Delta_1, \Delta_2; \v_{\alpha}) =
  H_{\infty}(\Delta_1, \Delta_2; \v_{\alpha})
  \sum_{\boldsymbol{n}_{1,2} \in \mathbb{N}_0 \cdot (1, \ldots, 1)}^{} 
  e^{- N \sum_{I=1,2} J(\boldsymbol{n}_I) \Delta_I} \mathcal{I}_{\boldsymbol{n}_1, \boldsymbol{n}_2}(\Delta_1, \Delta_2; \v_{\alpha}),
\end{equation}
where $(1, \ldots, 1) \in \mathbb{Z}^L$.
The charge $J(\boldsymbol{n}_I)$ is the sum over all components,
\begin{equation}
  J(\boldsymbol{n}_I) = \sum_{A=0}^{L-1} n_{I, A}
\end{equation}
and $\mathcal{I}_{\boldsymbol{n}_1, \boldsymbol{n}_2}$ are the giant graviton coefficients obtained from the full expansion by
\begin{equation}
  \mathcal{I}_{\boldsymbol{n}_1, \boldsymbol{n}_2} = \lim_{\text{Higgs}} \mathcal{I}_{\boldsymbol{n}_1, \boldsymbol{n}_2, 0, 0}.
\end{equation}

A common feature of giant graviton expansions is a large number of cancellations in the coefficients.
The effect is that in the previous expansion one may set $\boldsymbol{n}_1 = 0$ when formally expanding around $e^{- \Delta_1}$ first and vice versa, without changing the result.
These truncated versions are called \emph{simple-sum expansions} \cite{Gaiotto:2021xce, Fujiwara:2023bdc}.

While the unrestricted form of the giant graviton expansion in the presence of background flux has not been written down, we may leverage the simple-sum expansion to write the Hilbert series with $\boldsymbol{b} \ne 0$ as
\begin{equation}
  H_N^{\boldsymbol{b}}(\Delta_1, \Delta_2; \v_{\alpha}) =
  H_{\infty}(\Delta_1, \Delta_2; \v_{\alpha})
  \sum_{\boldsymbol{n}_{2} \in [\boldsymbol{b}]_+}^{} 
  e^{- N J(\boldsymbol{n}_2) \Delta_2} \mathcal{I}_{0, \boldsymbol{n}_2}(\Delta_1, \Delta_2; \v_{\alpha})
\end{equation}
when expanding both sides in $e^{-\Delta_1}$ first and
\begin{equation}
  H_N^{\boldsymbol{b}}(\Delta_1, \Delta_2; \v_{\alpha}) =
  H_{\infty}(\Delta_1, \Delta_2; \v_{\alpha})
  \sum_{\boldsymbol{n}_{1} \in [-\boldsymbol{b}]_+}^{} 
  e^{- N J(\boldsymbol{n}_1) \Delta_1} \mathcal{I}_{\boldsymbol{n}_1, 0}(\Delta_1, \Delta_2; \v_{\alpha})
\end{equation}
when expanding in $e^{-\Delta_2}$ first.
Here, we have denoted by $[\boldsymbol{b}]$ the set of background fluxes related to $\boldsymbol{b}$ by shift symmetry,\begin{equation}
  [\boldsymbol{b}] \equiv \{ \boldsymbol{b} + m (1, \ldots, 1) \in \mathbb{Z}^L \}
  \quad \text{and} \quad
  [\boldsymbol{b}]_{+} \equiv [\boldsymbol{b}] \cap \mathbb{N}_0^L.
\end{equation}

Our main result is that the asymptotics of the giant graviton coefficients $\mathcal{I}_{\boldsymbol{n}_1, \boldsymbol{n}_2}$ in a Cardy-like limit
\begin{equation}
  \Delta_1, \Delta_2 \to 0, \quad J(\boldsymbol{n}_I) \Delta_{I} \sim \text{const.}
\end{equation}
are governed by the equivariant volume of a deformation of $\mathbb{C}^2 / \mathbb{Z}_L$,
\begin{equation}
  \mathcal{I}_{\boldsymbol{n}_1, \boldsymbol{n}_2} \sim
  \operatorname{PE} [K^2 \mathbb{V}_{\mathbb{C}^2 / \mathbb{Z}_L}(\boldsymbol{\lambda}; \epsilon_1, \epsilon_2) - K^2 \mathbb{V}_{\mathbb{C}^2 / \mathbb{Z}_L}(0; \epsilon_1, \epsilon_2)],
\end{equation}
with parameters $\boldsymbol{\lambda}, \epsilon_1, \epsilon_2$ explained below.
Here, $\operatorname{PE}$ means the \emph{plethystic exponential}, defined for chemical potentials $\epsilon_I$ as
\begin{equation}
  \operatorname{PE}[f(\epsilon_I)] \equiv \exp
  \left[
  \sum_{l=1}^{\infty} \frac{1}{l} f(l \epsilon_I)
  \right].
\end{equation}
In addition to satisfying $\operatorname{PE}[f+g] = \operatorname{PE}[f] \operatorname{PE}[g]$, for $f(\epsilon_I) = a e^{b \epsilon_I}$ the plethystic exponential formally reduces to $\operatorname{PE}[a e^{b \epsilon_I}] = (1 - e^{b \epsilon_I})^{-a}$.
The addition of baryonic fluxes $\boldsymbol{b}$ resolves the geometry $\mathbb{C}^2 / \mathbb{Z}_L$.
When the Cardy limit is taken in such a way that $\sum_{\alpha=0}^{K-1} \v_{\alpha} = L (\Delta_1 - \Delta_2) / 2 \to 0$ but the individual $\v_{\alpha}$ stay finite, there is an additional contribution%
\footnote{%
  This is explicitly shown for $\boldsymbol{b} = 0$ in section~\ref{sect:n1-n2-nonzero}.
  When $\boldsymbol{b} \ne 0$, only the calculations for $\boldsymbol{n}_1 = 0$ or $\boldsymbol{n}_2 = 0$ can be done explicitly.
  These, however, also fit with \eqref{eq:extra-term} as shown in section~\ref{sec:gen-bf}.
}
\begin{equation}
  \label{eq:extra-term}
  \exp \left[
    \frac{1}{2 L}
    \left(
      \frac{J(\boldsymbol{n}_1)}{\Delta_2} + \frac{J(\boldsymbol{n}_2)}{\Delta_1}
    \right)
    \v_{\alpha}^{+} (C^{-1})^{\alpha \beta} \v_{\beta}^{+} 
  \right],
\end{equation}
where $C_{\alpha \beta}$ is the $K \times K$ generalised Cartan matrix of affine $\widehat{\mathfrak{su}}(K)$ with components
\begin{equation}
	C_{\alpha\beta} = 2\delta_{\alpha\beta}-\delta_{\alpha,\beta+1}-\delta_{\alpha,\beta-1}
\end{equation}
and $(C^{-1})^{\alpha \beta}$ is its inverse, which is well-defined on the $(K-1)$-dimensional subspace of vectors whose components sum to zero.
Only the transverse part
\begin{equation}
  \v_{\alpha}^{+} \equiv \v_{\alpha} - \frac{1}{K} \sum_{\gamma=0}^{K-1} \v_{\gamma},
  \qquad
  \sum_{\alpha=0}^{K-1} \v_{\alpha}^{+} = 0,
\end{equation}
enters, so that $(C^{-1})^{\alpha \beta}$ only ever acts on this subspace.
In \cite{Dorey:2025qht}, a description of the coefficients $\mathcal{I}_{\boldsymbol{n}_1, \boldsymbol{n}_2}$ in terms of fermionic forms was proposed, which give the giant graviton coefficients an interpretation as vacuum characters of a sum of vertex algebras of the form
\begin{equation}
  \mathcal{V}(m, K) = \mathcal{H} \oplus \widehat{\mathfrak{su}}(K)_m \oplus W_{m,K}^{\text{para}}
\end{equation}
at central charge $c = K m$, where $W_{m,K}^{\text{para}}$ is the $K$-parafermionic $W_m$-algebra.
When $J(\boldsymbol{n}_I) \Delta_I \ll 1$, the equivariant volume expression reduces to the ordinary Cardy formula at the correct central charge.
On the other hand, when $J(\boldsymbol{n}_I) \Delta_I \gg 1$, the leading contribution of that same expression matches the expected result from taking the Higgs-branch limit of the master volume of the corresponding seven-dimensional auxiliary space.%
\footnote{%
  In particular, the leading order contributions for the large $N$-then-Higgs-branch limit and the Higgs-branch-then-Cardy limit are identical.
  It is expected that the subleading structure differs, as is well-known in the context of comparing large $N$ and Cardy limits, even if the results were to commute with the Higgs-branch limit.
}

The limit of the giant graviton coefficients considered here differs from the ordinary Cardy limit and anomaly polynomial result of the 6d index obtained from M5-brane giant gravitons (as used for example in the related works \cite{Chen:2026fpe, Hristov:2026tde}) in that it is highly asymmetric: $\omega, \Delta_3, \Delta_4 \to \infty$ while $\Delta_1, \Delta_2 \to 0$.

We will present our results in three stages, each introducing new features: first, $K = 1$, $\boldsymbol{b} = 0$ and general $L$, then general $K$ as well and finally general $K$, $L$ and $\boldsymbol{b}$.

\paragraph{$K = 1$:} In this case, the giant graviton coefficient is plethystic and the asymptotics are straightforwardly calculated to be
\begin{equation}
  \log \mathcal{I}_{n_1, n_2} \sim 
  \frac{\Li_3 (e^{L (n_1 \Delta_1 + n_2 \Delta_2)}) - \zeta(3)}{L \Delta_1 \Delta_2}.
  \label{eq:asymp-pe}
\end{equation}
The equivariant volume is given by
\begin{equation}
  \mathbb{V}_{\mathbb{C}^2 / \mathbb{Z}_L}(\boldsymbol{\lambda}; \epsilon_1, \epsilon_2) =
  \frac{L}{\epsilon_2 (L \epsilon_1 - \epsilon_2)} \exp
  \left[
    - \frac{1}{L} ((L \epsilon_1 - \epsilon_2) \lambda_1 + \epsilon_2 \lambda_2)
  \right]
\end{equation}
which, after a coordinate change, becomes
\begin{equation}
  \mathbb{V}_{\mathbb{C}^2 / \mathbb{Z}_L}(\boldsymbol{\lambda}; \bar{\epsilon}_1, \bar{\epsilon}_2) =
  \frac{1}{L \bar{\epsilon}_1 \bar{\epsilon}_2} \exp (- \bar{\epsilon}_1 \lambda_1 - \bar{\epsilon}_2 \lambda_2).
\end{equation}
Identifying
\begin{equation}
  \bar{\epsilon}_1 =  \Delta_1,
  \quad
  \bar{\epsilon}_2 =  \Delta_2
  \quad \text{and} \quad
  \lambda_I \equiv -J(\boldsymbol{n}_I) = -L n_I
\end{equation}
and using a formal series expansion of the trilogarithm, one finds that \eqref{eq:asymp-pe} is indeed equivalent to
\begin{equation}
  \mathcal{I}_{n_1, n_2} \sim \operatorname{PE} [\mathbb{V}_{\mathbb{C}^2 / \mathbb{Z}_L}(\boldsymbol{\lambda}; \epsilon_1, \epsilon_2) - \mathbb{V}_{\mathbb{C}^2 / \mathbb{Z}_L}(0; \epsilon_1, \epsilon_2)]. \label{eq:HB-main-result}
\end{equation}

\paragraph{General $K$:} For general $K$ (but $\boldsymbol{b} = 0$), the giant graviton coefficients are given in terms of fermionic forms \cite{Dorey:2025qht}.
A slightly lengthier calculation yields a similar result,
\begin{equation}
  \begin{aligned}
  \log \mathcal{I}_{\boldsymbol{n}_1, \boldsymbol{n}_2} \sim{}&
  \frac{K^2}{L \Delta_1 \Delta_2}
  \left[
    \Li_3(e^{L(n_1 \Delta_1 + n_2 \Delta_2)/K}) - \zeta(3)
  \right] \\
  &+
  \frac{1}{2 L}
  \left(
    \frac{J(\boldsymbol{n}_1)}{\Delta_2} +
    \frac{J(\boldsymbol{n}_2)}{\Delta_1}
  \right)
  \v_{\alpha}^{+} (C^{-1})^{\alpha \beta} \v_{\beta}^{+},
  \end{aligned}
  \label{eq:In1n2-intro}
\end{equation}
where the latter term is as explained above.
Just as before, the exponential of the trilogarithmic part can be formally resummed into the plethystic exponential of the equivariant volume, using the same equivariant variables as above except now $\lambda_I \to \lambda_I / K$.

In order to prepare for $\boldsymbol{b} \ne 0$, we will now set $\boldsymbol{n}_1 = 0$%
\footnote{%
As mentioned in a previous footnote, the lack of an explicit description of the giant graviton coefficients for $\boldsymbol{b} \ne 0$ and both $\boldsymbol{n}_{1,2} \ne 0$ requires us to set one of the configuration numbers to zero.
We will comment on the general case at the end of section~\ref{sect:n1-n2-nonzero}.}
and note that the trilogarithmic term in \eqref{eq:In1n2-intro} can also be rewritten as a contribution from $L$ fixed points,
\begin{equation}
  \frac{K^2}{L \Delta_1 \Delta_2}
  \left[
    \Li_3(e^{L n_2 \Delta_2/K}) - \zeta(3)
  \right]
  =
  - K^2 \sum_{A=0}^{L-1} \frac{1}{w_A w_{A+1}}
  \left[
    \Li_3 (e^{L \phi_{\boldsymbol{n}_2}^{(A)}/ K}) - \zeta(3)
  \right],
\end{equation}
where
\begin{equation}
  w_A = - L \Delta_1 + (L - A) \beta
  \quad \text{and} \quad
  L \phi_{\boldsymbol{n}_2}^{(A)} =
  w_0 n_2 = L n_2 \Delta_2
\end{equation}
independently of $A$.
In terms of the equivariant volume, we lift the two Kähler deformation parameters to $L+1$ parameters $\lambda_A$, $A = 0, \ldots, L$, given by
\begin{equation}
  \lambda_A = - \frac{A n_2}{K}.
\end{equation}
This expresses the equivariant volume of the orbifold using the Kähler parameters of its resolution, specialised to the point where all exceptional curves have zero area, but does not change the overall value.

\paragraph{General $K$ and $\boldsymbol{b}$:}
For the general result with $\boldsymbol{b} \sim \mathcal{O}(\Delta^{-1}) \ne 0$ and $\boldsymbol{n}_1 = 0$, we sort the components $n_{2,A}$ in ascending order, such that $0 \leq n_{2,0} \leq n_{2,1} \leq \ldots \leq n_{2,L-1}$.
The asymptotics are then
\begin{equation}
  \log \mathcal{I}_{0, \boldsymbol{n}_2} \sim
  - K^2 \sum_{A=0}^{L-1} \frac{1}{w_A w_{A+1}}
  \left[
    \Li_3 (e^{L \phi_{\boldsymbol{n}_2}^{(A)}/K}) - \zeta(3)
  \right]
  +
  \frac{1}{2 L}
    \frac{J(\boldsymbol{n}_2)}{\Delta_1}
  \v_{\alpha}^{+} (C^{-1})^{\alpha \beta} \v_{\beta}^{+},
\end{equation}
where $w_A$ as before but
\begin{equation}
  L \phi_{\boldsymbol{n}_2}^{(A)} = \sum_{0 \leq B \leq A}^{} w_B (n_{2,B} - n_{2,B-1})
\end{equation}
with $n_{2,-1} \equiv 0$.
When all $n_{2,A} \equiv n_2$ are equal, this reduces back to $L \phi_{\boldsymbol{n}_2}^{(A)} = w_0 n_2 = L n_2 \Delta_2$.
Re-expressing in terms of equivariant volume, the trilogarithmic contribution becomes
\begin{equation}
  \left.\mathcal{I}_{0, \boldsymbol{n}_2}\right|_{\text{tri}}
    = \operatorname{PE}
    \left[
      K^2 \mathbb{V}_{\mathbb{C}^2 / \mathbb{Z}_L}(\boldsymbol{\lambda}; \epsilon_1, \epsilon_2) -
      K^2 \mathbb{V}_{\mathbb{C}^2 / \mathbb{Z}_L}(0; \epsilon_1, \epsilon_2)
    \right]
\end{equation}
with $(\epsilon_1,\epsilon_2)=(\beta,L\Delta_2)$ and
\begin{equation}
  \lambda_A = - \frac{1}{K} \sum_{0 \leq B < A} n_{2,B},
  \quad A = 0, \ldots, L.
\end{equation}
These Kähler deformation parameters $\boldsymbol{\lambda}$ have the effect of resolving the $\mathbb{C}^2 / \mathbb{Z}_L$ orbifold singularity, as we elaborate on in section~\ref{sec:geom-int}.
Concretely, the areas of the exceptional curves $C_j \simeq \mathbb{P}^1$, $j = 1, \ldots, L-1$, are given by
\begin{equation}
  t_j = 2 \lambda_j - \lambda_{j-1} - \lambda_{j+1} = (b_j - b_{j-1})/K \geq 0
\end{equation}
and become non-zero if the corresponding baryonic flux is present.

The rest of the paper is organised as follows. In section~\ref{sec:GG-asymptotics} we review the Higgs-branch limit and the resulting giant graviton expansion, and derive the polylogarithmic Cardy asymptotics of the plethystic coefficients. In section~\ref{sec:equiv-volume} we turn to the geometric side.
After reviewing equivariant volumes of K\"ahler cones, we show their relation to these coefficients and analyse the fixed-wrapping and large-wrapping regimes. Section~\ref{sec:nonplethystic} contains the affine fermionic form analysis for $K>1$. We derive the asymptotics for equal and unequal background fluxes, and give their geometric interpretation in terms of the resolved cone. We also comment on the relation between the 4d equivariant volume $\mathbb{V}_{\mathbb{C}^2 / \mathbb{Z}_L}$ and the seven-dimensional master volume of \cite{Chen:2024erz} in the Higgs-branch limit. Section~\ref{sec:outlook} discusses the implications and possible extensions beyond the Higgs-branch regime.

\section{Higgs-branch giant graviton expansion} \label{sec:GG-asymptotics}

In this section we first review the Higgs-branch limit of the three-dimensional superconformal index and the resulting finite-$N$ giant graviton expansion. In this limit, the superconformal index reduces to the Hilbert series of the Higgs-branch moduli space \cite{Razamat:2014pta, Hayashi:2024aaf}, and its finite-$N$ expansion defines the Higgs-branch giant graviton coefficients \cite{Dorey:2025qht}. We then analyse the Cardy asymptotics of these coefficients, first in the smooth ADHM case and then for the orbifold $\mathbb{C}^2/\mathbb{Z}_L$.

For a $3$d $\caln=4$ theory with superconformal index \eqref{eq:SCI-defn}, the Higgs-branch limit sets the Coulomb branch fugacities to zero while keeping a subset of Higgs-branch fugacities fixed. We define
\begin{align}
	y = e^{-\omega},\qquad q_I = e^{-\Delta_I}\,,\qquad I=1,\dots,4,
\end{align}
so the BPS constraint \eqref{eq:SCI-constraint} becomes
\begin{equation}
	y = q_1 q_2 q_3 q_4.\label{eq:3d-constraint-fugacities}
\end{equation}
The Higgs-branch limit is
\begin{align}
	q_3,q_4,y \to 0\quad\text{with}\quad q_1,q_2~\text{fixed}, \label{eq:Higgs-limit-qs}
\end{align}
in such a way to preserve the constraint \eqref{eq:3d-constraint-fugacities}.

In this limit, the index \eqref{eq:SCI-defn} reduces to the graded character of the Higgs-branch moduli space $\calm_H$, namely its Hilbert series \cite{Dorey:2025qht}
\begin{equation}
	\lim_{q_3,q_4,y\to 0} \cali (y,q_I) = \mathrm{Hilb}[\mathcal M_H](q_1,q_2) \equiv H_N(q_1,q_2), \label{eq:HB-to-Hilb}
\end{equation} and we denote the giant graviton index in this limit by
\begin{equation}
	\cali_{n_1, n_2}(q_1, q_2) = \lim_{\mathrm{Higgs}} \mathcal{I}_{n_1, n_2, 0, 0}(y, q_I).
\end{equation}

\subsection{ADHM quiver}

The simplest example we consider is ADHM theory with one flavour, which is self-mirror and has Higgs and Coulomb branch $\mathcal{M}_H = \mathcal{M}_C = \operatorname{Sym}_N \mathbb{C}^2$. The generating function of the Hilbert series is thus plethystic:
\begin{equation}
	H(\Lambda) := \sum_{N=0}^\infty \Lambda^N H_N(q_1,q_2) = \mathrm{PE}\!\left[\Lambda\,H_1(q_1,q_2)\right] =\prod_{k_1,k_2\ge 0}\frac{1}{1-\Lambda q_1^{k_1}q_2^{k_2}} , \label{eq:plethystic-C2}
\end{equation}
where $\Lambda$ is a fugacity conjugate to $N$. Evaluating then $H_N$ by a contour integral in $\Lambda$ and summing residues away from the origin yields the finite-$N$ giant graviton expansion in the Higgs-branch regime \cite{Dorey:2025qht}:
\begin{equation}
	H_N(q_1,q_2) = H_{\infty}(q_1,q_2) \sum_{n_1, n_2 = 0}^{\infty} q_1^{N n_1} q_2^{N n_2} \cali_{n_1, n_2}(q_1,q_2),
\end{equation}
with
\begin{equation}
	H_\infty(q_1,q_2) = \prod_{\substack{k_1, k_2 = 0 \\(k_1, k_2) \not= (0,0)}}^{\infty} \frac{1}{1 - q_1^{k_1} q_2^{k_2}} \label{eq:Hinfinity}
\end{equation}
and
\begin{equation}
	H_\infty(q_1,q_2) \cali_{n_1, n_2}(q_1,q_2) = \prod_{\substack{k_1 = -n_1, k_2 = -n_2 \\(k_1, k_2) \not= (0,0)}}^{\infty} \frac{1}{1 - q_1^{k_1} q_2^{k_2}}. \label{eq:Hinfinity-Zn1n2}
\end{equation}

From \eqref{eq:Hinfinity} and \eqref{eq:Hinfinity-Zn1n2}, the giant graviton coefficients are obtained as
\begin{equation}
	\cali_{n_1,n_2} = \prod_{(k_1,k_2) \in S_{n_1,n_2}\setminus S_0} \frac{1}{1-q_1^{k_1} q_2^{k_2}}, \label{eq:GG-index-ADHM-implicit}
\end{equation}
where
\begin{align}
	S_0         & = \{ (k_1,k_2) \in \mathbb{Z}^2~ |~ k_1 \geq 0, k_2 \geq 0, (k_1,k_2) \neq (0,0)\}         \\
	S_{n_1,n_2} & = \{(k_1,k_2) \in \mathbb{Z}^2 ~|~ k_1 \geq -n_1, k_2 \geq -n_2 , (k_1,k_2) \neq (0,0) \}.
\end{align}
The difference set $S_{n_1,n_2}\setminus S_0$ is a set of lattice points with at least one negative index. So \eqref{eq:GG-index-ADHM-implicit} splits into three disjoint pieces
\begin{equation}
	\cali_{n_1,n_2} (q_1,q_2)= \calb_1 (q_1,q_2) \calb_2 (q_1,q_2) \calb_3 (q_1,q_2) \label{eq:ADHM-GG-explicit}
\end{equation} where
\begin{equation}
	\calb_i (q_1,q_2) = \prod_{(k_1,k_2)\in R_i} \frac{1}{1-q_1^{k_1}\,q_2^{k_2}}, \qquad i=1,2,3, \label{eq:ADHM-blocks-uniform}
\end{equation}
with
\begin{align}
	R_1 & =\{(k_1,k_2)\mid k_1=-n_1,\dots,-1,\ \ k_2=0,1,2,\dots\} \label{eq:region-R1}    \\
	R_2 & =\{(k_1,k_2)\mid k_1=0,1,2,\dots,\ \ k_2=-n_2,\dots,-1\} \label{eq:region-R2}    \\
	R_3 & =\{(k_1,k_2)\mid k_1=-n_1,\dots,-1,\ \ k_2=-n_2,\dots,-1\}. \label{eq:region-R3}
\end{align}

%For $n_2 = 0$,
%\begin{equation}
%\cali_{n_1,0}(q_1,q_2) = \prod_{k_1 = 1}^{n_1} \prod_{k_2 = 0}^{\infty}\frac{1}{1 - q_1^{-k_1} q_2^{k_2}} = \prod_{k_1 = 1}^{n_1} \frac{1}{(q_1^{-k_1}; q_2)_{\infty}}. \label{eq:ADHM-GG-index}
%\end{equation}
%and similarly for $n_1=0$,
%\begin{equation}
%\cali_{0,n_2}(q_1,q_2) = \prod_{k_1 = 0}^{\infty} \prod_{k_2 = 1}^{n_2} \frac{1}{1 - q_1^{k_1} q_2^{-k_2}} =\prod_{k_2 = 1}^{n_2} \frac{1}{(q_2^{- k_2}; q_1)_{\infty}}. \label{eq:ADHM-GG-index-m2}
%\end{equation}

We now study the giant graviton free energy $\log \cali_{n_1,n_2}$ in a Cardy regime
\begin{equation}
	\Delta_1,\Delta_2 \to 0^+ \qquad \qquad q_1,q_2 \to 1^-, \label{eq:HB-Cardy}
\end{equation}
which probes the singular behaviour of the Higgs-branch generating function.

From \eqref{eq:GG-index-ADHM-implicit} we write\footnote{Polylogarithms are defined by $\Li_{s}(z) = \sum_{k=1}^{\infty} \frac{z^k}{k^s}$. They satisfy $z\partial_z \Li_s (z)=\Li_{s-1}(z)$ and $\Li_1(z) = - \log (1 - z)$.}
\begin{equation}
	\log \cali_{n_1,n_2}(q_1,q_2) = -\sum_{(k_1,k_2)\in S_{n_1,n_2}\setminus S_0}\log \left(1-q_1^{k_1}q_2^{k_2}\right) = \sum_{(k_1,k_2)\in S_{n_1,n_2}\setminus S_0}\Li_1 \left(q_1^{k_1}q_2^{k_2}\right) . \label{eq:log-to-Li1}
\end{equation}

It is convenient to treat the three blocks in \eqref{eq:ADHM-GG-explicit} separately. To extract the leading Cardy behaviour, we retain only the bulk Euler--Maclaurin term and replace each lattice sum by the integral over the corresponding continuum region\footnote{A tempting alternative approach is to use the expansion $-\log(1-x)=\sum_{\ell\ge1}x^\ell/\ell$, which converges for $|x|<1$. In the Higgs-branch giant graviton products, however, the factors take the form $x=q_1^{k_1}q_2^{k_2}$ with $k_1<0$ in the vertical strip (and similarly $k_2<0$ in the horizontal strip), and for sufficiently small $k_2$ one has $x>1$. Thus the $\ell$-expansion does not converge termwise on the full summation domain.}. The boundary terms omitted in the Euler--Maclaurin expansion are parametrically subleading in the Cardy regime. Moreover, for $\Delta_i>0$ the arguments of the polylogarithms cross the branch cut $[1,\infty)$, so the resulting expressions are understood by analytic continuation from the convergent region. %In particular, expressions such as $\Li_s(e^x)$ with $x>0$ are understood in this sense throughout.

For the vertical strip $\mathcal B_1$, we then obtain
\begin{align}
	\log \mathcal B_1 & \sim \int_0^{n_1} dk_1 \int_0^\infty dk_2  \Li_1 \left(e^{\Delta_1 k_1-\Delta_2 k_2}\right) \\
	                  & = \frac{1}{\Delta_2}\int_0^{n_1} dk_1  \Li_2 \left(e^{\Delta_1 k_1}\right)                  \\
	                  & = \frac{\Li_3(e^{\Delta_1 n_1})-\zeta(3)}{\Delta_1\Delta_2}. \label{eq:B1-cardy}
\end{align}

By symmetry $(1\leftrightarrow2)$, the horizontal strip $\mathcal B_2$ gives
\begin{equation}
	\log \mathcal B_2 \sim \frac{\Li_3(e^{\Delta_2 n_2})-\zeta(3)}{\Delta_1\Delta_2}. \label{eq:B2-cardy}
\end{equation}

Finally, the third block $\mathcal B_3$ contributes
\begin{align}
	\log \mathcal B_3 & \sim \int_0^{n_1} dk_1 \int_0^{n_2} dk_2 \Li_1 \left(e^{\Delta_1 k_1+\Delta_2 k_2}\right)                                                                                                          \\
	                  & = \frac{1}{\Delta_2}\int_0^{n_1} dk_1 \left[ \Li_2\left(e^{\Delta_1 k_1+\Delta_2 n_2}\right) -\Li_2\left(e^{\Delta_1 k_1}\right) \right]                                                           \\
	                  & = \frac{1}{\Delta_1\Delta_2} \left[ \Li_3\left(e^{\Delta_1 n_1+\Delta_2 n_2}\right) -\Li_3\left(e^{\Delta_2 n_2}\right) -\Li_3\left(e^{\Delta_1 n_1}\right) +\zeta(3) \right]. \label{eq:B3-cardy}
\end{align}

Adding these,\footnote{The closed-form Cardy formula \eqref{eq:log-In1n2-Cardy} can also be recovered from the fermionic form representation of the ADHM coefficients \cite{Dorey:2025qht} by a saddle-point analysis. There the effective action decouples mode by mode, which is special to the plethystic case. For $K>1$, the corresponding fermionic forms couple the different colour components through the Cartan data, and this loss of decoupling is the main technical obstruction to extending beyond the plethystic regime.}
\begin{equation}
	\log \cali_{n_1,n_2} (q_1,q_2) \sim \frac{\Li_3(e^{\mu})-\zeta(3)}{\Delta_1\Delta_2},  \label{eq:log-In1n2-Cardy}
\end{equation} where we introduced
\begin{equation}
	\mu = \Delta_1 n_1 + \Delta_2 n_2. \label{eq:mu-parameter-defn}
\end{equation}

Lastly, let us discuss the Euler--Maclaurin corrections more precisely. For example, for the strip contribution $\mathcal B_1$ the first corrections take the form
\begin{equation}
	\log \mathcal B_1 = \frac{\Li_3(e^{\mu_1})-\zeta(3)}{\Delta_1\Delta_2} +\frac{\Li_2(e^{\mu_1})-\zeta(2)}{2\Delta_2} +\frac{\Li_2(e^{\mu_1})-\zeta(2)}{2\Delta_1} +\cdots, \qquad \mu_1=\Delta_1 n_1, \label{eq:B1-EM-refined}
\end{equation}
where the first term is the bulk contribution retained above, and the next two terms are the leading boundary corrections\footnote{The omitted terms can be similarly shown to be subleading relative to the bulk contribution by estimating the difference between the exact lattice sum and its continuum integral for generic fugacity ratios.}.

Then, with $n_1$ fixed and $\Delta_{1,2}\to0^+$, we have $\mu_1=\Delta_1 n_1\to0$ and
\begin{equation}
	\Li_3(e^\mu)-\zeta(3)=\zeta(2)\mu+\mathcal O(\mu^2\log\mu), \qquad \Li_2(e^\mu)-\zeta(2)=\mathcal O(\mu\log\mu). \label{eq:Li-small-mu-EM}
\end{equation}
For $\mathcal B_1$, the bulk term behaves as
\begin{equation}
	\frac{\Li_3(e^{\mu_1})-\zeta(3)}{\Delta_1\Delta_2} = \frac{\pi^2}{6}\frac{n_1}{\Delta_2} +\mathcal O\!\left(\frac{\Delta_1}{\Delta_2}\log\Delta_1\right),
\end{equation}
whereas the first Euler--Maclaurin corrections behave as
\begin{equation}
	\frac{\Li_2(e^{\mu_1})-\zeta(2)}{2\Delta_2} = \mathcal O\!\left(\frac{\Delta_1}{\Delta_2}\log\Delta_1\right), \qquad \frac{\Li_2(e^{\mu_1})-\zeta(2)}{2\Delta_1} = \mathcal O(\log\Delta_1).
\end{equation}
Hence, in the limit $\Delta_1\sim\Delta_2\sim\delta\to0^+$, the bulk contribution is of order $\delta^{-1}$, while the first Euler--Maclaurin corrections are at most logarithmic.

If instead $\mu_1$ is held fixed and positive in the Cardy limit, then both $\Li_3(e^{\mu_1})-\zeta(3)$ and $\Li_2(e^{\mu_1})-\zeta(2)$ are of order one, so the bulk term scales as $1/(\Delta_1\Delta_2)$ whereas the first boundary corrections scale only as $1/\Delta_i$, and are therefore subleading. Finally, in the large-$\mu_1$ regime, the bulk term scales as $\mu_1^3/(\Delta_1\Delta_2)$, while the first boundary corrections scale at most as $\mu_1^2/\Delta_i$, and are again subleading. The same conclusion applies to $\mathcal B_2$ and $\mathcal B_3$. Thus the Euler--Maclaurin corrections do not modify the leading Cardy behaviour \eqref{eq:log-In1n2-Cardy}.

\subsection{\texorpdfstring{$A_{L-1}$}{A\_L-1} circular quiver}

We now move to the orbifold Higgs-branch Calabi--Yau cone $\mathbb{C}^2/\mathbb{Z}_L$.  In the plethystic case $K=1$, the Higgs-branch moduli space is $\operatorname{Sym}_N(\mathbb{C}^2/\mathbb{Z}_L)$, and the corresponding Hilbert series admits a finite-$N$ giant graviton expansion organised into $\mathbb{Z}_L$ sectors \cite{Dorey:2025qht}:
\begin{align}
	H_\infty(q_1,q_2)\,\cali_{n,n_1,n_2}(q_1,q_2) & =\prod_{k=0}^{L-1}\ \prod_{\substack{k_1\ge -n_1,\ k_2\ge -n_2 \\ (k,k_1,k_2)\neq(n,0,0)}} \frac{1}{1-q_1^{Lk_1+(k-n)}\,q_2^{Lk_2+(k-n)}}, \label{eq:orb-Hinf-Z} \\
	H_\infty(q_1,q_2)                             & =\prod_{k=0}^{L-1}\ \prod_{\substack{k_1,k_2\ge 0              \\ (k,k_1,k_2)\neq(0,0,0)}} \frac{1}{1-q_1^{Lk_1+k}\,q_2^{Lk_2+k}}. \label{eq:orb-Hinf}
\end{align}
Note that the corresponding coefficients carry, in addition to $(n_1,n_2)$, a discrete label $n\in\{0,1,\dots,L-1\}$. The role of $n$ is to single out a distinguished $\mathbb{Z}_L$ residue class in the product, through both the shift $(k-n)$ and the exclusion of the factor at $(k,k_1,k_2)=(n,0,0)$. For concreteness, we focus on the $n=0$ sector in what follows.\footnote{This choice should not be confused with the simple-sum truncation \cite{Fujiwara:2023bdc}, which amounts to expanding in one fugacity first and effectively restricting to sectors with $n_1=0$ or $n_2=0$.} The other sectors can be treated by the same method after accounting for the shifts in the summation bounds under cyclic relabelling.

The giant graviton coefficients then become
\begin{align}
	\cali_{n_1,n_2}(q_1,q_2) & = \prod_{k=0}^{L-1} \Bigg[ \prod_{k_1=-n_1}^{-1}\prod_{k_2=0}^{\infty} \frac{1}{1-q_1^{Lk_1+k}\,q_2^{Lk_2+k}} \prod_{k_1=0}^{\infty}\prod_{k_2=-n_2}^{-1} \frac{1}{1-q_1^{Lk_1+k}\,q_2^{Lk_2+k}} \nonumber \\ &\qquad\quad\;\;\prod_{k_1=-n_1}^{-1}\prod_{k_2=-n_2}^{-1} \frac{1}{1-q_1^{Lk_1+k}\,q_2^{Lk_2+k}} \Bigg], \label{eq:orbifold-3blocks}
\end{align}
which is the orbifold analogue of the three-block factorisation \eqref{eq:ADHM-GG-explicit}, and we similarly write it as
\begin{equation}
	\cali_{n_1,n_2}(q_1,q_2)= \calb^{(L)}_1 \calb^{(L)}_2 \calb^{(L)}_3, \qquad \calb^{(L)}_i =\prod_{k=0}^{L-1}\ \prod_{(k_1,k_2)\in R_i} \frac{1}{1-q_1^{Lk_1+k}\,q_2^{Lk_2+k}}, \qquad i=1,2,3. \label{eq:orb-blocks-uniform}
\end{equation}

Keeping again only the bulk Euler--Maclaurin term, we replace the orbifold lattice sums by integrals over the corresponding continuum regions. Proceeding exactly as in the smooth case, we obtain
\begin{align}
	\log \mathcal{B}^{(L)}_1 &\sim \frac{1}{L\Delta_1\Delta_2}\left(\Li_3(e^{L\Delta_1 n_1})-\zeta(3)\right), \label{eq:orb-B1} \\
	\log \mathcal{B}^{(L)}_2 &\sim \frac{1}{L\Delta_1\Delta_2}\left(\Li_3(e^{L\Delta_2 n_2})-\zeta(3)\right), \label{eq:orb-B2} \\
	\log \mathcal{B}^{(L)}_3 &\sim \frac{1}{L\Delta_1\Delta_2} \left( \Li_3(e^{L(\Delta_1 n_1+\Delta_2 n_2)}) -\Li_3(e^{L\Delta_1 n_1}) -\Li_3(e^{L\Delta_2 n_2}) +\zeta(3) \right) .  \label{eq:orb-B3}
\end{align}
Adding the three contributions gives the leading orbifold Cardy formula
\begin{equation}
	\log \cali_{n_1,n_2}(q_1,q_2) \sim \frac{\Li_3(e^{\mu L})-\zeta(3)}{L\,\Delta_1\Delta_2}, \label{eq:orb-cardy-li3}
\end{equation} where $\mu$ is the parameter defined in \eqref{eq:mu-parameter-defn}.

\section{Equivariant volume and giant graviton asymptotics} \label{sec:equiv-volume}

We now switch gears and turn to the geometric side of the story. We begin by briefly recalling the toric setting and the relation between equivariant volume and master volume for general K\"ahler cones, following \cite{Martelli:2023oqk}, and then specialise to the Higgs-branch cone geometries. Finally, we show that in the Cardy regime the leading asymptotics of the Higgs-branch giant graviton coefficients admit a plethystic representation in terms of the equivariant volume of the Higgs-branch cone.

Let $(M_{2m},\omega)$ be a Kähler $2m$-manifold with a Hamiltonian $T^m$-action generated by the
Reeb vector
\begin{equation}
	\xi = \sum_{i=1}^m \epsilon_i\,\partial_{\varphi_i},
\end{equation}
with the Hamiltonian
\begin{equation}
	H = \sum_{i=1}^m \epsilon_i\,y_i\,,\qquad
	i_\xi\omega = -dH.
\end{equation}
Here, $\varphi_i$ are angular coordinates on the $T^m$ orbits with periodicities $\varphi_i\sim \varphi_i+2\pi$, and $y_i$ are the moment map coordinates\footnote{The $y_i$ in this section are moment-map coordinates and should not be confused with the colour fugacities $y_\alpha$ used
	later.} satisfying
\begin{equation}
	i_{\partial_{\varphi_i}} \omega = - dy_i.
\end{equation}

The equivariant volume of $M$ is defined as
\begin{equation}
	\Veq(\lambda_I,\epsilon_i)
	\;=\;\frac{1}{(2\pi)^m}
	\int_{M_{2m}} e^{-H}\,\frac{\omega^m}{m!}
	\;=\;(-1)^m\int_{M_{2m}}e^{-H-\omega/2\pi}\,.
	\label{eq:V-def}
\end{equation}
K\"ahler deformations are described by parameters
$\lambda_I$ that shift the facets of the moment polyhedron $\calp$. The equivariant volume integral can alternatively be written as
\begin{equation}
	\Veq = \int_{\calp} e^{ -\sum_i \epsilon_i y_i } \, dy_1 dy_2 \cdots dy_m.
\end{equation}

The master volume $\calv$ is the degree-\((m-1)\) term in the
$\lambda_I$ expansion of $\Veq$\footnote{Though we won't be using it, let us note that a distinguished role is played also by the supersymmetric action $S_{\rm SUSY}$ in the description of on-shell solutions and near-horizon geometries, which corresponds to the degree-$(m-2)$ term in the expansion \eqref{eq:Veq-expansion} \cite{Martelli:2023oqk}.}:
\begin{equation}
	\Veq(\lambda_I,\epsilon_i) = \sum^\infty_{k=0} \Veq^{(k)} (\lambda_I,\epsilon_i). \label{eq:Veq-expansion}
\end{equation}
More precisely,
\begin{align}
	\calv(\lambda_I,\epsilon_i) & = (2\pi)^m \Veq^{(m-1)} (\lambda_I,\epsilon_i) \label{eq:calv-Veq-relation} \\ \Veq^{(m-1)} (\lambda_I,\epsilon_i) &= \frac{1}{(m-1)!}\sum_{a_1,\dots,a_{m-1}}
	\lambda_{a_1}\cdots\lambda_{a_{m-1}}
	\left.\frac{\partial^{m-1}\Veq}{\partial\lambda_{a_1}\cdots\partial\lambda_{a_{m-1}}}\right|_{\lambda=0},
	\label{eq:Veq-expansion-terms}
\end{align}
and this agrees with the geometric definition \cite{Gauntlett:2019roi, Martelli:2023oqk}
\begin{equation}
	\calv (\lambda_I,\epsilon_i)
	=
	\int_Y \eta \wedge \frac{\omega_B^{m-1}}{(m-1)!}\,,
	\label{eq:master-geom}
\end{equation}
where $\eta$ is the contact 1-form dual to the Reeb vector, and $Y$ is the link of the cone, with the metric\footnote{Note that the metric on the full cone is $ds^2 = dr^2 + r^2 ds^2_Y$.}
\begin{equation}
	ds^2_Y=\eta^2+ds^2_{2m-2},
\end{equation}
and the transverse Kähler form $\omega_B$, which is given by a linear combination of K\"ahler parameters $\lambda_I$. We refer to \cite{Martelli:2023oqk} for the proof and more details of this construction.

We now specialise to the Higgs-branch cone geometries, namely $\mathbb{C}^2$ and its orbifolds $\mathbb{C}^2/\mathbb{Z}_L$.

For $\mathbb{C}^2$ with the standard $T^2$ action and K\"ahler deformations $\lambda_{1,2}$ shifting the two moment-map facets, the equivariant volume is
\begin{equation}
	\Veq_{\mathbb{C}^2} (\lambda_1,\lambda_2;\epsilon_1,\epsilon_2) = \frac{1}{\epsilon_1\epsilon_2} \exp\left[-\left(\epsilon_1 \lambda_1 + \epsilon_2 \lambda_2\right)\right].  \label{eq:Veq-C2}
\end{equation}
Expanding in $\lambda_{1,2}$ gives
\begin{equation}
	\Veq_{\mathbb{C}^2} = \frac{1}{\epsilon_1 \epsilon_2} - \left( \frac{\lambda_1}{\epsilon_2} + \frac{\lambda_2}{\epsilon_1}\right) + \calo(\lambda^2),
\end{equation} so that $\Veq^{(1)}$ reproduces the master volume of the link $Y_3 \simeq S^3$.

For the orbifold cone $\mathbb{C}^2/\mathbb{Z}_L$ of type $A_{L-1}$,
\begin{align}
	\Veq_{\mathbb{C}^2/\mathbb{Z}_L}(\lambda_1,\lambda_2;\epsilon_1,\epsilon_2) & = \frac{L}{\epsilon_2 (L\epsilon_1-\epsilon_2)} \exp\left[ -\frac{1}{L}\left((L\epsilon_1-\epsilon_2)\lambda_1+\epsilon_2\lambda_2\right)\right] \label{eq:Veq-orb} \\ &= \frac{1}{L\bar{\epsilon}_1\bar{\epsilon}_2}\exp\left(-\bar{\epsilon}_1 \lambda_1 -\bar{\epsilon}_2 \lambda_2\right) , \label{eq:VC2-ZL}
\end{align}
where the second expression is obtained via the coordinate change which relates the Reeb components as\footnote{The parameters $(\epsilon_1,\epsilon_2)$ and $(\bar\epsilon_1,\bar\epsilon_2)$ describe the same Reeb vector $\xi$, but expressed in two different bases of angular generators. For the orbifold cone $\mathbb{C}^2/\mathbb{Z}_L$ it is useful to distinguish (i) the globally-defined $2\pi$-periodic angles $(\varphi_1,\varphi_2)$ generating the $T^2$ action on the quotient and (ii) a covering space parametrisation in terms of angles $(\phi_1,\phi_2)$ on $\mathbb{C}^2$. Writing $\xi=\epsilon_i\partial_{\varphi_i}=\bar\epsilon_i\partial_{\phi_i}$, the two sets of components are related by the induced linear change of basis \eqref{eq:epsilon-gauge-transfo}, see appendix~\ref{app:equiv-vol}.}
\begin{equation}
	\bar{\epsilon}_1 = \epsilon_1 - \frac{\epsilon_2}{L} \qquad \qquad \bar{\epsilon}_2 = \frac{\epsilon_2}{L} .\label{eq:epsilon-gauge-transfo}
\end{equation}

Equivariant volumes of toric K\"ahler cones (and their orbifolds), including \eqref{eq:Veq-C2}, \eqref{eq:Veq-orb}, and \eqref{eq:VC2-ZL}, follow from the fixed-point localisation formula of \cite{Martelli:2023oqk}. For completeness, we also reproduce these expressions directly from \eqref{eq:V-def} in appendix~\ref{app:equiv-vol}.

Let us now connect the two strands of the analysis: the Cardy regime \eqref{eq:HB-Cardy} asymptotics of the Higgs-branch giant graviton coefficients derived in section~\ref{sec:GG-asymptotics}, and the equivariant volume of K\"ahler cones reviewed above.%, by verifying \eqref{eq:HB-main-result}. 

From \eqref{eq:VC2-ZL}, identifying $(\bar{\epsilon}_1,\bar{\epsilon}_2)=(\Delta_1, \Delta_2)$, we obtain
\begin{equation}
\sum_{\ell=1}^{\infty}\frac{1}{\ell}
\left[
\mathbb V_{\mathbb{C}^2/\mathbb{Z}_L}(\lambda_1,\lambda_2;\ell\epsilon_1,\ell\epsilon_2)-\mathbb V_{\mathbb{C}^2/\mathbb{Z}_L}(0,0;\ell\epsilon_1,\ell\epsilon_2) \right]
= \sum_{\ell=1}^{\infty}\frac{e^{-\ell(\Delta_1\lambda_1+\Delta_2\lambda_2)}-1}{\ell^3 L\Delta_1\Delta_2},\label{eq:C2-ell-sum-identity}
\end{equation}
which matches precisely the Cardy-regime giant graviton free energy \eqref{eq:orb-cardy-li3} upon the geometric identification $\lambda_i=-Ln_i$.\footnote{Note that in the orbifold coefficient, the allowed fugacity weights lie on the shifted sublattice $(\alpha,\beta)=(Lk_1+k,Lk_2+k)$. Hence, increasing a wrapping number $n_i$ by one moves the boundary of the counting region by $L$ units in the corresponding weight. Since the K\"ahler parameters $\lambda_i$ in the equivariant volume measure precisely this displacement, the geometric identification is $\lambda_i=-Ln_i$.} For $L=1$ this reduces to the smooth-case relation following from \eqref{eq:log-In1n2-Cardy} and \eqref{eq:Veq-C2}. Thus, within the Higgs-branch Cardy regime, the giant graviton coefficients admit a resummed representation in terms of the equivariant volume of the Higgs-branch cone. Exponentiating then gives the plethystic form \eqref{eq:HB-main-result}. The subtraction of $\Veq(0,0;\ell\epsilon_1,\ell\epsilon_2)$ isolates the contribution induced by the Kähler deformation parameters $\lambda_I$. In particular, this ensures that the exponent vanishes at $n_1=n_2=0$, as required by the normalisation of the giant graviton coefficients \cite{Imamura:2022aua}.

The Cardy formula \eqref{eq:orb-cardy-li3} admits two distinct asymptotic regimes, depending on the behaviour of the combination \eqref{eq:mu-parameter-defn}. If the wrapping numbers are kept fixed while $\Delta_i\to0^+$, then $\mu\to0$, leading to a fixed-wrapping Cardy limit. If instead the wrapping numbers scale so that $\mu \gg 1$, one obtains a large-wrapping limit. We now discuss these two regimes in turn.

\subsection{Fixed-wrapping Cardy asymptotics}

We first study the regime in which the wrapping numbers are kept fixed while $\Delta_i \to 0^+$. In this regime, the asymptotic behaviour of \eqref{eq:orb-cardy-li3} is governed by the small-argument expansion of the trilogarithm,
\begin{equation}
	\Li_3(e^x) \sim \zeta(3) + \frac{\pi^2}{6} x + \calo(x^2 |\log x|), \qquad x \to 0^+.
\end{equation}

Restricting to the simple-sum case with $n_1=0$ and $n_2=m$, we take the fixed-wrapping Cardy limit at a fixed generic ratio of $\Delta_1$ and $\Delta_2$. Then $\mu=\Delta_2m\to0$, and the subtraction by $\zeta(3)$ in \eqref{eq:orb-cardy-li3} removes the constant term in the small-$\mu$ expansion. We obtain
\begin{equation}
	\log \cali_{0,m} \sim \frac{\pi^2}{6} \frac{m}{\Delta_1},
\end{equation}
which is independent of the orbifold order $L$, since the factor of $L$ in the argument of the trilogarithm cancels against the overall prefactor in \eqref{eq:orb-cardy-li3}.

In the plethystic case $K=1$, this agrees with the vertex algebra interpretation of the Higgs-branch giant graviton coefficients \cite{Dorey:2025qht}. In the unrefined specialisation $q=q_1q_2=1$, the coefficient reduces to a product of $L$ vacuum characters of the direct sum of a Heisenberg algebra $\mathcal{H}$ and the $W$-algebra $W_m$,
\begin{equation}
	V=\mathcal{H}\oplus W_m,
\end{equation}
with central charge $c(V)=m$ for each copy and modular variable $L \Delta_1$.

\subsection{Large-wrapping asymptotics}

We now consider the large-wrapping regime in which the wrapping numbers $n_i$ are taken large together with the Cardy limit so that $\mu \gg 1$. In this regime, the asymptotic behaviour of the Higgs-branch giant graviton coefficients is governed by the large-argument expansion of the trilogarithm in \eqref{eq:orb-cardy-li3}.

Using the standard asymptotic expansion of the polylogarithm for large positive argument \cite{wood:1992}, we have\footnote{For real $x>0$, $\Li_3(e^x)$ denotes the upper boundary value on the positive real branch cut.}
\begin{equation}
	\Li_3(e^{x}) \sim -\frac{x^3}{6} + \frac{i\pi x^2}{2} +\frac{\pi^2 x}{3} + \mathcal{O}\left(e^{-x}\right), \qquad x \to+\infty.  \label{eq:Li3-asymp}
\end{equation}
Substituting this into \eqref{eq:orb-cardy-li3} gives
\begin{align}
	\log \cali_{n_1,n_2}(q_1,q_2) & \sim \frac{1}{L\,\Delta_1\Delta_2} \left[-\frac{(L\mu)^3}{6} +\frac{i\pi(L\mu)^2}{2} +\frac{\pi^2}{3}L\mu +\mathcal{O}(1) \right] \\ &= -\frac{L^2\mu^3}{6\,\Delta_1\Delta_2}  +\frac{i\pi L \mu^2}{2\Delta_1\Delta_2} +\frac{\pi^2 \mu}{3 \Delta_1 \Delta_2}+\mathcal{O}\left(\frac{1}{L \,\Delta_1\Delta_2}\right), \qquad \mu\to+\infty . \label{eq:large-wrap-orb}
\end{align}

As mentioned in the introduction, the giant-graviton/volume-minimisation correspondence relates the \emph{leading} large-wrapping asymptotics of the giant graviton coefficients to the master volume of the corresponding off-shell family $Y_{2m-1}$. In the present Higgs-branch setting, the Cardy-regime formula allows us to access a systematic tower of subleading terms.

Let us now compare the leading term in \eqref{eq:large-wrap-orb} with the earlier giant graviton/master-volume proposal. For AdS$_4\times \text{SE}_7$, the large wrapping asymptotics conjectured in \cite{Chen:2024erz} take the form
\begin{equation}
	\log \cali_{(n_I)}(\omega,\Delta_I) \sim -\frac{\omega^2}{4(2\pi)^4}  \, \calv_{Y_7}(\Delta_I,\lambda_I) \label{eq:GG-master-volume-AdS4}
\end{equation}
and were explicitly verified in the $\text{SE}_7=S^7$ case.

For $K=1$, the parent cone is $X_4=(\mathbb C^2/\mathbb Z_L)\times\mathbb C^2$. Denoting its deformation parameters by $\Lambda_I$, the master volume is
\begin{equation}
\calv_{Y_7} = \frac{(2\pi)^4}{6L} \frac{\left(\sum_{I=1}^4\Delta_I\Lambda_I\right)^3}{\Delta_1\Delta_2\Delta_3\Delta_4},
\end{equation}
which is the $K=1$ case of the formula we derive in section~\ref{sec:parent-master-volume}. Matching the classical wrapping weight gives $(\Lambda_1,\Lambda_2,\Lambda_3,\Lambda_4) =(Ln_1,Ln_2,0,0)$, and so
\begin{equation}
\calv_{Y_7} = \frac{(2\pi)^4L^2\mu^3}{6\Delta_1\Delta_2\Delta_3\Delta_4}, \qquad
\mu=n_1\Delta_1+n_2\Delta_2.
\end{equation}
In the Higgs-branch limit, $\omega-\Delta_3-\Delta_4$ remains finite and $\Delta_3/\Delta_4\to1$, so
\begin{equation}
\frac{\omega^2}{\Delta_3\Delta_4}\longrightarrow 4,
\qquad
-\frac{\omega^2}{4(2\pi)^4}\calv_{Y_7} \longrightarrow -\frac{L^2\mu^3}{6\Delta_1\Delta_2},
\end{equation}
which reproduces the cubic term in (\ref{eq:large-wrap-orb}). So the leading large-wrapping behaviour of the Higgs-branch giant graviton coefficients is indeed reproduced by the Higgs-branch reduction of (\ref{eq:GG-master-volume-AdS4}).

To identify the subleading linear term in (\ref{eq:large-wrap-orb}), we use the K\"ahler-parameter expansion of the equivariant volume of the Higgs-branch cone:
\begin{equation}
\Veq_{\mathbb{C}^2/\mathbb{Z}_L} = \Veq^{(0)}_{\mathbb{C}^2/\mathbb{Z}_L} +\Veq^{(1)}_{\mathbb{C}^2/\mathbb{Z}_L} + \mathcal{O}(\lambda^2),
\end{equation}
where the degree-one term $\Veq^{(1)}_{\mathbb{C}^2/\mathbb{Z}_L}$ is, up to the overall normalisation in \eqref{eq:calv-Veq-relation}, the master volume of the three-dimensional link $Y_3$ of the Higgs-branch cone. For $\mathbb{C}^2/\mathbb{Z}_L$, this link is the lens space
\begin{equation}
	Y_3 \simeq S^3/\mathbb{Z}_L ,
\end{equation}
so the linear term in (\ref{eq:large-wrap-orb}) is 
\begin{equation}
  \frac{\pi^2\mu}{3\Delta_1\Delta_2}
  = \frac{1}{12}\calv_{Y_3}.
\end{equation}

Thus, in the Higgs-branch Cardy regime, the leading large-wrapping asymptotics of the giant graviton coefficients are still governed by the master volume of the parent seven-dimensional geometry, in agreement with the earlier proposal, while the subleading corrections are determined by the full equivariant-volume resummation. %is organised by the equivariant volume of the Higgs-branch cone.

\section{\texorpdfstring{$U(N)^L$}{U(N)\textasciicircum L} gauge theory with \texorpdfstring{$K$}{K} flavours} \label{sec:nonplethystic}

We now consider the Higgs-branch limit of the superconformal index of the $U(N)^L$ quiver gauge theory with $K > 1$ flavours.
It was found in \cite{Dorey:2025qht} that the giant graviton coefficients admit a representation in terms of affine fermionic forms associated with the $\widehat{\mathfrak{su}}(K)$ affine Lie algebra.
Concretely, for baryonic background flux $\boldsymbol{m} \in [\boldsymbol b]_+$, the giant graviton coefficients $\mathcal{I}_{0,\boldsymbol{m}}$ are given by
\begin{equation}
	\mathcal{I}_{0,\boldsymbol{m}} = \frac{n(\boldsymbol{m}, q, \vec{y})}{n(0, q, \vec{y})},
\end{equation}
where $q = q_1 q_2 = e^{- \Delta_1 - \Delta_2} \equiv e^{- \beta}$ and $y_{\alpha} = e^{- \beta L_{\alpha} / 2 - \v_{\alpha}}$.%
\footnote{%
  The coefficients $\mathcal{I}_{\boldsymbol{m},0}$ are given by replacing $\v_{\alpha}$ with $- \v_{\alpha}$, which effectively interchanges $\Delta_1$ and $\Delta_2$.
}
The natural chemical potentials appearing here are not just $\v_{\alpha}$, but $\beta L_{\alpha}/2 + \v_{\alpha}$.
The quantities $L_{\alpha}$, $\alpha = 0, \ldots, K-1$ sum up to $L$ and are the flavour multiplicities of the dual quiver to figure~\ref{fig:quiver-diag}, which effectively interchanges $K$ and $L$ and can be obtained from the $K_A$ by a known recipe \cite{deBoer:1996mp}.

In the following, it is useful to label the background fluxes $m_A$ by $m_{\alpha, i}$ instead, where $i = 1, \ldots, L_{\alpha}$ and $\alpha$ indicates the node in the quiver associated to the background flux.
Since the leading saddle point answer depends only on $L$ and not on individual $L_{\alpha}$, we will not go into more detail about the mirror map.

With $L (\Delta_1 - \Delta_2)/2 = \sum_{\alpha = 0}^{K-1} \v_{\alpha}$ and $L = \sum_{\alpha = 0}^{K-1} L_{\alpha}$, the fugacities $y_{\alpha}$ obey the constraint
\begin{equation}
	y \equiv \prod_{\alpha = 0}^{K-1} y_{\alpha} = e^{- L \Delta_1}.
\end{equation}
Here, $y$ is the fugacity for the imaginary root of the $\widehat{\mathfrak{su}}(K)$ affine Lie algebra.

\medskip
The affine fermionic form is \emph{Hatayama's fermionic form} \cite{Hatayama:2001gy} applied to an affine Lie algebra,\footnote{In comparison to the notation in \cite{Dorey:2025qht}, we have changed the summation over partition-vectors $\tau \in \mathcal{P}^K$ to a summation over multiplicities $\mu_k$ of the transposed partitions $\tau^T$, which are related by $\mu_k((\tau^{\alpha})^T) = \tau_k^{\alpha} - \tau_{k+1}^{\alpha}$. Relatedly, $n(\boldsymbol{m}, q, \vec{y})$ is now directly a function of the baryonic background flux $\boldsymbol{m}$ rather than the transposed flux $\boldsymbol{m}^T$.}
\begin{equation}
	n(\boldsymbol{m},q,\vec y) = \sum_{\{\mu_k\}}
	q^{\frac{1}{2} \boldsymbol{\mu} \A \boldsymbol{\mu} - (\boldsymbol{\mu}, \boldsymbol{\tilde{\mu}})}
	\prod_{k = 1}^{\infty}
	\prod_{\alpha=0}^{K-1} y_\alpha^{k \mu_k^{\alpha}}
	\left[\sum_{l=1}^{\infty} \min(k, l) (\tilde{\mu}_{\alpha, l} - C_{\alpha \beta} \mu_l^{\beta}), \mu_k^{\alpha}\right]_q \!\!,
	\label{eq:fermionic-form-general}
\end{equation}
where the sum is over sequences $\mu_k \in \mathbb{N}_0^K$ with finite support (multiplicities of some partition) and
\begin{equation}
	\boldsymbol{\mu} \A \boldsymbol{\mu} = \sum_{(\alpha, k)} \sum_{(\beta, l)} \A_{\alpha \beta}^{kl} \mu_k^{\alpha} \mu_l^{\beta}, \quad
	\A_{\alpha \beta}^{k l} \equiv C_{\alpha \beta} \min(k, l).
\end{equation}
Composite indices of the form $(\alpha, k)$ always refer to Greek indices $\alpha, \beta, \ldots$ ranging from $0$ to $K-1$ and Latin indices $k, l, \ldots$ ranging from $1$ to $\infty$.
The variables $\tilde{\mu}_{\alpha, k} \equiv \operatorname{mult}_k(\boldsymbol{m}_{\alpha})$ are the multiplicities of the partitions $\boldsymbol{m}_{\alpha} \in \mathcal{P}$ and
\begin{equation}
	(\boldsymbol{\mu}, \boldsymbol{\tilde{\mu}}) = \sum_{\alpha = 0}^{K-1} \sum_{k, l = 1}^{\infty} \min(k, l) \mu_k^{\alpha} \tilde{\mu}_{\alpha, l}.
\end{equation}
The $q$-binomial coefficient $[n, m]_q$ is defined in terms of the $q$-Pochhammer symbol $(a;q)_m \equiv \prod_{i=0}^{m-1} (1 - a q^i)$ as
\begin{equation}
	[n,m]_q:=\binom{n+m}{m}_q =\frac{(q;q)_{n+m}}{(q;q)_n(q;q)_m}
	= \frac{(q^{n+1}; q)_{m}}{(q;q)_{m}} \qquad n\in\mathbb Z,\quad m\in\mathbb N_0. \label{eq:qbinom-nonpleth}
\end{equation}

Our goal in this section is to extract their asymptotics in the Cardy-like limit $\Delta_1, \Delta_2 \to 0$. This means that simultaneously the deformation parameter $q$ goes to unity and also the fugacity $y$ of the imaginary root.

In fermionic sum representations of coset WZW characters, it is well-known that the Cardy-like asymptotics $q \to 1$ are governed by a sum of (Rogers') dilogarithms $L(z)$ \cite{Kedem:1992jv,Kedem:1993ze, Kuniba:1992sp}.
That this indeed recovers the effective central charge of the algebra according to Cardy's formula is asserted by swathes of dilogarithm identities of the form \cite{kirillov1989identities,Kirillov:1994en}
\begin{equation}
	\label{eq:dilog-ident}
	\sum_{(\alpha, k)} L(f_k^{(\alpha)}) = \frac{\pi^2}{6} c,
\end{equation}
where $c$ is the effective central charge.

For Hatayama's fermionic form, one can generally take the $q \to 1$ limit safely,
in which case $\mathcal I_{0,\boldsymbol m}$ degenerates to the vacuum character of the algebra
\begin{equation}
  \label{eq:vertex-alg}
	\prod_{\alpha = 0}^{K-1} \prod_{i = 1}^{L_{\alpha}} \mathcal{V}(m_{\alpha, i}, K)
\end{equation}
where any factor
\begin{equation}
	\mathcal{V}(m,K) \equiv \mathcal{H} \oplus \widehat{\mathfrak{su}}(K)_m \oplus W_{m,K}^{\text{para}},
\end{equation}
is taken at central charge $c = K m$ \cite{Dorey:2025qht}.
However, we will compute the limit $\Delta_1, \Delta_2 \to 0$ while simultaneously scaling $m$ such that $\Delta_{1} m$ and $\Delta_2 m$ are held constant (or approach zero).
Then, $\mathcal{I}_{0,\boldsymbol{m}}$ does diverge as $q$ approaches one, as we confirm later on.

To obtain the leading order asymptotics, we write the full fermionic form $n(\boldsymbol{m}, q, \vec{y})$ as
\begin{equation}
	n(\boldsymbol{m}, q, \vec{y}) = \sum_{\{\mu_k\}}^{} e^{S[\mu]}
\end{equation}
with the effective action
\begin{equation}
	\label{eq:saddlept-action-324}
	\begin{aligned}
		S[\mu] \approx
		 & - \frac{\beta}{2} \boldsymbol{\mu} \A \boldsymbol{\mu}
		+ \beta (\boldsymbol{\mu}, \boldsymbol{\tilde{\mu}}) +
		\sum_{(\alpha, k)}
		k \mu_k^{\alpha} \log y_{\alpha}                          \\
		 & + \sum_{(\alpha, k)}
		\int_{0}^{\mu_{k}^{\alpha}} \!\d t \left[\log (1 - q^{t} w_k^{(\alpha)})
			- \log (1 - q^t)\right].
	\end{aligned}
\end{equation}
Here and in the following, we define
\begin{equation}
	v_{k}^{(\alpha)} = q^{\mu_k^{\alpha}}
	\quad \text{and} \quad
	w_{k}^{(\alpha)} = q^{\sum_{l=1}^{\infty} \min(k,l) (\tilde{\mu}_{\alpha, l} - C_{\alpha \beta} \mu_l^{\beta})},
\end{equation}
as well as
\begin{equation}
	f_k^{(\alpha)} = \frac{1 - v_k^{(\alpha)}}{1 - v_{k}^{(\alpha)} w_k^{(\alpha)}}
	\quad \text{and} \quad
	g_k^{(\alpha)} = \frac{(1 - v_k^{(\alpha)}) w_k^{(\alpha)}}{1 - v_{k}^{(\alpha)} w_k^{(\alpha)}},
\end{equation}
which are related to $v$ and $w$ by the arguments of the dilogarithmic five-term identity.
In terms of these new variables, the saddle point equations become (as derived in \cref{sec:saddlept-eq-deriv})
\begin{align}
	\label{eq:f-eom-210}
	f_k^{(\alpha)} & =
	y_{\alpha}^k
	\left(
	\prod_{i=1}^{L_{\alpha}} q^{- \min(k, m_{\alpha,i})}
	\right)
	\prod_{(\beta, l)}
	(1 - f_l^{(\beta)})^{\A_{\alpha \beta}^{kl}}
	\quad \text{and}                                                             \\
	g_k^{(\alpha)} & = y_{\alpha}^k \prod_{(\beta, l)}
	(1 - g_l^{(\beta)})^{\A_{\alpha \beta}^{kl}}.
  \label{eq:g-eom-b36}
\end{align}
The equations \eqref{eq:f-eom-210} and \eqref{eq:g-eom-b36} are equivalent.
However, for the saddle point value of the action, it is useful to retain both formulations:
$S$ reduces to a sum of dilogarithms conceptually similar to \cref{eq:dilog-ident} (we derive this in \cref{sec:saddlept-action-deriv}):
\begin{equation}
	\label{eq:saddlept-158}
	\begin{aligned}
		S = & \frac{1}{\beta} \sum_{(\alpha, k)}^{} 
		\left[\Li_2(f_k^{(\alpha)}) - \Li_2(g_k^{(\alpha)})\right] +             \\
		    & \frac{1}{2 \beta} \Li_1 (\boldsymbol{f}) \A \Li_1 (\boldsymbol{f})
		- \frac{1}{2 \beta} \Li_1 (\boldsymbol{g}) \A \Li_1 (\boldsymbol{g}).
	\end{aligned}
\end{equation}
This similarity can be made even more precise:
define the function
\begin{equation}
  \label{eq:cm-557}
  \frac{\pi^2}{6} c[\boldsymbol{m}] \equiv
  \sum_{(\alpha, k)}^{} \left[
    L(f_k^{(\alpha)})
    - \frac{\pi i}{2} D_k^{(\alpha)} \log (1 - f_k^{(\alpha)})
 \right],
\end{equation}
where
\begin{equation}
  \pi i D_k^{(\alpha)}
  = \log f_k^{(\alpha)} - \sum_{(\beta, l)}^{} \A_{\alpha \beta}^{kl} \log (1 - f_l^{(\beta)})
\end{equation}
encodes the chemical potential and the functions $f_k^{(\alpha)} = f_{k}^{(\alpha)}(\boldsymbol{m})$ satisfy the saddle point equation \eqref{eq:f-eom-210} with background flux $\boldsymbol{m}$.
Then,
\begin{equation}
  \label{eq:s-c-604}
  S = \frac{\pi^2}{6 \beta} (c[\boldsymbol{m}] - c[0]),
\end{equation}
since $g_k^{(\alpha)}$ solves the same equations of motion as $f_k^{(\alpha)}(0)$.
The function $c[\boldsymbol{m}]$ appears in the context of the thermodynamic Bethe ansatz (TBA) and encodes the effective central charge of the system.\footnote{See for example the discussion in \cite{Kuniba:1992sp} for non-affine Lie algebras.}
It would be interesting to see whether one can extract a dilogarithm identity for affine Lie algebras from \cref{eq:cm-557} analogous to \cref{eq:dilog-ident}, potentially when $y_{\alpha} = 1$.

In the source-free case $\boldsymbol{m} = 0$, it is clear that the leading order contribution \eqref{eq:s-c-604} vanishes.
In particular, this means that $\log n(0, q, \vec{y})$ is subleading and therefore that
\begin{equation}
	\log \mathcal{I}_{0, \boldsymbol{m}} = \log \frac{n(\boldsymbol{m}, q, \vec{y})}{n(0, q, \vec{y})} \approx
	\log n(\boldsymbol{m}, q, \vec{y}) \approx S.
\end{equation}
The saddle point equations also showcase how the limit $q \to 1$ at $\boldsymbol{m} \ne 0$ has the same effect as sending $\boldsymbol{m} \to 0$ at $q \ne 1$.
When $q=1$, the fermionic form becomes the vacuum character of $\mathcal{V}$ and still exhibits a Cardy-like divergence as the imaginary root $y = e^{-L \Delta_1} \to 1$.
This can be recovered from the results in this section by setting $\beta m \ll 1$.

While in the plethystic case the trilogarithmic dependence on the background flux $m$ was seen explicitly through an integral of the form
\begin{equation}
	S \sim \frac{1}{\Delta} \int_{0}^{m} \! \d k \Li_2 (e^{\Delta k}) \sim \frac{1}{\Delta^2} \Li_3 (e^{\Delta m}),
\end{equation}
such a formulation is not manifest here. Rather, the action takes the form
\begin{equation}
	S \sim \frac{1}{\beta} \sum_{k=1}^{\infty} \Li_2 (f_k) \sim \frac{1}{\beta} \int_{0}^{\infty} \! \d k \Li_2(f_k),
\end{equation}
where the $m$-dependence is buried in the equations of motion for $f_k$.
To make the $m$-dependence manifest, we convert \cref{eq:s-c-604} to an integral over $c$,
\begin{equation}
  \label{eq:s-m-int-942}
  S = \frac{\pi^2}{6 \beta} \int_{c[0]}^{c[\boldsymbol{m}]} \! \d c =
  \frac{\pi^2}{6 \beta} \int_{0}^{\boldsymbol{m}} \sum_{\alpha = 0}^{K-1} \sum_{i=1}^{L_{\alpha}} \frac{\partial c}{\partial m_{\alpha, i}} \d m_{\alpha, i}.
\end{equation}
From \eqref{eq:saddlept-158}, the differential is equal to
\begin{equation}
  \frac{\pi^2}{6}
	\d c =
	\Li_1(\boldsymbol{f}) \cdot
	\d (\log \boldsymbol{f}) +
	\Li_1(\boldsymbol{f}) \A \, \d (\Li_1 \boldsymbol{f}) =
	\Li_1 (\boldsymbol{f}) \cdot \d (\log \boldsymbol{f} + \A \Li_1 \boldsymbol{f}),
\end{equation}
where the dot denotes scalar multiplication over the indices $k$ and $\alpha$.
By \cref{eq:f-eom-210} (and using the fact that the fugacities $y_{\alpha}$ are independent of the sources $\boldsymbol{m}_{\alpha}$), this becomes
\begin{equation}
  \frac{\pi^2}{6}
	\d c =
	\beta \sum_{\alpha = 0}^{K-1}
	\sum_{i=1}^{L_{\alpha}}
	\left(\sum_{k = m_{\alpha,i}}^{\infty}
	\Li_1(f_k^{(\alpha)})
	\right)
	\, \d m_{\alpha, i}.
\end{equation}

We will focus on the case that all $m_{\alpha,i} \equiv m$ are equal, treating the most general case in \cref{sec:gen-bf}.
Since the resulting action depends only on the end points, we may choose any integration path.
Here, we integrate along the diagonal path from $0$ to $\boldsymbol{m}$ by choosing an integration variable $z \in [0, m]$ such that $m_{\alpha,i} = z$.

Then,
\begin{equation}
	S = \int_{0}^{m} \! \d z
	\sum_{\alpha=0}^{K-1} L_{\alpha} \sum_{k=z}^{\infty} \Li_1(f_k^{(\alpha)}).
\end{equation}
At this point, it is necessary to better understand the profile of the solutions $f_k^{(\alpha)}$ to the equations of motion.
This is discussed in more detail in \cref{sec:saddle-eom}.
The upshot is that one may approximate $L_{\alpha}$ by $L / K$ (the difference is subleading) and also that
\begin{equation}
	\label{eq:li-int-309}
	\sum_{\alpha = 0}^{K-1}  \sum_{k=z}^{\infty}  \Li_1(f_k^{(\alpha)}) \approx
	\sum_{\alpha = 0}^{K-1}  \int_{z}^{\infty} \! \d k \, \Li_1(f_k^{(\alpha)}) \approx
	- \frac{K}{\log y}
	\left(\sum_{\alpha = 0}^{K-1}
	\Li_2(f_z^{(\alpha)})
	+ \text{bdy} \right).
\end{equation}
Furthermore, we argue in \cref{sec:deviations} that $f_z^{(\alpha)}$ is equal up to subleading corrections to the centre-of-mass mode $\bar{f}_z$, given by
\begin{equation}
	\log \bar{f}_z \equiv \frac{1}{K} \sum_{\alpha=0}^{K-1} \log f^{(\alpha)}_z =
	\frac{z}{K} \log y + \frac{L}{K} \beta \min(z,z) =
	L \Delta_2 z / K.
\end{equation}
Then,
\begin{equation}
	\frac{L}{K} \frac{(-K)}{\log y} \int_{0}^{m} \! \d z \, \sum_{\alpha=0}^{K-1} \Li_2(f_z^{(\alpha)})
	\approx
	\frac{1}{\Delta_1} \int_{0}^{m} \! \d z \, K \Li_2(\bar{f}_z)
  =
	\frac{K^2}{L \Delta_1 \Delta_2} \left[\Li_3(e^{L \Delta_2 m / K}) - \zeta(3) \right].
\end{equation}

The boundary term contribution in \cref{eq:li-int-309} is of the form
\begin{equation}
	\label{eq:bdy-contrib}
	\text{bdy} = \frac{1}{2} \v_{\alpha}^{+} (C^{-1})^{\alpha \beta} \v_{\beta}^{+}
\end{equation}
and picks up a simple factor of $m$ when integrated over $z$.
The inverse of the generalised Cartan matrix $C_{\alpha \beta}$ is well-defined here: only the transverse part $\v_{\alpha}^{+}$ enters, which sums to zero and therefore lies in the non-degenerate subspace of $C_{\alpha \beta}$ (see Claim~2 in section~\ref{sec:saddle-eom}).
The boundary contribution should be interpreted as an anomaly related to the modular transformation of $\widehat{\mathfrak{su}}(K)_m$ \cite{Kac:1984mq}.
In a stricter Cardy-like limit, where all $\v_{\alpha} \to 0$ individually, this term vanishes.

In total, the leading order divergence at equal background flux $m_{\alpha,i} = m$ is
\begin{equation}
  \label{eq:log-I-res-950}
	\lim_{\Delta_1, \Delta_2 \to 0} \log \mathcal{I}_{0, \boldsymbol{m}} \sim
	\frac{K^2}{L \Delta_1 \Delta_2} \left[\Li_3(e^{L \Delta_2 m / K}) - \zeta(3) \right] +
	\frac{m}{2 \Delta_1} \v_{\alpha}^{+} (C^{-1})^{\alpha \beta} \v_{\beta}^{+}.
\end{equation}

\subsection{Equations of motion}
\label{sec:saddle-eom}

Taking the logarithm of \cref{eq:f-eom-210}, the saddle point equations become
\begin{equation}
  \label{eq:saddle-eom-log}
	\log f_k^{(\alpha)} = k \log y_{\alpha} +
	\beta \sum_{i=1}^{L_{\alpha}} \min(k, m_{\alpha,i})
	- \sum_{l=1}^{\infty} \sum_{\beta = 0}^{K-1}
	\min(k,l) C_{\alpha \beta} \Li_1(f_l^{(\beta)}).
\end{equation}

At large $k$, the solutions $\log f_k^{(\alpha)}$ approach linear functions of slope
\begin{equation}
	\label{eq:asymp-slope}
	\partial_k \log f_k^{(\alpha)}\big|_{k \to \infty} = \log y_{\alpha} = - \beta L_{\alpha} / 2 - \v_{\alpha}.
\end{equation}
Also, the centre-of-mass mode satisfies the simplified equation
\begin{equation}
	\log \bar{f}_k \equiv \frac{1}{K} \sum_{\alpha=0}^{K-1} \log f_k^{(\alpha)} = k \frac{\log y}{K} + \frac{\beta}{K} \sum_{\alpha=0}^{K-1} \sum_{i=1}^{L_{\alpha}} \min(k, m_{\alpha, i}),
\end{equation}
which is a piecewise linear function in $k$.

It is helpful to also consider the saddle point equation as converted to a differential equation.
Taking the second derivative in $k$,\footnote{Strictly speaking, we mean the discrete second derivative $\partial_k^2 f_k = f_{k-1} + f_{k+1} - 2 f_k$.}
\begin{equation}
	\label{eq:saddlept-deriv}
	\partial_k^2 \log f_k^{(\alpha)} = - \beta \sum_{i=1}^{L_{\alpha}} \delta_{k, m_{\alpha,i}}
	+
	C_{\alpha \beta} \Li_1(f_k^{(\beta)}).
\end{equation}

In \cref{sec:deviations}, we present numerical evidence for the behaviour of $f_k^{(\alpha)}$ at $k = z$ along the integration path where all $m_{\alpha, i} = z$.
Splitting
\begin{equation}
	\log f_k^{(\alpha)} = \log \bar{f}_k + \log \delta f_k^{(\alpha)}
\end{equation}
into a centre-of-mass mode $\bar{f}_k$ as before and transverse deviations $\delta f_k^{(\alpha)}$,
we argue that
\begin{align}
	\log \delta f_k^{(\alpha)} |_{k = z}                  & = \mathcal{O}(\beta)
	\quad \text{and}                                                                 \\
	\partial_k \log \delta f_k^{(\alpha)} |_{k = z^{\pm}} & = \mathcal{O}(\beta).
	\label{eq:fluc-deriv-O}
\end{align}
Here, $z^{\pm}$ refers to approaching $k=z$ from above or below, since the first derivative jumps due to the delta function in \cref{eq:saddlept-deriv}.

\medskip
This assumption enables us to prove the remaining claims from the previous section.
There, we made use of two approximations:
\begin{enumerate}
	\item The difference between flavours $L_{\alpha}$ and $L/K$ at node $\alpha$ is subleading.
	\item The integral of $\Li_1$ in \cref{eq:li-int-309} is approximately a dilogarithm term and the boundary contribution \eqref{eq:bdy-contrib}.
\end{enumerate}

\paragraph{Claim 1.}
Since the difference vector $L_{\alpha} - L/K$ lies in the image of the Cartan matrix, there exists a vector $w^{\beta}$ such that
\begin{equation}
	L_{\alpha} - \frac{L}{K} = w^{\beta} C_{\beta \alpha}.
\end{equation}
Then, by the saddle point equations,
\begin{equation}
	\int_{z}^{\infty} \!\d k \,
	\sum_{\alpha=0}^{K-1}
	\left(
	L_{\alpha} - \frac{L}{K}
	\right)
	\Li_1(f_k^{(\alpha)})
	=
	\int_{z}^{\infty} \!\d k \,
	w^{\beta} C_{\beta \alpha} \Li_1(f_k^{(\alpha)})
	=
	\sum_{\alpha=0}^{K-1}
	\int_{z}^{\infty} \!\d k \,
	w^{\alpha} \partial_k^2 \log f_k^{(\alpha)}.
\end{equation}
Since $k > z$, $\partial_k^2 \log \bar{f}_k = 0$ and this is equal to
\begin{equation}
	\int_{z}^{\infty} \!\d k \,
	\sum_{\alpha=0}^{K-1} w^{\alpha} \partial_k^2
	\log \delta f_k^{(\alpha)}
	=
	\sum_{\alpha=0}^{K-1}
	w^{\alpha}
	\partial_k \log \delta f_{k}^{(\alpha)} \big|_{k=z}^{\infty}.
\end{equation}
The boundary at $k = \infty$ has
\begin{equation}
	\partial_k \log \delta f_k^{(\alpha)} \big|_{k = \infty} = - \beta L_{\alpha} / 2 - \v_{\alpha} - \frac{\log y}{K} = -\v_{\alpha}^{+} - \frac{\beta}{2} \left( L_{\alpha} - \frac{L}{K} \right) = \mathcal{O}(1)
\end{equation}
as $\Delta_1, \Delta_2 \to 0$.
As per \cref{eq:fluc-deriv-O}, the boundary contribution at $k = z$ is even smaller, of order $\mathcal{O}(\beta)$.
Therefore, both terms are subleading compared to the $1 / \log y \sim \Delta_1^{-1}$ divergence in \eqref{eq:li-int-309}.

\paragraph{Claim 2.}
Along the integration path, $m_{\alpha,i} \equiv z$ and $k > z$, so
\begin{equation}
	\partial_k \log \bar{f}_k \big|_{k > z} = \frac{\log y}{K}.
\end{equation}
Therefore,
\begin{equation}
	\int_{z}^{\infty} \! \d k \, \Li_1(f_k^{(\alpha)}) =
	\frac{K}{\log y} \int_{z}^{\infty} \! \d k \, \Li_1(f_k^{(\alpha)}) \partial_k \log \bar{f}_k.
\end{equation}
We split $\log \bar{f}_k$ up into
\begin{equation}
	\log \bar{f}_k = \log f_k^{(\alpha)} - \log \delta f_k^{(\alpha)}.
\end{equation}
The first term integrates to
\begin{equation}
	- \frac{K}{\log y} \Li_2(f_z^{(\alpha)}),
\end{equation}
where the boundary at infinity does not contribute since by \cref{eq:asymp-slope} $\Li_2(f_k^{(\alpha)}) \to 0$ as $k \to \infty$ if $\operatorname{Re} \beta, \operatorname{Re} \v_{\alpha} > 0$.

For the second term, which we have called the boundary contribution previously, we use the fact that the actual expression \eqref{eq:li-int-309} sums over $\alpha$.
When $k > z$, we can ignore the delta function contribution in \cref{eq:saddlept-deriv} and also replace $\partial_k^2 \log f_k^{(\alpha)}$ with $\partial_k^2 \log \delta f_k^{(\alpha)} = \partial_k^2 (\log f_k^{(\alpha)} - \log \bar{f}_k)$.
Since the vector $\log \delta f_k^{(\alpha)}$ lies in the image of $C_{\alpha \beta}$, this determines $\Li_1(f_k^{(\alpha)})$ up to a component along the kernel $(1, \ldots, 1)$ of $C_{\alpha \beta}$.
That component drops out of the sum below, since $\sum_{\alpha=0}^{K-1} \partial_k \log \delta f_k^{(\alpha)} = 0$.
The total sum therefore becomes
\begin{equation}
	\begin{aligned}
		\sum_{\alpha=0}^{K-1} \Li_1(f_k^{(\alpha)}) (\partial_k \log \delta f_k^{(\alpha)})
		 & =
		(\partial_k^2 \log \delta f_k^{(\beta)}) (C^{-1})^{\beta \alpha} \partial_k \log \delta f_k^{(\alpha)} \\
		 & =
		\frac{1}{2} \partial_k
		\left[
			(\partial_k \log \delta f_k^{(\alpha)}) (C^{-1})^{\alpha \beta} (\partial_k \log \delta f_k^{(\beta)})
			\right].
	\end{aligned}
\end{equation}
The boundary of the integral over $k \in (z, \infty)$ at $k = \infty$ has
\begin{equation}
	\partial_k \log \delta f_k^{(\alpha)} \big|_{k = \infty} = - \beta L_{\alpha} / 2 - \v_{\alpha} - \frac{\log y}{K} = -\v_{\alpha}^{+} - \frac{\beta}{2} \left( L_{\alpha} - \frac{L}{K} \right) \to -\v_{\alpha}^{+}
\end{equation}
as $\Delta_1, \Delta_2 \to 0$.
Therefore, the contribution is
\begin{equation}
	-\frac{K}{2 \log y} \v_{\alpha}^{+} (C^{-1})^{\alpha \beta} \v_{\beta}^{+}.
\end{equation}
As in the first claim, the boundary at $k = z$ is subleading due to \cref{eq:fluc-deriv-O}.

\subsection{General background flux}
\label{sec:gen-bf}

Until now, we focused on giant graviton coefficients $\mathcal{I}_{0, \boldsymbol{m}}$ with equal background fluxes $m_{\alpha,i} = m$.
However, turning on baryonic charge $\boldsymbol{b}$ in the Hilbert series results in unequal $m_{\alpha,i}$.
If $\boldsymbol{b} = \mathcal{O}(1)$, the differences wash out at $m \sim \beta^{-1}$:
\begin{equation}
\beta m_{\alpha,i} = \beta (m + b_{\alpha,i}) = \beta m + \mathcal{O}(\beta).
\end{equation}
To get a non-trivial result, it is therefore necessary to simultaneously rescale the background fluxes $\boldsymbol{b}$ such that $\beta b_{\alpha,i}$ stays macroscopic in the Cardy limit $\Delta_1, \Delta_2 \to 0$.

\begin{figure}[htbp]
  \centering
  \begin{tikzpicture}[x=1cm, y=1cm, every node/.style={font=\scriptsize}]
    \coordinate (p0) at (0,0);
    \coordinate (p1) at (2.3,1);
    \coordinate (p2) at (4.6,0);
    \coordinate (p3) at (6.9,1);
    \coordinate (p4) at (9.2,0);
    \coordinate (p5) at (11.5,1);
    \draw[thick] (p0) -- node[midway,above,sloped] {$z_0$} (p1);
    \draw[thick] (p1) -- node[midway,below,sloped] {$z_1$} (p2);
    \draw[thick] (p2) -- node[midway,above,sloped] {$z_2$} (p3);
    \draw[thick,densely dotted] (p3) --
      node[midway,below,sloped] {$z_3,\ldots,z_{L-2}$} (p4);
    \draw[thick] (p4) -- node[midway,above,sloped] {$z_{L-1}$} (p5);
    \foreach \p in {p0,p1,p2,p3,p4,p5}
      \fill (\p) circle (2pt);
    \node[below=6pt] at (p0) {$(0,\ldots,0)$};
    \node[above=6pt] at (p1) {$(m,\ldots,m)$};
    \node[below=6pt] at (p2) {$(m,m+b_1,\ldots,m+b_1)$};
    \node[above=6pt] at (p3) {$(m,m+b_1,m+b_2,\ldots,m+b_2)$};
    \node[below=6pt] at (p4)
      {$(m,m+b_1,\ldots,m+b_{L-2},m+b_{L-2})$};
    \node[above=6pt] at (p5) {$(m,m+b_1,\ldots,m+b_{L-1})$};
  \end{tikzpicture}
  \caption{Piecewise integration path for the background fluxes. Without baryonic charge $\boldsymbol{b}$, the path consists of just the first segment.}
  \label{fig:background-flux-path}
\end{figure}
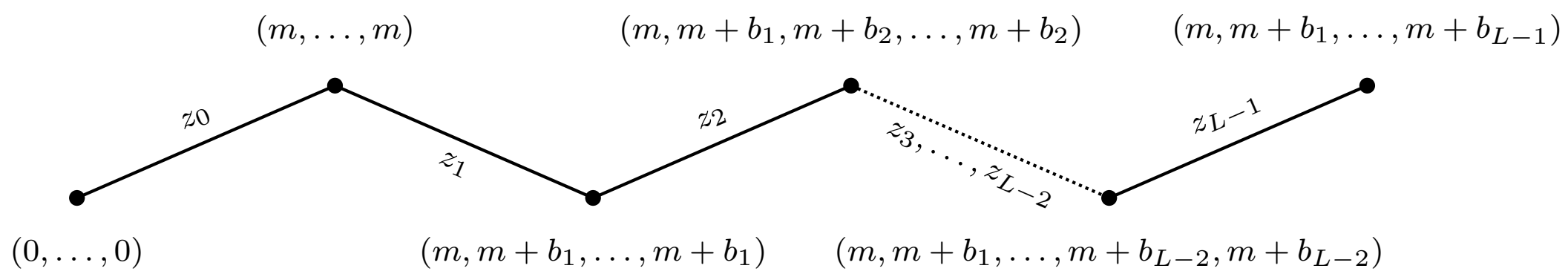

To simplify the notation in the following, we will revert $(\alpha, i)$ into a single index $A = 0, \ldots, L-1$ in such a way that the $b_A$ are sorted in ascending order,
\begin{equation}
  0 = b_0 \leq b_1 \leq \ldots \leq b_{L-1}
\end{equation}
and $m_A = m + b_A$.
We denote the corresponding $(\alpha, i)$ by $\alpha_A$ and $i_A$.
Previously, we chose in \cref{eq:s-m-int-942} the integration path $(z, \ldots, z)$ from $0$ to $\boldsymbol{m}$.
Now, we parametrise the background flux via
\begin{equation}
  (z_0, z_0 + z_1, z_0 + z_1 + z_2, \ldots, z_0 + \ldots + z_{L-1})
\end{equation}
and go along piecewise, first integrating $\int_{0}^{m} \! \d z_0$, then $\int_{0}^{b_1 - b_0} \! \d z_1$, $\int_{0}^{b_2 - b_1} \! \d z_2$ and so on, until $\int_{0}^{b_{L-1} - b_{L-2}} \! \d z_{L-1}$.
The whole path can be seen in figure~\ref{fig:background-flux-path}.

The integral over $z_0$ gives the same contribution $S_0$ as \cref{eq:log-I-res-950} previously, since along this part of the integration path the functions $\log f_k^{(\alpha)}$ solve the saddle point equations with equal fluxes $m_{\alpha,i} = z_0$.
The next integral has an integrand of the form
\begin{equation}
  \frac{\pi^2}{6} \d c = \beta
  \sum_{A=1}^{L-1} \left(
    \sum_{k=m + z_1}^{\infty} \Li_1(f_k^{(\alpha_A)})
  \right)
  \d z_1
\end{equation}
where, importantly, the summation ignores $A = 0$.
Here, $\log f_k^{(\alpha)}$ solves the saddle point equations with fluxes $(m, m + z_1, \ldots, m + z_1)$.
The contribution to the saddle point action $S$ is then
\begin{equation}
  S_1 = \int_{0}^{b_1} \! \d z_1 \, 
  \sum_{\alpha=0}^{K-1} L_{\alpha}'
  \sum_{k=m + z_1}^{\infty} \Li_1(f_k^{(\alpha)})
\end{equation}
where the new flavours are $L_{\alpha}' = L_{\alpha} - \delta_{\alpha, \alpha_0}$.
The functions $\log f_k^{(\alpha)}$ have sources at two locations, $m$ and $m + z_1$.
As before, it is necessary for the following argument to assume that the deviations from the centre-of-mass mode are of order $\mathcal{O}(\beta)$.
We comment on this in \cref{sec:deviations}.
Then, we may again replace $L_{\alpha}'$ by $(L-1)/K$ and integrate the first polylogarithm to a dilogarithm,
\begin{equation}
  \sum_{\alpha=0}^{K-1} \sum_{k=m+z_1}^{\infty} \Li_1(f_k^{(\alpha)}) \approx
  - \frac{K}{\log y} \left(
    \sum_{\alpha=0}^{K-1} \Li_2(f_{m+z_1}^{(\alpha)}) + \frac{1}{2} \v_{\alpha}^{+} (C^{-1})^{\alpha \beta} \v_{\beta}^{+}
  \right).
\end{equation}
Integrating again,
\begin{equation}
\frac{L-1}{K} \frac{(-K)}{\log y} \int_{0}^{b_1} \! \d z_1 \sum_{\alpha=0}^{K-1} \Li_2(f_{m+z_1}^{(\alpha)})
\approx
\frac{L-1}{L} \frac{1}{\Delta_1} \int_{0}^{b_1} \! \d z_1 K \Li_2(\bar{f}_{m+z_1}).
\end{equation}
In that regime,
\begin{equation}
  \begin{aligned}
  \log \bar{f}_{m+z_1} 
  &= \frac{(m + z_1)}{K} \log y + \frac{1}{K} \beta \min(m + z_1, m) + \frac{L-1}{K} \beta \min (m + z_1, m + z_1) \\
  &=
  L \Delta_2 m / K + w_1 z_1 / K
  \end{aligned}
\end{equation}
with
\begin{equation}
  \label{eq:def-w_A}
  w_A \equiv - L \Delta_1 + (L-A) \beta = L \Delta_2 - A \beta.
\end{equation}
Therefore,
\begin{equation}
  \label{eq:s1-contrib-a23}
  \begin{aligned}
    S_1 =& \frac{(L-1) K^2}{L \Delta_1 w_1} 
  \left[
    \Li_3(e^{L \Delta_2 m / K + w_1 b_1 / K})
    - \Li_3(e^{L \Delta_2 m / K})
  \right] \\
  &+
  \frac{L-1}{2 L \Delta_1} b_1 \v_{\alpha}^{+} (C^{-1})^{\alpha \beta} \v_{\beta}^{+}.
  \end{aligned}
\end{equation}
This is similar to the contribution $S_0$.
When $m = 0$, it behaves like a theory with $L-1$ effective flavours and background flux given by $b_1$ instead.
When $b_1 = 0$ on the other hand, \eqref{eq:s1-contrib-a23} vanishes.
To further simplify the notation in the exponents, we define%
\footnote{Remember that $m_A = m + b_A$.}
\begin{equation}
  L \phi_{\boldsymbol{m}}^{(A)} \equiv \sum_{0 \leq B \leq A} w_B (m_B - m_{B-1})
	=
	w_A m_A+\beta\sum_{0 \leq B < A} m_B,
\end{equation}
where we understand $m_{-1} \equiv 0$ and $\phi_{\boldsymbol{m}}^{(-1)} \equiv 0$.
The complete action $S = \sum_{A=0}^{L-1} S_A$ can be obtained in a similar fashion to $S_1$ after integrating over all $L$ segments in the above way. The result is
\begin{equation}
  \label{eq:action-gen-flux-c}
  \begin{aligned}
    S =&
  \sum_{A=0}^{L-1} \frac{(L-A) K^2}{L \Delta_1 w_A} \left[
    \Li_3(e^{ L \phi_{\boldsymbol{m}}^{(A)} / K})
    -
    \Li_3(e^{L \phi_{\boldsymbol{m}}^{(A-1)} / K})
  \right] \\
    &+
    \frac{1}{2 \Delta_1} \v_{\alpha}^{+} (C^{-1})^{\alpha \beta} \v_{\beta}^{+}
    \underbrace{\left(
      \sum_{A=0}^{L-1} 
      \frac{L-A}{L} (m_A - m_{A-1})
  \right)}_{= \frac{1}{L} \sum_{A=0}^{L-1} m_{A}}.
  \end{aligned}
\end{equation}
Collecting terms with fixed $\phi_{\boldsymbol{m}}^{(A)}$,
\begin{equation}
  \label{eq:action-gen-flux}
    S =
  - \sum_{A=0}^{L-1} \frac{K^2}{w_A w_{A+1}} \left[
    \Li_3(e^{ L \phi_{\boldsymbol{m}}^{(A)} / K})
    -
    \zeta(3)
  \right] 
    +
    \frac{J(\boldsymbol{m})}{2 L \Delta_1} \v_{\alpha}^{+} (C^{-1})^{\alpha \beta} \v_{\beta}^{+}
\end{equation}
where as in the introduction
\begin{equation}
  J(\boldsymbol{m}) = \sum_{A=0}^{L-1} m_A.
\end{equation}
Just as before, this is independent of how the baryonic fluxes $m_A$ are distributed over the colours $\alpha_A$. We note that, as $\boldsymbol{b} \to 0$ and all $m_A = m$, the action reduces to the equal-flux result \eqref{eq:log-I-res-950} through the identity
\begin{equation}
  - \sum_{A=0}^{L-1} \frac{1}{w_A w_{A+1}} = \frac{1}{L \Delta_1 \Delta_2}.
\end{equation}

\paragraph{Central charge.}
We now compare this result to the expected central charge of the full vertex algebra $\mathcal{V}$ in \cref{eq:vertex-alg}.
This is easiest to see going back to the expression \eqref{eq:action-gen-flux-c}.
In the limit $\beta m_{\alpha,i} \ll 1$,
\begin{equation}
    \Li_3(e^{ L \phi_{\boldsymbol{m}}^{(A)} / K})
    -
    \Li_3(e^{L \phi_{\boldsymbol{m}}^{(A-1)} / K})
    =
    \frac{\pi^2}{6} \frac{1}{K} w_A (m_A - m_{A-1}) + o(\beta m_A).
\end{equation}
The first line of the action \eqref{eq:action-gen-flux-c} then becomes
\begin{equation}
  \frac{\pi^2}{6} \frac{1}{L \Delta_1} \underbrace{K \sum_{A=0}^{L-1} (L - A) (m_A - m_{A-1})}_{\equiv c(\boldsymbol{m})},
\end{equation}
where again $L \Delta_1$ is the chemical potential of the imaginary root of the algebra.
$c(\boldsymbol{m})$ can indeed be rearranged to
\begin{equation}
  c(\boldsymbol{m}) =
  K \sum_{A=0}^{L-1} m_A = K \sum_{\alpha=0}^{K-1} \sum_{i=1}^{L_{\alpha}} m_{\alpha, i},
\end{equation}
the central charge of $\mathcal{V}$.

\subsection{Geometric interpretation}
\label{sec:geom-int}

From a geometric perspective, the effect of non-zero baryonic flux $\boldsymbol{b}$ in the trilogarithmic part of the action \eqref{eq:action-gen-flux} is to resolve the type $A_{L-1}$ singularity of the cone $\mathbb{C}^2 / \mathbb{Z}_L$.
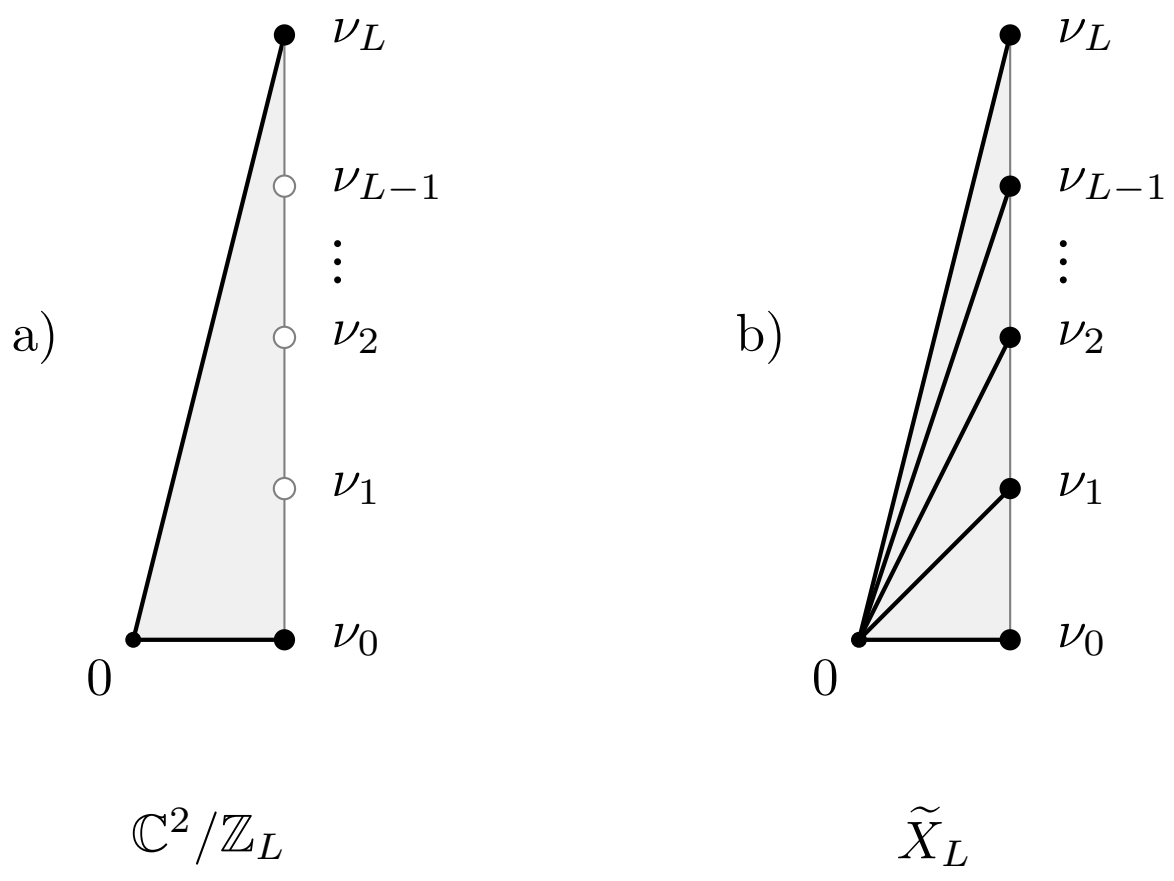
\begin{figure}
  \centering
  \begin{tikzpicture}[x=1cm,y=1cm,font=\small]
    \begin{scope}
      \node at (-0.65,2) {\phantomsubcaption\label{fig:ale-toric-singular}\subref*{fig:ale-toric-singular})};
      \fill[gray!12] (0,0) -- (1,0) -- (1,4) -- cycle;
      \draw[thick] (0,0) -- (1,0);
      \draw[thick] (0,0) -- (1,4);
      \draw[gray] (1,0) -- (1,4);
      \foreach \j in {1,2,3} {
        \draw[gray,fill=white] (1,\j) circle (2pt);
      }
      \fill (1,0) circle (2pt) node[right=5pt] {$\nu_0$};
      \node[right=5pt] at (1,1) {$\nu_1$};
      \node[right=5pt] at (1,2) {$\nu_2$};
      \begin{scope}[shift={(1,2.5)},xshift=10pt]
        \foreach \dy in {-0.12,0,0.12}
          \fill (0,\dy) circle (0.65pt);
      \end{scope}
      \node[right=5pt] at (1,3) {$\nu_{L-1}$};
      \fill (1,4) circle (2pt) node[right=5pt] {$\nu_L$};
      \fill (0,0) circle (1.5pt) node[below left] {$0$};
      \node at (0.5,-1.3) {$\mathbb{C}^2/\mathbb{Z}_L$};
    \end{scope}
    \begin{scope}[xshift=4.8cm]
      \node at (-0.65,2) {\phantomsubcaption\label{fig:ale-toric-resolved}\subref*{fig:ale-toric-resolved})};
      \fill[gray!12] (0,0) -- (1,0) -- (1,4) -- cycle;
      \draw[gray] (1,0) -- (1,4);
      \foreach \j in {0,...,4} {
        \draw[thick] (0,0) -- (1,\j);
        \fill (1,\j) circle (2pt);
      }
      \node[right=5pt] at (1,0) {$\nu_0$};
      \node[right=5pt] at (1,1) {$\nu_1$};
      \node[right=5pt] at (1,2) {$\nu_2$};
      \begin{scope}[shift={(1,2.5)},xshift=10pt]
        \foreach \dy in {-0.12,0,0.12}
          \fill (0,\dy) circle (0.65pt);
      \end{scope}
      \node[right=5pt] at (1,3) {$\nu_{L-1}$};
      \node[right=5pt] at (1,4) {$\nu_L$};
      \fill (0,0) circle (1.5pt) node[below left] {$0$};
      \node at (0.5,-1.3) {$\widetilde X_L$};
    \end{scope}
  \end{tikzpicture}
  \caption{Toric diagrams of $\mathbb{C}^2/\mathbb{Z}_L$ and its minimal resolution $\widetilde X_L$.
  They consist of a set of vectors $\nu_j \in \mathbb{Z}^2$.
  The singular cone $\mathbb{C}^2 / \mathbb{Z}_L$ corresponds to the vectors $\nu_0 = (1,0)$ and $\nu_L = (1,L)$, while the resolved cone $\widetilde{X}_L$ corresponds to the set of vectors $\nu_j = (1, j)$ for $j = 0, \ldots, L$.
 }
  \label{fig:ale-toric-resolution}
\end{figure}
We first describe the equivariant volume of the resolution in geometric variables and then identify these variables with the background fluxes and chemical potentials of the giant graviton coefficients.

\begin{figure}
  \centering
  \begin{tikzpicture}[x=0.9cm,y=0.9cm,font=\small]
    \begin{scope}
      \node at (-1,3.4) {\phantomsubcaption\label{fig:ale-dual-cone}\subref*{fig:ale-dual-cone})};
      \fill[gray!12] (0,0) -- (0,3) -- (4,3) -- (4,-2) -- cycle;
      \draw[->,gray] (-0.55,0) -- (4.35,0) node[right,black] {$X$};
      \draw[->,gray] (0,-2.25) -- (0,3.4) node[above,black] {$Y$};
      \draw[gray,dashed] (0,-2.25) -- (0,3);
      \draw[gray,dashed] (-0.55,0.55) -- (2.25,-2.25);
      \draw[gray,dashed] (-0.55,0.275) -- (4,-2);
      \draw[thick] (0,0) -- (0,3);
      \draw[thick] (0,0) -- (4,-2);
      \node[left] at (0,2) {$F_0$};
      \node[below] at (3,-1.5) {$F_2$};
      \fill (0,0) circle (2pt) node[below left] {$0$};
      \node[above right] at (0,0) {$F_1$};
      \node at (2,1.7) {$\mathcal C$};
      \node at (1.8,-2.9) {$\boldsymbol\lambda=(0,0,0)$};
    \end{scope}
    \begin{scope}[xshift=6.2cm]
      \node at (-1,3.4) {\phantomsubcaption\label{fig:ale-deformed-polygon}\subref*{fig:ale-deformed-polygon})};
      \fill[gray!12] (0,1) -- (0,3) -- (4,3) -- (4,-2) -- (2,-1) -- cycle;
      \draw[->,gray] (-0.55,0) -- (4.35,0) node[right,black] {$X$};
      \draw[->,gray] (0,-2.25) -- (0,3.4) node[above,black] {$Y$};
      \draw[gray,dashed] (0,-2.25) -- (0,3);
      \draw[gray,dashed] (-0.55,1.55) -- (3.25,-2.25);
      \draw[gray,dashed] (-0.55,0.275) -- (4,-2);
      \draw[thick] (0,3) -- (0,1) -- (2,-1) -- (4,-2);
      \node[left] at (0,2) {$F_0$};
      \node[above right] at (0.6,0.4) {$F_1$};
      \node[below] at (3,-1.5) {$F_2$};
      \fill (0,1) circle (2pt) node[left] {$(0,1)$};
      \fill (2,-1) circle (2pt) node[below left] {$(2,-1)$};
      \node[below left] at (0,0) {$0$};
      \node at (2,1.7) {$\mathcal P_{\boldsymbol\lambda}$};
      \node at (1.8,-2.9) {$\boldsymbol\lambda=(0,1,0)$};
    \end{scope}
  \end{tikzpicture}
  \caption{Base manifolds for $L=2$, with the interior regions shaded grey.
  \subref{fig:ale-dual-cone}) The undeformed dual cone $\mathcal C = \mathcal{P}_0$, where the facet $F_1$ degenerates to a point. \subref{fig:ale-deformed-polygon}) The moment polygon $\mathcal{P}_{\boldsymbol{\lambda}}$ for $\boldsymbol\lambda=(0,1,0)$. All three facets impose non-trivial constraints on the region.}
  \label{fig:ale-moment-polygons}
\end{figure}
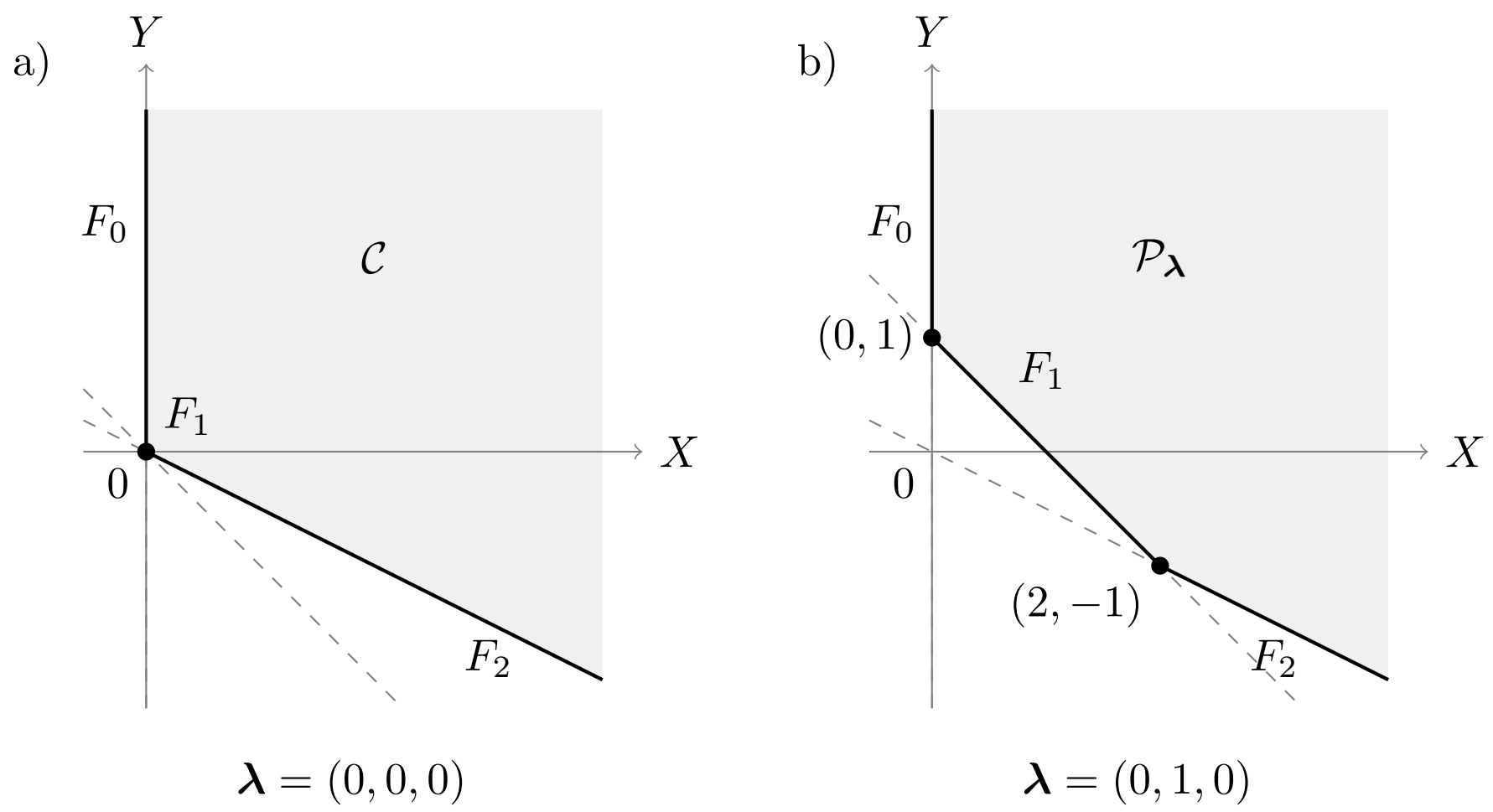

Denote the resolution of $\mathbb{C}^2 / \mathbb{Z}_L$ by $\widetilde{X}_L$. The singular cone is characterised by its \emph{toric diagram} in figure~\ref{fig:ale-toric-singular}, consisting of two vectors $\nu_0 = (1,0)$ and $\nu_L = (1,L)$. It is resolved by adding the additional vectors%
\footnote{%
The adjacent vectors satisfy $\det(\nu_j,\nu_{j+1})=1$. The local orbifold order is the absolute value of this determinant, so the corresponding toric charts are smooth.
}
\begin{equation}
	\nu_j=(1,j),
	\qquad j=1,\ldots,L-1
\end{equation}
(see figure~\ref{fig:ale-toric-resolved}).
The spaces $\mathbb{C}^2 / \mathbb{Z}_L$ and $\widetilde{X}_L$ can be understood as $T^2$-fibrations over certain base manifolds. For $\mathbb{C}^2 / \mathbb{Z}_L$, this base manifold is the \emph{dual cone} $\mathcal{C}$ defined by
\begin{equation}
  \mathcal{C} = \{ \boldsymbol{X} \in \mathbb{R}^2 : \boldsymbol{X} \cdot \nu_j \geq 0, \quad j = 0, \ldots, L \}.
  \label{eq:Cdefn}
\end{equation}
Turning on Kähler deformations $\boldsymbol{\lambda} = (\lambda_0, \ldots, \lambda_L)$, the base manifold for $\widetilde{X}_L$ is the \emph{moment polygon} $\mathcal{P}_{\boldsymbol{\lambda}}$ obtained by shifting the lines $\boldsymbol{X} \cdot \nu_j = 0$,
\begin{equation}
  \mathcal{P}_{\boldsymbol{\lambda}} = \{ \boldsymbol{X} \in \mathbb{R}^2 : \boldsymbol{X} \cdot \nu_j \geq \lambda_j, \quad j = 0, \ldots, L \}, \label{eq:nonpe-resolved-polygon}
\end{equation}
see figure~\ref{fig:ale-moment-polygons} for $L=2$.
For any $\mathcal{P}_{\boldsymbol{\lambda}}$, the sets $F_j = \{ \boldsymbol{X} \in \mathcal{P}_{\boldsymbol{\lambda}} : \boldsymbol{X} \cdot \nu_j = \lambda_j \}$ are called the \emph{facets} of $\mathcal{P}_{\boldsymbol{\lambda}}$.
In particular, for $\mathcal{C} = \mathcal{P}_{0}$ the facets $F_1, \ldots, F_{L-1}$ degenerate to a point with only $F_0$ and $F_L$ staying non-trivial.%
\footnote{%
  On the (non-trivial) facets of $\mathcal{P}_{\boldsymbol{\lambda}}$, the $T^2$-fibration degenerates to an $S^1$-fibration and on the vertices of $\mathcal{P}_{\boldsymbol{\lambda}}$ to a single point.
  The K\"ahler class of the deformation of $\widetilde{X}_L$ is determined by the volume of the \emph{exceptional curves} $C_j \simeq \mathbb{P}^1$ for $j = 1, \ldots, L-1$, which are the $S^1$-fibrations over the corresponding facets $F_j$.
  We derive in appendix~\ref{app:resolved-volume} that their normalised areas satisfy
\begin{equation}
	t_j
	\equiv
	\frac{1}{2\pi}\int_{C_j}\omega
	=
	2\lambda_j-\lambda_{j-1}-\lambda_{j+1}.
	\label{eq:nonpe-exceptional-areas}
\end{equation}
When all $t_j > 0$, the geometry is the full smooth resolution. In this way, the $\lambda_j$ provide a (redundant) parametrisation of the K\"ahler class, up to shifts of $\mathcal{P}_{\boldsymbol{\lambda}}$ by $\boldsymbol{X} \to \boldsymbol{X} - \boldsymbol{X}_0$.
}

For the angular parameters choose the quotient torus basis $\varphi_i$ with periodicities $\varphi_i \sim \varphi_i + 2 \pi$. Denote the corresponding equivariant weights parametrising the Reeb vector by $\epsilon_{1,2}$. The equivariant volume, normalised as in eq.~\eqref{eq:V-def}, becomes
\begin{equation}
\Veq_{\widetilde X_L} (\boldsymbol\lambda;\epsilon_1,\epsilon_2)
=
\frac{1}{(2 \pi)^2} \int_{\widetilde{X}_L}^{} e^{-H} \frac{\omega^2}{2}
=
\int_{\calp_{\boldsymbol\lambda}} e^{-\epsilon_1X-\epsilon_2Y}\,\d X\,\d Y .
\end{equation}
The remaining integral localises to contributions from $L$ fixed points $\boldsymbol{X}_A = (X_A, Y_A)$, $A = 0, \ldots, L-1$ \cite{Martelli:2023oqk}, corresponding to the intersections of neighbouring facets $F_A$ and $F_{A+1}$,
\begin{equation}
  \mathbb{V}_{\widetilde{X}_L}(\boldsymbol{\lambda}; \epsilon_1, \epsilon_2) =
  \sum_{A=0}^{L-1} \frac{e^{-\epsilon_1 X_A - \epsilon_2 Y_A}}{e^{\mathbb{T}}|_{\boldsymbol{X}_A}} =
  \sum_{A=0}^{L-1} \frac{e^{-\epsilon_1 X_A - \epsilon_2 Y_A}}{[(A+1)\epsilon_1 - \epsilon_2] [\epsilon_2 - A \epsilon_1]},
\end{equation}
where $e^{\mathbb{T}}|_{\boldsymbol{X}_A}$ is the equivariant Euler class of the tangent bundle at $\boldsymbol{X}_A$.
The locations of the fixed points $\boldsymbol{X}_A = (X_A, Y_A)$ in terms of the Kähler deformation parameters are
\begin{equation}
  \begin{aligned}
    X_A &= (A+1) \lambda_A - A \lambda_{A+1}, \\
    Y_A &= \lambda_{A+1} - \lambda_A. \label{eq:nonpe-resolved-vertices}
  \end{aligned}
\end{equation}
For real weights, the integral converges when $0<\epsilon_2<L\epsilon_1$; we do this calculation of the polygon integral directly in appendix~\ref{app:resolved-volume}.

We now relate the equivariant volume to the trilogarithmic part $S_{\text{tri}}$ of the action \eqref{eq:action-gen-flux}.
As in the plethystic case, we find that one may identify $e^{S_{\text{tri}}}$ with the plethystic exponential of the equivariant volume,
\begin{equation}
  \begin{aligned}
    S_{\text{tri}} &=
  \log 
\operatorname{PE} [K^2 \mathbb{V}_{\widetilde{X}_L}(\boldsymbol{\lambda}; \epsilon_1, \epsilon_2) - K^2 \mathbb{V}_{\widetilde{X}_L}(0; \epsilon_1, \epsilon_2)] \\
  &\equiv
	K^2\sum_{\ell=1}^{\infty}\frac1\ell
	\Bigl[
	\Veq_{\widetilde X_L}
	(\boldsymbol\lambda;\ell\epsilon_1,\ell\epsilon_2)
	-
	\Veq_{\widetilde X_L}
	(0;\ell\epsilon_1,\ell\epsilon_2)
	\Bigr],
	\label{eq:nonpe-resolved-resummation}
  \end{aligned}
\end{equation}
with a suitable choice for the equivariant and Kähler deformation parameters.
Concretely, using the ordered fluxes $m_A=m+b_A$ introduced before, set
\begin{equation}
	\epsilon_1=\beta,
	\qquad
	\epsilon_2=L\Delta_2,
	\quad \text{and} \quad
	\lambda_j=-\frac1K\sum_{0 \leq A < j} m_A,
	\quad j=0,\ldots,L.
	\label{eq:nonpe-resolved-map}
\end{equation}
By \eqref{eq:epsilon-gauge-transfo}, these equivariant weights correspond to $(\bar\epsilon_1,\bar\epsilon_2)=(\Delta_1,\Delta_2)$ in the covering-space basis.

Under this identification, the exceptional-curve areas become
\begin{equation}
	t_j
	=
	\frac{m_j-m_{j-1}}K
	=
	\frac{b_j-b_{j-1}}K
	\geq0.
	\label{eq:nonpe-flux-areas}
\end{equation}
Thus the differences between neighbouring baryonic fluxes determine the exceptional-curve areas. Strictly increasing $b_A$ give positive areas for every curve. Equal neighbouring fluxes contract the corresponding curves. Changing the common flux $m$ translates the polygon in the $Y$ direction and leaves these areas unchanged.

Substituting these definitions into the formula for the equivariant volume, one finds
\begin{equation}
  \frac{1}{e^{\mathbb{T}}|_{\boldsymbol{X}_A}} = - \frac{1}{w_A w_{A+1}}
  \quad \text{and} \quad
  - \epsilon_1 X_A - \epsilon_2 Y_A = L \phi_{\boldsymbol{m}}^{(A)} / K,
\end{equation}
so
\begin{equation}
  \mathbb{V}_{\widetilde{X}_L}(\boldsymbol{\lambda}; \ell \epsilon_1, \ell \epsilon_2) =
  - \sum_{A=0}^{L-1} \frac{1}{\ell^2 w_A w_{A+1}} e^{\ell L \phi_{\boldsymbol{m}}^{(A)}/K}.
\end{equation}
Therefore the plethystic exponential indeed formally resums to
\begin{equation}
  \begin{aligned}
  &K^2\sum_{\ell=1}^{\infty}\frac1\ell
	\Bigl[
	\Veq_{\widetilde X_L}
	(\boldsymbol\lambda;\ell\epsilon_1,\ell\epsilon_2)
	-
	\Veq_{\widetilde X_L}
	(0;\ell\epsilon_1,\ell\epsilon_2)
	\Bigr] \\
  =&
  -\sum_{\ell=1}^{\infty}
  \sum_{A=0}^{L-1} 
  \frac{K^2}{\ell^3 w_A w_{A+1}}
	\Bigl[
    e^{\ell L \phi_{\boldsymbol{m}}^{(A)}/K} - 1
	\Bigr] \\
  =&
  - 
  \sum_{A=0}^{L-1} 
  \frac{K^2}{w_A w_{A+1}}
  \left[
    \Li_3(e^{L \phi_{\boldsymbol{m}}^{(A)}/K})
    - \zeta(3)
  \right]
  =
  S_{\text{tri}}.
  \end{aligned}
\end{equation}

\subsection{Comparison with the parent master volume}
\label{sec:parent-master-volume}

Let us now return to equal fluxes $m_A=m$, corresponding to vanishing external baryonic flux $\boldsymbol b=0$, and compare the leading large-wrapping term in \eqref{eq:log-I-res-950} with the parent master-volume prediction \eqref{eq:GG-master-volume-AdS4}. The parent geometry is the full Calabi--Yau fourfold
\begin{equation}
X_4=\mathbb C^2/\mathbb Z_L\times\mathbb C^2/\mathbb Z_K,
\end{equation}
with covering-space equivariant weights $(\Delta_1,\Delta_2)$ and $(\Delta_3,\Delta_4)$ for the two factors, respectively. Its seven-dimensional link determines the off-shell family $Y_7$ entering the proposal of \cite{Chen:2024erz}.

In the product quotient-torus basis, the moment cone is
\begin{equation}
X\geq0,\qquad X+LY\geq0,\qquad U\geq0,\qquad U+KV\geq0.
\end{equation}
Its four primitive inward normals are therefore
\begin{equation}
\begin{aligned}
v_1&=(1,0,0,0), & v_2&=(1,L,0,0),\\
v_3&=(0,0,1,0), & v_4&=(0,0,1,K).
\end{aligned}
\end{equation}
The covering moment coordinates are $y_1=X$, $y_2=X+LY$, $y_3=U$ and $y_4=U+KV$. Writing the Hamiltonian as $H=\sum_I\Delta_I y_I$ then gives the quotient-basis equivariant vector
\begin{equation}
\boldsymbol\epsilon^{(4)} =(\Delta_1+\Delta_2,L\Delta_2, \Delta_3+\Delta_4,K\Delta_4) =\sum_{I=1}^4\Delta_Iv_I, \label{eq:Reeb-relation-general-SE7}
\end{equation}
as expected from \cite{Chen:2024erz}.

Denote the parent deformation parameters by $\Lambda_I$. Substituting the four normals above and \eqref{eq:Reeb-relation-general-SE7} into the master-volume formula \cite{Chen:2024erz}, in the conventions of that reference, gives\footnote{Here the orbifold factor $1/(KL)$ follows from $\det(v_1,v_2,v_3,v_4)=KL$.}
\begin{equation}
\calv_{Y_7} = \frac{(2\pi)^4}{6KL} \frac{\left(\sum_I\Delta_I\Lambda_I\right)^3} {\Delta_1\Delta_2\Delta_3\Delta_4}. \label{eq:parent-orbifold-master-volume}
\end{equation}

For the equal-flux coefficient $\mathcal I_{0,\boldsymbol m}$, the classical weight in the giant graviton expansion is
\begin{equation}
e^{-N\Delta_2\sum_A m_A}=e^{-NLm\Delta_2}.
\end{equation}
Comparing with the parent weight $e^{-N\sum_I\Lambda_I\Delta_I}$ identifies
\begin{equation}
\Lambda_2=Lm,\qquad \Lambda_1=\Lambda_3=\Lambda_4=0.
\end{equation} 
Thus \eqref{eq:parent-orbifold-master-volume} becomes
\begin{equation}
\calv_{Y_7} = \frac{(2\pi)^4L^2m^3\Delta_2^2}{6K\Delta_1\Delta_3\Delta_4}.
\end{equation}
In the Higgs-branch limit, $\omega-\Delta_3-\Delta_4$ remains finite and $\Delta_3/\Delta_4\to1$, so
\begin{equation}
	\frac{\omega^2}{\Delta_3\Delta_4}\longrightarrow 4, 
\end{equation}
and hence the parent prediction reduces to
\begin{equation}
-\frac{\omega^2}{4(2\pi)^4}\calv_{Y_7} \longrightarrow -\frac{L^2\Delta_2^2m^3}{6K\Delta_1}. \label{eq:parent-master-HB}
\end{equation}
On the field-theory side, expanding the trilogarithmic term in \eqref{eq:log-I-res-950} at $L\Delta_2m/K\gg1$, using \eqref{eq:Li3-asymp}, gives
\begin{equation}
\frac{K^2}{L\Delta_1\Delta_2} \left[\Li_3(e^{L\Delta_2m/K})-\zeta(3)\right] \sim -\frac{K^2}{6L\Delta_1\Delta_2} \left(\frac{L\Delta_2m}{K}\right)^3 = -\frac{L^2\Delta_2^2m^3}{6K\Delta_1}.
\end{equation}
The transverse-colour contribution in \eqref{eq:log-I-res-950} grows only linearly with $m$ at fixed chemical potentials and does not modify this cubic term. Thus the leading large-wrapping term of the equal-flux Cardy result agrees with the Higgs-branch reduction of the parent master-volume prediction for general $K$ and $L$.

\subsection{Two non-zero configuration numbers} \label{sect:n1-n2-nonzero}

When there is no baryonic background flux ($\boldsymbol{b} = 0$), the giant graviton coefficients can be computed in terms of fermionic forms even when both $\boldsymbol{n}_I \neq 0$ \cite{roehl2026}.
Let $\boldsymbol{n}_1 = (n_1, \ldots, n_1) \in [-\boldsymbol{b}]_+$ and $\boldsymbol{n}_2 = (n_2, \ldots, n_2) \in [\boldsymbol{b}]_+$.
We will without loss of generality assume that $n_1 \leq n_2$. The giant graviton coefficient is then given by
\begin{equation}
  \label{eq:In1n2-split}
  \mathcal{I}_{\boldsymbol{n}_1, \boldsymbol{n}_2} =
  \mathcal{I}_{\boldsymbol{n}_1', 0} \mathcal{I}_{0, \boldsymbol{n}_2'} \prod_{\alpha=0}^{K-1} \prod_{k=-L_{\alpha} n_1}^{-1} \frac{1}{1 - e^{- \beta k}},
\end{equation}
where the components $(\boldsymbol{n}_1')_{\alpha,i}$ and $(\boldsymbol{n}_2')_{\alpha,i}$ are
\begin{equation}
  (\boldsymbol{n}'_{1})_{\alpha} = (\underbrace{1, \ldots, 1}_{L_{\alpha} n_1})
  \quad \text{and} \quad
  (\boldsymbol{n}'_{2})_{\alpha} = (\underbrace{n_2 - n_1, \ldots, n_2 - n_1}_{L_{\alpha}}, \underbrace{1, \ldots, 1}_{L_{\alpha} n_1}).
\end{equation}

This requires knowledge of the general result with non-zero baryonic background flux for a single wrapping number. Importantly, the number of entries is not equal to $L_{\alpha}$, so occurrences of $L$ in \eqref{eq:action-gen-flux} have to be distinguished in the derivation into those originating in the fugacities $L \Delta_I$ and those coming from the length of $\boldsymbol{n}_{1,2}'$, which we will call $M$ in the following.

A more precise derivation leads to the trilogarithmic contribution to \eqref{eq:action-gen-flux} of the form
\begin{equation}
  \begin{aligned}
    S_{\text{tri}} &\approx -\sum_{A=0}^{M-1}\frac{K^2}{w_Aw_{A+1}}
    \left[\Li_3(e^{L\phi_{\boldsymbol m}^{(A)}/K})-\zeta(3)\right],\\
    w_A&=\log y+(M-A)\beta,
    \qquad
    L\phi_{\boldsymbol m}^{(A)}=w_Am_A+\beta\sum_{0 \leq B<A}m_B.
  \end{aligned}
\end{equation}
Here $\log y=-L\Delta_2$ for the first factor and $\log y=-L\Delta_1$ for the second.

The first factor in \eqref{eq:In1n2-split} corresponds to $M = \sum_{\alpha=0}^{K-1}  L_{\alpha} n_1 = L n_1$ and entries $m_A = 1$, $A = 0, \ldots, M-1$. Using the fact that $n_1 = \mathcal{O}(\beta^{-1})$ and suppressing subleading contributions to the chemical potentials, this leads to
\begin{equation}
  \left.\log\mathcal I_{\boldsymbol n'_1,0}\right|_{\text{tri}}
  \approx \frac{K^2}{\beta L\Delta_2}
  \left[\Li_3(e^{\beta L n_1/K})-\zeta(3)\right].
\end{equation}
The second factor in \eqref{eq:In1n2-split} has $M = L (n_1 + 1)$ with sorted fluxes
\begin{equation}
  m_A =
  \begin{cases}
    1 & A = 0, \ldots, L n_1 - 1, \\
    n_2 - n_1 & A = L n_1, \ldots, L (n_1 + 1) - 1.
  \end{cases}
\end{equation}
This leads to two contributions
\begin{equation}
  \begin{aligned}
    \left.\log\mathcal I_{0,\boldsymbol n'_2}\right|_{\text{tri}}
    \approx{}&\frac{K^2}{L\Delta_1\Delta_2}
    \left[\Li_3(e^{L(n_1\Delta_1+n_2\Delta_2)/K})-\zeta(3)\right]\\
    &-\frac{K^2}{\beta L\Delta_2}
    \left[\Li_3(e^{\beta L n_1/K})-\zeta(3)\right].
  \end{aligned}
\end{equation}
The final factor in \eqref{eq:In1n2-split} is subleading, $\mathcal{O}(\beta^{-1})$. Together, the leading order contribution is therefore
\begin{equation}
  \left.\log\mathcal I_{\boldsymbol n_1,\boldsymbol n_2}\right|_{\text{tri}}
  \approx\frac{K^2}{L\Delta_1\Delta_2}
  \left[\Li_3(e^{L(n_1\Delta_1+n_2\Delta_2)/K})-\zeta(3)\right],
\end{equation}
just as expected for two non-zero wrapping numbers.

The charges of $\boldsymbol{n}_1'$ and $\boldsymbol{n}_2'$ satisfy
\begin{equation}
  J(\boldsymbol{n}_1') = L n_1 = J(\boldsymbol{n}_1)
  \quad \text{and} \quad
  J(\boldsymbol{n}_2') = L (n_2 - n_1) + L n_1 = J(\boldsymbol{n}_2),
\end{equation}
so the additional term from \eqref{eq:action-gen-flux} in this context simply becomes
\begin{equation}
    \frac{1}{2 L}
    \left(
      \frac{J(\boldsymbol{n}_1)}{\Delta_2} + \frac{J(\boldsymbol{n}_2)}{\Delta_1}
    \right)
    \v_{\alpha}^{+} (C^{-1})^{\alpha \beta} \v_{\beta}^{+}.
\end{equation}

Turning on $\boldsymbol{b}$ while leaving both $\boldsymbol{n}_1$ and $\boldsymbol{n}_2$ non-zero is slightly more complicated. First, when $\boldsymbol{b} \ne 0$ there is no direct master volume proposal to compare it to. More pressingly, in this case the fermionic form derivation in \cite{Dorey:2025qht} does not work anymore or needs to be modified. It is therefore currently not possible to derive the results directly from a saddle-point calculation. We note, however, that the choice
\begin{equation}
  \label{eq:lambda-general}
 \lambda_j=-\frac{1}{K}\left( \sum_{B<j}n_{2,B}+\sum_{B\ge j}n_{1,B} \right)  
\end{equation}
in the equivariant volume combines the two one-sided identifications and also reproduces the mixed-wrapping trilogarithmic result at vanishing baryonic flux.
This therefore suggests as a natural candidate
\begin{equation}
\log\mathcal I_{\boldsymbol n_1,\boldsymbol n_2}
\stackrel{?}{\sim}
  -
  \sum_{A=0}^{L-1}
  \frac{K^2}{w_A w_{A+1}}
  \left[
    \Li_3(e^{L \phi_{\boldsymbol{m}}^{(A)}/K})
    - \zeta(3)
  \right]
+
\frac{1}{2 L}
\left(
  \frac{J(\boldsymbol{n}_1)}{\Delta_2} + \frac{J(\boldsymbol{n}_2)}{\Delta_1}
\right)
\v_{\alpha}^{+} (C^{-1})^{\alpha \beta} \v_{\beta}^{+}
\end{equation}
with
$(\epsilon_1, \epsilon_2) = (\beta, L \Delta_2)$
and $L \phi_{\boldsymbol{m}}^{(A)}/K = - \epsilon_1 X_A - \epsilon_2 Y_A$ depending on $\lambda_j$ as before, now using \eqref{eq:lambda-general}.

\section{Discussion} \label{sec:outlook}

In this paper we have studied the Cardy asymptotics of the Higgs-branch giant graviton expansion in both the plethystic and non-plethystic cases. In the plethystic case, we obtained a trilogarithmic expression that captures both the fixed-wrapping and large-wrapping regimes. At fixed wrapping, the leading behaviour is governed by the effective central charge of the associated vertex algebra. At large wrapping, we recovered the Higgs-branch reduction of the previously proposed master-volume behaviour of the parent $Y_7$, together with systematic subleading terms. We showed that the equivariant volume of the Higgs-branch cone governs the wrapping dependence of these asymptotics, including the corrections to the leading master-volume term.

In the non-plethystic case, the equivariant volume still governs the trilogarithmic contribution. For unequal background fluxes, the relevant geometry is the resolution of $\mathbb C^2/\mathbb Z_L$. Differences between neighbouring ordered fluxes determine the areas of compact curves, while a common shift translates the moment polygon without changing those areas. The equal-flux result is recovered when all these curves contract and the geometry reduces to the translated singular cone. A special property of the non-plethystic case is that the asymptotics also contain an additional term quadratic in the relative colour potentials, which generically contributes at leading Cardy order. Its form agrees with the modular factor of the affine $\widehat{\mathfrak{su}}(K)_m$ characters appearing in the vertex-algebra description \cite{Kac:1984mq}. It grows only linearly with the wrapping number, however, and therefore leaves the leading cubic master-volume behaviour unchanged. 

An important question is how much of this structure survives away from the Higgs-branch limit. Our calculation first restricts the index to the Higgs-branch sector and then takes its Cardy asymptotics. The agreement with the Higgs-branch reduction of the parent master-volume formula is encouraging, but it does not determine the fate of the additional terms. A useful next test would be to obtain a subleading large-wrapping term in the full giant graviton coefficient and compare its Higgs-branch limit with our result. 

Our results also have an application to the topologically twisted index at zero external flux. For ADHM theory, the B-twisted index coincides, up to a conventional fugacity prefactor, with the Higgs-branch Hilbert series \cite{Closset:2016arn}. Its giant graviton expansion therefore also gives an expansion of this twisted index, to which our coefficient asymptotics apply. More generally, the superconformal and twisted indices can be obtained by different gluings of hemisphere blocks.
One possibility is to perform a giant graviton expansion of the blocks themselves, as suggested in \cite{Chen:2024erz}, and then apply the twisted gluing prescription.
This would be the giant graviton analogue of recent work relating the Cardy expansions of the superconformal and refined twisted indices to sums of products of squashed-three-sphere saddle contributions \cite{Bobev:2026lvl}.
Equivariant volumes also enter in proposals for M2-brane sphere partition functions, where they determine the constant-map contributions to the topological-string free energy \cite{Cassia:2025aus, Cassia:2025jkr}.

A broader motivation for seeking a giant graviton decomposition for the twisted index is $\mathcal I$-extremisation. For static supersymmetric magnetically charged AdS$_4$ black holes, extremising the large-$N$ twisted index over chemical potentials at fixed magnetic flux gives the black-hole entropy \cite{Benini:2015eyy}. Its geometric counterpart uses a supersymmetric action constructed from the master volume and its derivatives, subject to flux quantisation \cite{Gauntlett:2019roi}. Both the master volume and the supersymmetric action can be extracted from the K\"ahler expansion of the corresponding equivariant volume \cite{Martelli:2023oqk,Colombo:2023fhu}. A giant graviton expansion in these magnetic backgrounds would therefore allow us to ask whether the geometric entropy function can be derived from the sum over wrapping sectors.

Another direction is to understand what individual giant graviton sectors describe in supergravity. The master-volume correspondence suggests that the K\"ahler deformations describe the backreaction of particular giant graviton configurations \cite{Chen:2024erz}. A possible way to investigate this interpretation is through gravitational localisation, which has been used to compute supersymmetric AdS$_5$ black-hole on-shell actions from toric data \cite{Colombo:2025ihp}. Its extension to include the internal geometry was proposed as a way to connect with the giant graviton/master-volume relation. A related proposal expresses black-saddle actions, including higher-derivative corrections, through equivariant integrals of the anomaly polynomial \cite{Cassani:2026teb}. It would be interesting to understand whether similar approaches can provide a gravitational explanation of the Higgs-branch giant graviton asymptotics.

\acknowledgments

We would like to thank Heng-Yu Chen, Nick Dorey, and Boan (Andy) Zhao for helpful discussions. 
The work of PLR has been partially supported by STFC consolidated grant ST/X000664/1, a research studentship from the Cambridge Philosophical Society and the National Research Foundation of Korea (NRF) grant RS-2024-00405629.

\appendix
\crefalias{section}{appendix}

\section{Equivariant volumes of simple cones} \label{app:equiv-vol}

For $\mathbb{C}^2$, we use complex coordinates $(z_1,z_2)$ with the standard $T^2$ action
\begin{equation}
	(z_1,z_2)\;\mapsto\;(e^{i\varphi_1}z_1,e^{i\varphi_2}z_2) , \qquad (\varphi_1,\varphi_2) \in \mathbb{R}^2/2\pi \mathbb{Z}^2.
\end{equation}
The Kähler form and moment maps are
\begin{equation}
	\omega
	= \frac{i}{2}\bigl(dz_1\wedge d\bar z_1 + dz_2\wedge d\bar z_2\bigr)
	= dy_1\wedge d\varphi_1 + dy_2\wedge d\varphi_2,\qquad
	y_i = \frac{|z_i|^2}{2}\ge0\,.
\end{equation}
The Reeb vector is
\begin{equation}
	\xi = \epsilon_1\,\partial_{\varphi_1}
	+ \epsilon_2 \, \partial_{\varphi_2},
\end{equation}
and the symplectic cone is just $\mathbb R^2_{\ge0}$ with coordinates
$(y_1,y_2)$.

Non-compact facets correspond to $y_1=0$ and $y_2=0$, so there are two Kähler parameters $\lambda_1,\lambda_2$, which shift the moment maps $y_a\rightarrow y_a+\lambda_a$.

So, we obtain
\begin{align}
	\Veq_{\mathbb C^2}(\lambda_1,\lambda_2;\epsilon_1,\epsilon_2)
	 & = \frac{1}{(2\pi)^2}
	\int_{\mathbb C^2} e^{-H} \frac{\omega^2}{2!}
	\\ &= \frac{1}{(2\pi)^2}
	\int_0^{2\pi}\d \varphi_1\int_0^{2\pi}\d\varphi_2
	\int_0^\infty \d y_1\int_0^\infty \d y_2\;
	e^{-\epsilon_1(y_1+\lambda_1)-\epsilon_2(y_2+\lambda_2)}
	\\
	 & = \frac{1}{\epsilon_1\epsilon_2}\;
	e^{-(\epsilon_1\lambda_1+\epsilon_2\lambda_2)}.
	\label{eq:VC2-exact}
\end{align}
Expanding the exponential in $\lambda_{1,2}$ gives
\begin{equation}
	\Veq_{\mathbb C^2}(\lambda_1,\lambda_2;\epsilon_1,\epsilon_2)
	= \frac{1}{\epsilon_1\epsilon_2}
	\;-\;
	\left(\frac{\lambda_1}{\epsilon_2}
	+ \frac{\lambda_2}{\epsilon_1}\right)
	\;+\;\mathcal O(\lambda^2).
	\label{eq:VC2-expansion}
\end{equation}
Thus,
\begin{equation}
	\Veq^{(0)} = \frac{1}{\epsilon_1\epsilon_2},\qquad
	\Veq^{(1)} = -\left(\frac{\lambda_1}{\epsilon_2}
	+ \frac{\lambda_2}{\epsilon_1}\right),
\end{equation} where the first term corresponds to the Sasakian volume, while the second term gives the master volume
\begin{equation}
	\calv_{Y_3} = {(2\pi)^2} \Veq^{(1)} = -4\pi^2 \left(\frac{\lambda_1}{\epsilon_2}
	+ \frac{\lambda_2}{\epsilon_1}\right)
\end{equation} of the link $Y_3 \simeq S^3$, and encodes the deformation of the transverse Kähler class by
$\lambda_a$.

%Finally, the supersymmetric action is obtained from \eqref{eq:Ssusy-defn} as
%\begin{equation}
%	S_{\rm SUSY}(\epsilon_1,\epsilon_2)
%	= -\sum_{a=1}^2 \frac{\partial}{\partial\lambda_a}
%	\calv_{Y_3}(\lambda,\epsilon)
%	= 4\pi^2\left(\frac{1}{\epsilon_2}+\frac{1}{\epsilon_1}\right)
%	\label{eq:SSUSY-C2}.
%\end{equation}

Now we move to the orbifold case. A standard way to write the $\mathbb{C}^2/\mathbb{Z}_L$ action is
\begin{equation}
	(z_1,z_2) \rightarrow (\omega^q z_1, \omega^{-1} z_2), \qquad \omega = e^{2\pi i/L}, \label{eq:C2-ZL-action}
\end{equation} with $\gcd(q,L)=1$.

One can understand \eqref{eq:C2-ZL-action} in an elementary way as follows. A generator $g$ of $\mathbb{Z}_L$ satisfies $g^L = \mathrm{id}$, and thus acts on $\mathbb{C}^2$ as
\begin{equation}
	g\, : \, (z_1,z_2) \rightarrow (e^{\frac{2\pi i a}{L}} z_1, e^{\frac{2\pi i b}{L}}z_2),
\end{equation} with $a$ and $b$ satisfying the condition $\gcd (a,L)=\gcd(b,L)=1$ so that the action is faithful and the quotient has only an isolated singularity\footnote{The necessary condition for a faithful action is $\gcd(a,b,L)=1$. Note that if $\gcd(a,b,L)=d >1$, then $g^{L/d}$ acts trivially, so the map has a non-trivial kernel and one really is quotienting by $\mathbb{Z}_{L/d}$, not all of $\mathbb{Z}_L$. The stronger condition $\gcd (a,L)=\gcd(b,L)=1$ is then obtained by demanding that there are no fixed loci except at the origin.}. Then, $\tilde{g}=g^{-b^{-1}}$ is also a generator of $\mathbb{Z}_L$ and acts as
\begin{equation}
	\tilde{g}\,:\, (z_1,z_2) \rightarrow (e^{-\frac{2\pi i b^{-1}a}{L}}z_1, e^{-2\pi i/L} z_2),
\end{equation}
which gives \eqref{eq:C2-ZL-action} upon defining $q\equiv-ab^{-1}\pmod L$.

Alternatively, using the toric description \cite{Martelli:2023oqk}, one can arrive at \eqref{eq:C2-ZL-action} by starting from the fan vectors\footnote{For $q=1$, $v^1=\nu_0$ and $v^2=\nu_L$ in the notation of the main text.}
\begin{equation}
	v^1 = (1,0), \qquad v^2 = (q,L),
\end{equation} with the corresponding action
\begin{equation}
	(z_1,z_2) \rightarrow (e^{2\pi i Q_1} z_1, e^{2\pi i Q_2} z_2),
\end{equation}
and then solving the GLSM condition
\begin{equation}
	\sum_a Q_a v^a \in \mathbb{Z}^2.
\end{equation}

The coordinates on the covering space $\mathbb{C}^2$ are given as\footnote{For the orbifold calculation, we denote the covering-space angles by $\phi_i$ and reserve $\varphi_i$ for the quotient torus.}
\begin{equation}
	z_1 = \sqrt{2y_1} e^{i\phi_1} \qquad z_2=\sqrt{2y_2} e^{i\phi_2}, \qquad \phi_i \sim \phi_i + 2\pi,
\end{equation}
so \eqref{eq:C2-ZL-action} gives
\begin{equation}
	(\phi_1,\phi_2) \rightarrow (\phi_1 + \frac{2\pi q}{L},\phi_2 - \frac{2\pi}{L}), \label{eq:C2-ZL-identification}
\end{equation} which shows that $(\phi_1,\phi_2)$ are not good coordinates on the quotient torus. The fan vectors determine the phase rotations acting on the quotient torus such that 
\begin{equation}
\begin{pmatrix} 
\varphi_1\\ \varphi_2\end{pmatrix} = \phi_1 v^1+\phi_2 v^2 = \begin{pmatrix} \phi_1+q\phi_2\\ L\phi_2 \end{pmatrix}
\pmod{2\pi\mathbb Z^2}.
\end{equation} Inverting this then gives\footnote{Note that there are many equivalent valid choices for the angles $\varphi_1,\varphi_2$. We picked \eqref{eq:orb-local-angular-coordinates} to match with the conventions of \cite{Martelli:2023oqk}, as determined by the relative signs of the gauge invariant combinations in \eqref{eq:global-epsilon-combinations}.}
\begin{equation}
	\phi_1 = \varphi_1 - \frac{q}{L} \varphi_2 \qquad \phi_2 = \frac{1}{L} \varphi_2, \label{eq:orb-local-angular-coordinates}
\end{equation}
so that the shift $\varphi_2 \rightarrow \varphi_2 - 2\pi$ produces the orbifold identification \eqref{eq:C2-ZL-identification}, and $(\varphi_1,\varphi_2)$ form a basis of cycles for the orbifold. % \footnote{Note that the orbifold $\mathbb{C}^2/\mathbb{Z}_L$ can be thought of as a lattice generated by the vectors $\tilde{v}_1=(2\pi, 0)$, $\tilde{v}_2=(0,2\pi)$, and $\tilde{v}_3 = (\frac{2\pi q}{L},-\frac{2\pi}{L})$. } %The vector fields $\partial_{\varphi_i}$ are then globally defined $U(1)$ generators on $\mathbb{C}^2/\mathbb{Z}_L$.

The K\"ahler form
\begin{equation}
	\omega = \d y_1 \wedge \d \phi_1 + \d y_2 \wedge \d \phi_2 \qquad\qquad  \frac{\omega^2}{2!} = \d y_1 \wedge \d \phi_1 \wedge \d y_2 \wedge \d\phi_2
\end{equation} and the moment map conditions
\begin{equation}
	i_{\partial_{\phi_i}} \omega = - \d y_i
\end{equation} are unchanged locally on the cover but it is useful to express them in terms of coordinates on the quotient torus. Using \eqref{eq:orb-local-angular-coordinates}, we get
\begin{equation}
	\omega = \d y_1 \wedge \d \varphi_1 - \frac{q}{L} \d y_1 \wedge \d\varphi_2  + \frac{1}{L} \d y_2 \wedge \d\varphi_2,
\end{equation} implying
\begin{equation}
	i_{\partial_{\varphi_i}} \omega = - \d Y_i, \qquad\quad Y_1 = y_1 \qquad Y_2 = \frac{y_2-qy_1}{L}.  \label{eq:orbifold-Y-coordinates}
\end{equation}

The Reeb vector on the quotient can be written in terms of both angular coordinates as
\begin{equation}
	\xi = \epsilon_i \partial_{\varphi_i} =\bar{\epsilon}_i \partial_{\phi_i}.
\end{equation} Solving
\begin{equation}
	i_\xi \omega = -\d H  = \epsilon_1 i_{\partial_{\varphi_1}} \omega + \epsilon_2 i_{\partial_{\varphi_2}} \omega  = -\epsilon_1 \d Y_1 - \epsilon_2 \d Y_2 ,
\end{equation}
we then obtain the Hamiltonian
\begin{equation}
	H = \epsilon_1 Y_1 + \epsilon_2 Y_2 = \bar{\epsilon}_1 y_1 + \bar{\epsilon}_2 y_2,
\end{equation}
where we find via \eqref{eq:orbifold-Y-coordinates} that the components of the Reeb vector in the covering and quotient space angular coordinates are related as
\begin{equation}
	\bar{\epsilon}_1 = \epsilon_1 - \frac{q \epsilon_2}{L}\qquad \qquad  \bar{\epsilon}_2 = \frac{\epsilon_2}{L}, \label{eq:epsilon-bar-identify}
\end{equation} or equivalently,
\begin{equation}
	\epsilon_1 = \bar{\epsilon}_1 + q \bar{\epsilon}_2 \qquad\qquad \epsilon_2 = L \bar{\epsilon}_2. \label{eq:global-epsilon-combinations}
\end{equation}

We can now calculate the equivariant volume,
\begin{align}
	\mathbb{V}_{\mathbb C^2/\mathbb Z_L}(\lambda_1,\lambda_2;{\epsilon}_1,{\epsilon}_2) & = \frac{1}{(2\pi)^2} \int e^{-H} \frac{\omega^2}{2!} \\ &= \frac{1}{(2\pi)^2} \int \d\phi_1 \d\phi_2 \int \d y_1 \d y_2 e^{-H} \\ &= \frac{1}{(2\pi)^2} \int^{2\pi}_0 \int^{2\pi}_0 \frac{\d\varphi_1 \d\varphi_2}{L} \int_0^\infty \int_0^\infty \d y_1 \d y_2 e^{-H} \\ &= \frac{1}{L} \int^\infty_0 \int^\infty_0 \d y_1 \d y_2 \exp(-\bar\epsilon_1 (y_1 + \lambda_1) -\bar\epsilon_2 (y_2 + \lambda_2) ) \\ &= \frac{e^{-\bar{\epsilon}_i \lambda_i}}{L\bar{\epsilon}_1\bar{\epsilon}_2} . \label{eq:orb-equiv-volume}
\end{align}

Finally, let us note that in our case the Higgs-branch cone is the supersymmetric Calabi--Yau orbifold $\mathbb{C}^2/\mathbb{Z}_L$ of type $A_{L-1}$, so the $\mathbb{Z}_L$ action must preserve the holomorphic two-form $\d z_1\wedge \d z_2$.\footnote{Equivalently, the Calabi--Yau condition requires the orbifold group to act through $SU(2)\subset U(2)$. For the diagonal action $(z_1,z_2)\mapsto(\omega^q z_1,\omega^{-1}z_2)$, this means $\det(\omega^q,\omega^{-1})=\omega^{q-1}=1$.} Since
\begin{equation}
	\d z_1 \wedge \d z_2 \mapsto \omega^{q-1} \d z_1\wedge \d z_2,
\end{equation}
this requires $q\equiv 1 \pmod L$, and hence $q=1$.

\subsection{Resolution of \texorpdfstring{$\mathbb C^2/\mathbb Z_L$}{C\textasciicircum 2/Z\textunderscore L}} \label{app:resolved-volume}

We now derive the exceptional-curve areas and the equivariant volume used in the geometric interpretation of the non-plethystic result. 

On the interior of the polygon \eqref{eq:nonpe-resolved-polygon}, the moment coordinates $(X,Y)$ and the $2\pi$-periodic quotient angles $(\varphi_1,\varphi_2)$ give
\begin{equation}
\omega =\d X\wedge\d\varphi_1+\d Y\wedge\d\varphi_2. \label{eq:app-resolved-symplectic}
\end{equation}
The compact facet $F_j$ with normal $\nu_j=(1,j)$ joins $\boldsymbol X_{j-1}$ to $\boldsymbol X_j$. Substituting the vertex coordinates \eqref{eq:nonpe-resolved-vertices} gives
\begin{equation}
\boldsymbol X_j-\boldsymbol X_{j-1}=d_j(j,-1),
\quad \text{where} \quad
d_j=2\lambda_j-\lambda_{j-1}-\lambda_{j+1}, \quad j=1,\ldots,L-1.
\end{equation}
We first take $d_j>0$, so that all compact edges have positive length. Parametrise the $j$th edge by
\begin{equation}
(X,Y)=\boldsymbol X_{j-1}+s(j,-1),
\qquad 0\leq s\leq d_j.
\end{equation}
Along this edge, the circle generated by $\partial_{\varphi_1}+j\partial_{\varphi_2}$ collapses. The remaining angular coordinate is
\begin{equation}
\psi=j\varphi_1-\varphi_2, \qquad \psi\sim\psi+2\pi.
\end{equation} At each endpoint, the edge meets a neighbouring edge, so the remaining $\psi$-circle also contracts to a single point. Hence the surface above the compact edge consists of a circle over an interval, with the circle contracted at both ends. This gives the two-sphere $C_j\simeq\mathbb{P}^1$. 

Restricting \eqref{eq:app-resolved-symplectic} to $C_j$ gives
\begin{equation}
\left.\omega\right|_{C_j} =\d s\wedge(j\,\d\varphi_1-\d\varphi_2) =\d s\wedge\d\psi.
\end{equation}
Its normalised area is therefore
\begin{equation}
t_j =\frac{1}{2\pi}\int_{C_j}\omega =\frac{1}{2\pi} \int_0^{d_j}\d s\int_0^{2\pi}\d\psi =d_j =2\lambda_j -\lambda_{j-1}-\lambda_{j+1}. \label{eq:app-resolved-areas}
\end{equation}
The Hamiltonian and symplectic volume form are
\begin{equation}
H=\epsilon_1X+\epsilon_2Y, \qquad \frac{\omega^2}{2}
=\d X\wedge\d\varphi_1\wedge
\d Y\wedge\d\varphi_2,
\end{equation}
so the equivariant volume of the resolved space is given by
\begin{equation}
\Veq_{\widetilde X_L}(\boldsymbol\lambda;\epsilon_1,\epsilon_2) = \frac{1}{(2\pi)^2} \int_{\widetilde X_L}e^{-H}\frac{\omega^2}{2} = \int_{\calp_{\boldsymbol\lambda}} e^{-\epsilon_1 X-\epsilon_2 Y}\,\d X\,\d Y. \label{eq:app-resolved-polygon-integral}
\end{equation}
At fixed $Y$, the polygon inequalities require $X\geq X_{\min}(Y)$, where
\begin{equation}
X_{\min}(Y) =\max_{0\leq j\leq L}(\lambda_j-jY).
\end{equation}
The volume integral then splits as
\begin{align}
\Veq_{\widetilde X_L}
&=
\frac{1}{\epsilon_1}
\Biggl[
e^{-\epsilon_1\lambda_0}
\int_{Y_0}^{\infty}e^{-w_0Y}\,\d Y \nonumber\\
&\qquad +\sum_{j=1}^{L-1}
e^{-\epsilon_1\lambda_j} \int_{Y_j}^{Y_{j-1}}e^{-w_jY}\,\d Y
+ e^{-\epsilon_1\lambda_L} \int_{-\infty}^{Y_{L-1}}e^{-w_LY}\,\d Y \Biggr],
\end{align}
where $w_j=\epsilon_2-j\epsilon_1$, $j=0,\ldots,L$, with $0<\epsilon_2<L\epsilon_1$ for convergence. Evaluating the integrals gives
\begin{equation}
\Veq_{\widetilde X_L}
=
\frac{1}{\epsilon_1} \left[ \frac{Q_0}{w_0} +\sum_{j=1}^{L-1} \frac{Q_j-Q_{j-1}}{w_j} -\frac{Q_{L-1}}{w_L} \right], \label{eq:app-resolved-strip-integrals}
\end{equation}
where $Q_A=e^{-\epsilon_1X_A-\epsilon_2Y_A}$, $A=0,\ldots,L-1$. 

Collecting the two terms containing each $Q_A$, its coefficient becomes
\begin{equation}
\frac{1}{\epsilon_1}
\left(\frac{1}{w_A}-\frac{1}{w_{A+1}}\right)
=
-\frac{1}{w_Aw_{A+1}} = \frac{1}{[(A+1)\epsilon_1-\epsilon_2][\epsilon_2-A\epsilon_1]},
\end{equation}
where we used $w_{A+1}=w_A-\epsilon_1$. Hence
\begin{equation}
\Veq_{\widetilde X_L} (\boldsymbol\lambda;\epsilon_1,\epsilon_2) 
=
\sum_{A=0}^{L-1} \frac{e^{-\epsilon_1X_A-\epsilon_2Y_A}}{[(A+1)\epsilon_1-\epsilon_2][\epsilon_2-A\epsilon_1]}. \label{eq:app-resolved-volume}
\end{equation}

\section{Saddle point equations}
\label{sec:saddlept-eq-deriv}

In this appendix, we derive the saddle point equations \eqref{eq:f-eom-210} and \eqref{eq:g-eom-b36} by varying the effective action \eqref{eq:saddlept-action-324} with respect to $\mu_k^{\alpha}$, reproduced here for convenience:
\begin{equation}
	\label{eq:saddlept-action-324-2}
	\begin{aligned}
		S[\mu] \approx
		 & - \frac{\beta}{2} \boldsymbol{\mu} \A \boldsymbol{\mu}
		+ \beta (\boldsymbol{\mu}, \boldsymbol{\tilde{\mu}}) +
		\sum_{k=1}^{\infty} \sum_{\alpha = 0}^{K-1}
		k \mu_k^{\alpha} \log y_{\alpha}                          \\
		 & + \sum_{k=1}^{\infty}  \sum_{\alpha = 0}^{K-1}
		\int_{0}^{\mu_{k}^{\alpha}} \!\d t \left[\log (1 - q^{t} w_k^{(\alpha)})
			- \log (1 - q^t)\right].
	\end{aligned}
\end{equation}
First, note that
\begin{equation}
  \frac{\partial}{\partial\mu_k^{\alpha}}
  \left[-\frac{\beta}{2}\boldsymbol\mu\A\boldsymbol\mu
  +\beta(\boldsymbol\mu,\boldsymbol{\tilde\mu})\right]
  =-\log w_k^{(\alpha)}.
\end{equation}
Differentiating the integral with respect to its upper limit gives
\begin{equation}
  \log (1-v_k^{(\alpha)}w_k^{(\alpha)}) - \log(1-v_k^{(\alpha)})
  =-\log f_k^{(\alpha)}.
\end{equation}
For the dependence on $w_k^{(\alpha)}$, we note that
\begin{equation}
  \frac{\partial\log w_l^{(\gamma)}}{\partial\mu_k^{\alpha}}
  =\beta\A_{\gamma\alpha}^{lk}
  =\beta\A_{\alpha\gamma}^{kl},
\end{equation}
so
\begin{equation}
  \int_0^{\mu_l^{\gamma}}\!\d t\,
  \frac{\partial}{\partial\mu_k^{\alpha}}
  \log\bigl(1-q^t w_l^{(\gamma)}\bigr)
  =-\beta\A_{\alpha\gamma}^{kl}
  \int_0^{\mu_l^{\gamma}}\!\d t\,
  \frac{q^t w_l^{(\gamma)}}{1-q^t w_l^{(\gamma)}}
  =
  - \beta \A_{\alpha \gamma}^{kl}
  \int_{0}^{\mu_l^{\gamma}} \!\d t \,
  \Li_0 (q^t w_l^{(\gamma)}).
\end{equation}
Integrating the polylogarithm leads to
\begin{equation}
  \A_{\alpha \gamma}^{kl} \Li_1(q^t w_l^{(\gamma)}) \Big|_{t=0}^{\mu_l^{\gamma}}
  =
  -\A_{\alpha\gamma}^{kl}
  \log\frac{1-v_l^{(\gamma)}w_l^{(\gamma)}}{1-w_l^{(\gamma)}} =
  \A_{\alpha \gamma}^{kl} \log (1 - g_l^{(\gamma)}).
\end{equation}
In total,
\begin{equation}
  \label{eq:speq-der-b50}
  \frac{\partial S[\mu]}{\partial \mu_k^{\alpha}}=-\log w_k^{(\alpha)}+k\log y_\alpha-\log f_k^{(\alpha)}
  +\sum_{(\gamma,l)}\A_{\alpha\gamma}^{kl}
  \log (1 - g_l^{(\gamma)}) \stackrel{!}{=} 0.
\end{equation}
Using 
\begin{equation}
  g_k^{(\alpha)}=f_k^{(\alpha)}w_k^{(\alpha)},
\end{equation}
\cref{eq:speq-der-b50} can be directly rewritten as
\begin{equation}
  \log g_k^{(\alpha)}=k\log y_\alpha
  +\sum_{(\gamma,l)}\A_{\alpha\gamma}^{kl}
  \log\bigl(1-g_l^{(\gamma)}\bigr),
\end{equation}
the exponentiated form of which is \cref{eq:g-eom-b36}.
If one instead substitutes
\begin{equation}
  \log w_k^{(\alpha)}
  =-\beta\sum_{l=1}^{\infty}\min(k,l)\tilde\mu_{\alpha,l}
  -\sum_{(\gamma,l)}\A_{\alpha\gamma}^{kl}\log v_l^{(\gamma)}
  \quad \text{and} \quad
  v_k^{(\alpha)}
  =\frac{1-f_k^{(\alpha)}}{1-g_k^{(\alpha)}},
\end{equation}
the saddle point equation becomes
\begin{equation}
  \log f_k^{(\alpha)}=k\log y_\alpha
  +\beta\sum_{l=1}^{\infty}\min(k,l)\tilde\mu_{\alpha,l}
  +\sum_{(\gamma,l)}\A_{\alpha\gamma}^{kl}
  \log\bigl(1-f_l^{(\gamma)}\bigr).
\end{equation}
Using
\begin{equation}
  \sum_{l=1}^{\infty}\min(k,l)\tilde\mu_{\alpha,l}
  \equiv
  \sum_{l=1}^{\infty} \min(k,l) \operatorname{mult}_l(\boldsymbol{m}_{\alpha})
  =\sum_{i=1}^{L_\alpha}\min(k,m_{\alpha,i}),
\end{equation}
this recovers \cref{eq:f-eom-210}.

\section{Saddle point action}
\label{sec:saddlept-action-deriv}

In this appendix we derive \eqref{eq:saddlept-158} for the saddle point value of the action. We start with the integrals in \eqref{eq:saddlept-action-324}.

First,
\begin{equation}
	\int_{0}^{\mu_k^{\alpha}} \!\d t
	\log (1 - q^t w_k^{(\alpha)}) =
	- \int_{0}^{\mu_k^{\alpha}} \!\d t
	\Li_1(w_k^{(\alpha)} e^{- \beta t}) =
	\frac{1}{\beta} \left[\Li_2(v_k^{(\alpha)} w_k^{(\alpha)}) - \Li_2 (w_k^{(\alpha)})\right]
\end{equation}
and
\begin{equation}
	- \int_{0}^{\mu_k^{\alpha}} \!\d t
	\log (1 - q^t) =
	\frac{1}{\beta} \left[-\Li_2(v_k^{(\alpha)}) + \Li_2(1)\right].
\end{equation}
By the dilogarithmic five-term identity \cite{Kirillov:1994en},
\begin{equation}
	\Li_2(v_k^{(\alpha)} w_k^{(\alpha)}) - \Li_2 (w_k^{(\alpha)}) -\Li_2(v_k^{(\alpha)}) + \Li_2(1) =
	\Li_2(f_k^{(\alpha)}) - \Li_2(g_k^{(\alpha)}) + \log (v_k^{(\alpha)}) \log(f_k^{(\alpha)}).
\end{equation}
Adding these contributions to the remaining terms in \eqref{eq:saddlept-action-324}, we obtain
\begin{align}
	S ={} & - \frac{\beta}{2} \boldsymbol{\mu} \A \boldsymbol{\mu}
	+ \beta (\boldsymbol{\mu}, \tilde{\boldsymbol{\mu}})
	+ \sum_{(\alpha, k)}^{} 
	k \mu_k^{\alpha} \log y_{\alpha} \nonumber                          \\
	      & + \frac{1}{\beta} \sum_{(\alpha, k)}^{} 
	\left[
		\Li_2\bigl(f_k^{(\alpha)}\bigr)
		- \Li_2\bigl(g_k^{(\alpha)}\bigr)
		+ \log v_k^{(\alpha)} \log f_k^{(\alpha)}
		\right],
\end{align}
where we again used the composite indices $(\alpha, k)$ with $\alpha = 0, \ldots, K-1$ and $k = 1, \ldots, \infty$.
The definitions of $v_k^{(\alpha)}$ and $w_k^{(\alpha)}$ imply
\begin{equation}
	\log v_k^{(\alpha)} = -\beta \mu_k^{\alpha},
	\qquad
	\log w_k^{(\alpha)}
	= \beta \sum_{(\gamma, l)}^{} \A_{\alpha\gamma}^{kl}\mu_l^{\gamma}
	- \beta \sum_{l=1}^{\infty}\min(k,l)\tilde\mu_{\alpha,l}.
\end{equation}
Consequently,
\begin{equation}
	\beta (\boldsymbol{\mu}, \tilde{\boldsymbol{\mu}})
	- \beta \boldsymbol{\mu} \A \boldsymbol{\mu}
	= \frac{1}{\beta}
  \sum_{(\alpha, k)}^{} 
	\log v_k^{(\alpha)}\log w_k^{(\alpha)}.
\end{equation}
Moreover, the source-free saddle point equation \eqref{eq:g-eom-b36} gives
\begin{equation}
	k\log y_\alpha
	= \log g_k^{(\alpha)}
	- \sum_{(\gamma, l)}^{} 
	\A_{\alpha\gamma}^{kl}\log\bigl(1-g_l^{(\gamma)}\bigr).
\end{equation}
Using these two relations in the action yields
\begin{align}
	S ={} & \frac{1}{\beta}
  \sum_{(\alpha, k)}^{} 
	\left[
		\Li_2\bigl(f_k^{(\alpha)}\bigr)
		-\Li_2\bigl(g_k^{(\alpha)}\bigr)
		\right]
	+\frac{\beta}{2}\boldsymbol{\mu}\A\boldsymbol{\mu} \nonumber \\
	      & +\frac{1}{\beta}
        \sum_{(\alpha, k)}^{} \sum_{(\gamma, l)}^{} 
	\log v_k^{(\alpha)}\A_{\alpha\gamma}^{kl}
	\log\bigl(1-g_l^{(\gamma)}\bigr),
\end{align}
where we used $g_k^{(\alpha)}=f_k^{(\alpha)}w_k^{(\alpha)}$ to cancel the remaining logarithms. Finally,
\begin{equation}
	\log v_k^{(\alpha)}
	=\log\bigl(1-f_k^{(\alpha)}\bigr)
	-\log\bigl(1-g_k^{(\alpha)}\bigr),
\end{equation}
and
\begin{equation}
	\frac{\beta}{2}\boldsymbol{\mu}\A\boldsymbol{\mu}
	=\frac{1}{2\beta}
  \sum_{(\alpha, k)}^{} \sum_{(\gamma, l)}^{} 
	\log v_k^{(\alpha)}\A_{\alpha\gamma}^{kl}\log v_l^{(\gamma)}.
\end{equation}
Since $\A_{\alpha\beta}^{kl}$ is symmetric, the terms therefore combine to give
\begin{equation}
  \begin{aligned}
	S ={} & \frac{1}{\beta}
  \sum_{(\alpha, k)}^{} 
	\left[
		\Li_2\bigl(f_k^{(\alpha)}\bigr)
		-\Li_2\bigl(g_k^{(\alpha)}\bigr)
	\right]         \\
	      & +\frac{1}{2\beta}
        \sum_{(\alpha, k)}^{} \sum_{(\gamma, l)}^{} 
	\A_{\alpha\gamma}^{kl}
	\left[
		\Li_1\bigl(f_k^{(\alpha)}\bigr)\Li_1\bigl(f_l^{(\gamma)}\bigr)
		-\Li_1\bigl(g_k^{(\alpha)}\bigr)\Li_1\bigl(g_l^{(\gamma)}\bigr)
		\right],
  \end{aligned}
\end{equation}
which is \eqref{eq:saddlept-158}.

\section{Deviations from the centre of mass}
\label{sec:deviations}

Split the solutions to the saddle point equations \eqref{eq:f-eom-210} into
\begin{equation}
	\log f_k^{(\alpha)} = \log \bar{f}_k + \log \delta f_k^{(\alpha)},
\end{equation}
with the centre-of-mass mode $\bar{f}_k$ again defined by
\begin{equation}
	\log \bar{f}_k \equiv \frac{1}{K} \sum_{\alpha=0}^{K-1} \log f_k^{(\alpha)} = k \frac{\log y}{K} + \frac{\beta}{K} \sum_{\alpha=0}^{K-1} \sum_{i=1}^{L_{\alpha}} \min(k, m_{\alpha, i}).
\end{equation} 
We now set all $m_{\alpha,i} = z$.
In this appendix, we provide numerical evidence that the transverse deviations $\log \delta f_k^{(\alpha)}$ and their derivatives satisfy
\begin{align}
  \label{eq:bound-dev-1}
	\log \delta f_k^{(\alpha)} |_{k = z}                  & = \mathcal{O}(\beta)
	\quad \text{and}                                                                 \\
	\partial_k \log \delta f_k^{(\alpha)} |_{k = z^{\pm}} & = \mathcal{O}(\beta),
  \label{eq:bound-dev-2}
\end{align}
where by derivative we mean the discrete difference between $k+1$ and $k$.
First, however, it is necessary to talk about the analytic properties of the logarithmic version \eqref{eq:saddle-eom-log} of the saddle point equation.
In particular, there may be different saddles depending on the chosen logarithmic branch sheets.

In the case where all sources and fugacities are equal (i.e.~$L_{\alpha} = L/K$, $y_{\alpha} = y^{1/K}$ and $m_{\alpha,i} = z$), there is a particularly natural solution
\begin{equation}
  \log f_k^{(\alpha)} = \log \bar{f}_k = - \frac{L \Delta_1}{K} k + \frac{L (\Delta_1 + \Delta_2)}{K} \min(k, z),
\end{equation}
where $\log y = - L \Delta_1$.
We assume that this is the dominant solution and that the dominant solutions with unequal sources and fugacities are obtained by ``minimally'' deforming away from this.%
\footnote{In particular, if $\log f_k^{(\alpha)}$ were continued to a function $\log f^{(\alpha)}(x)$, $x = \beta k \in \mathbb{R}_{\geq 0}$, the chosen solution should be continuous.}

Additionally, we expect that convergence requires $\operatorname{Re} \Delta_2 < 0$.
This is motivated from the giant graviton expansion: the coefficients $\mathcal{I}_{0, \boldsymbol{m}}$ are the contributions of M5-brane giant gravitons, which are obtained by taking a corresponding six-dimensional superconformal index and analytically continuing it from $\operatorname{Re} \tilde{\Delta}_1 \equiv \operatorname{Re} (- \Delta_2) > 0$ to $\operatorname{Re} (- \Delta_2) < 0$.

We will therefore assume the following parameter range for the numerical solution and analytically continue the saddle point value to generic parameters:
\begin{equation}
  \label{eq:eq-bounds-224}
  \operatorname{Re} \v_{\alpha} > 0, \quad
  \Delta_1, \Delta_2 \in \mathbb{R}
  \quad \text{and} \quad
  - \Delta_1 < \Delta_2 < 0.
\end{equation}
The last requirement is chosen such that $\beta = \Delta_1 + \Delta_2 > 0$.

\begin{figure}[p]
  \centering
  \includegraphics[width=0.92\textwidth]{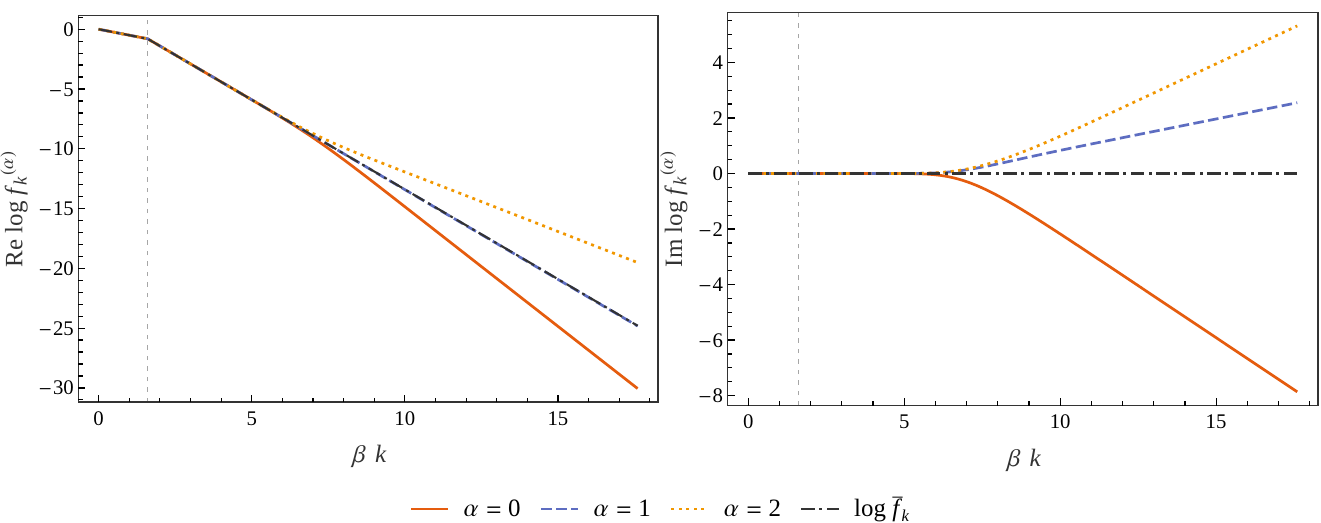}
  \caption{Real and imaginary parts of the numerical profiles $\log f_k^{(\alpha)}$ and the centre-of-mass profile $\log \bar{f}_k$ as functions of $\beta k$.
  The chosen parameters are $K = 3$, $L_{\alpha} = (2,1,0)$, $\Delta_1 = 3/100$, $\Delta_2 = -1/100$, $\beta z = 8/5$ as well as $\v_{\alpha} = L (\Delta_1 - \Delta_2) / 2 K + 10^{-4} \cdot (150 i,-45 i,-105 i)$.
  }
  \label{fig:k3-full-log-profile}

  \includegraphics[width=0.92\textwidth]{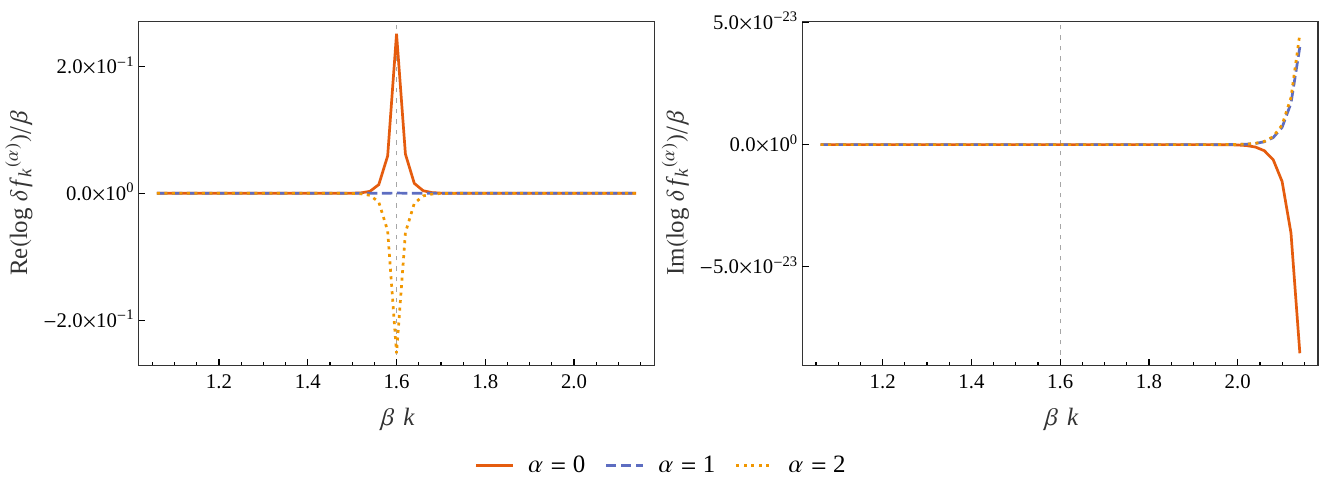}
  \caption{Real and imaginary parts of the transverse deviations $\log \delta f_k^{(\alpha)}/\beta$ near the source, as functions of $\beta k$, for the $K=3$ example. The parameters are given in figure \ref{fig:k3-full-log-profile}.}
  \label{fig:k3-source-deviation}

  \includegraphics[width=0.92\textwidth]{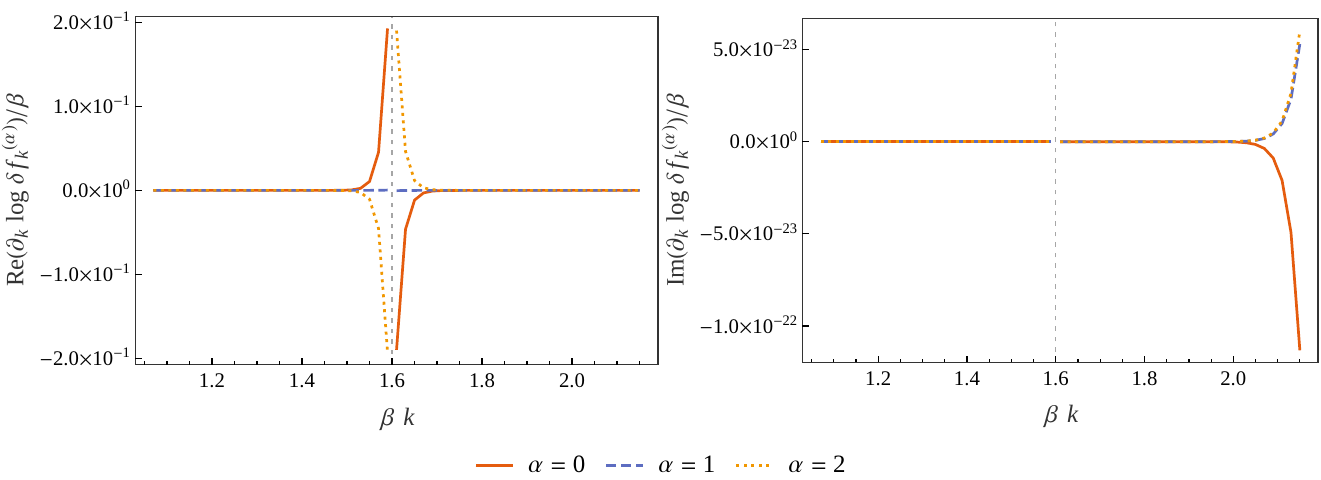}
  \caption{Real and imaginary parts of the discrete derivatives $\partial_k \log \delta f_k^{(\alpha)}/\beta$ near the source, as functions of $\beta k$, for the $K=3$ example. The parameters are given in figure \ref{fig:k3-full-log-profile}.}
  \label{fig:k3-source-deviation-deriv}
\end{figure}

\begin{figure}[p]
  \centering
  \includegraphics[width=0.92\textwidth]{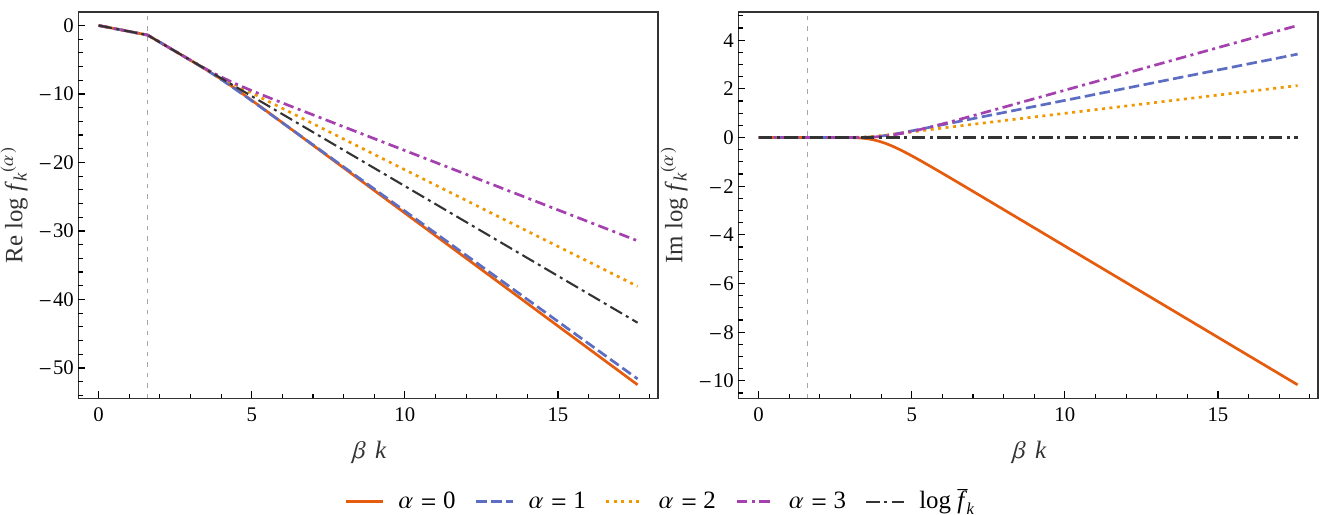}
  \caption{Real and imaginary parts of the numerical profiles $\log f_k^{(\alpha)}$ and the centre-of-mass profile $\log \bar{f}_k$ as functions of $\beta k$.
  The chosen parameters are $K = 4$, $L_{\alpha} = (3,3,1,0)$, $\Delta_1 = 3/100$, $\Delta_2 = -1/100$, $\beta z = 8 / 5$ as well as $\v_{\alpha} = L (\Delta_1 - \Delta_2) / 2 K + 10^{-4} \cdot (10, -5, -2, -3) + 10^{-3} \cdot (15, -5, - 3, -7)i$.
}
  \label{fig:full-log-profile}

  \includegraphics[width=0.92\textwidth]{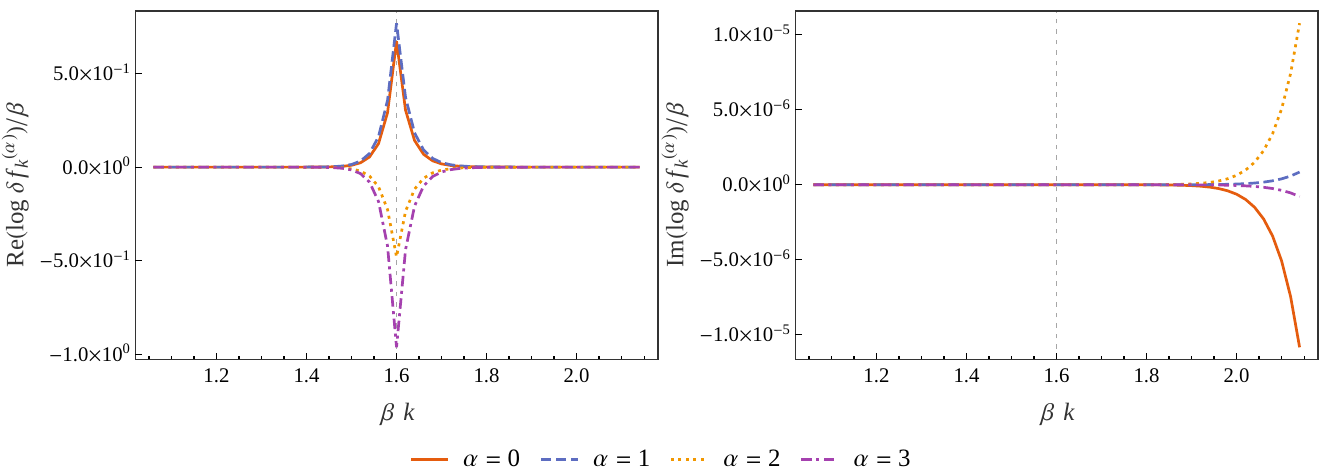}
  \caption{Real and imaginary parts of the transverse deviations $\log \delta f_k^{(\alpha)}/\beta$ near the source, as functions of $\beta k$. The parameters are given in figure \ref{fig:full-log-profile}.}
  \label{fig:source-deviation}

  \includegraphics[width=0.92\textwidth]{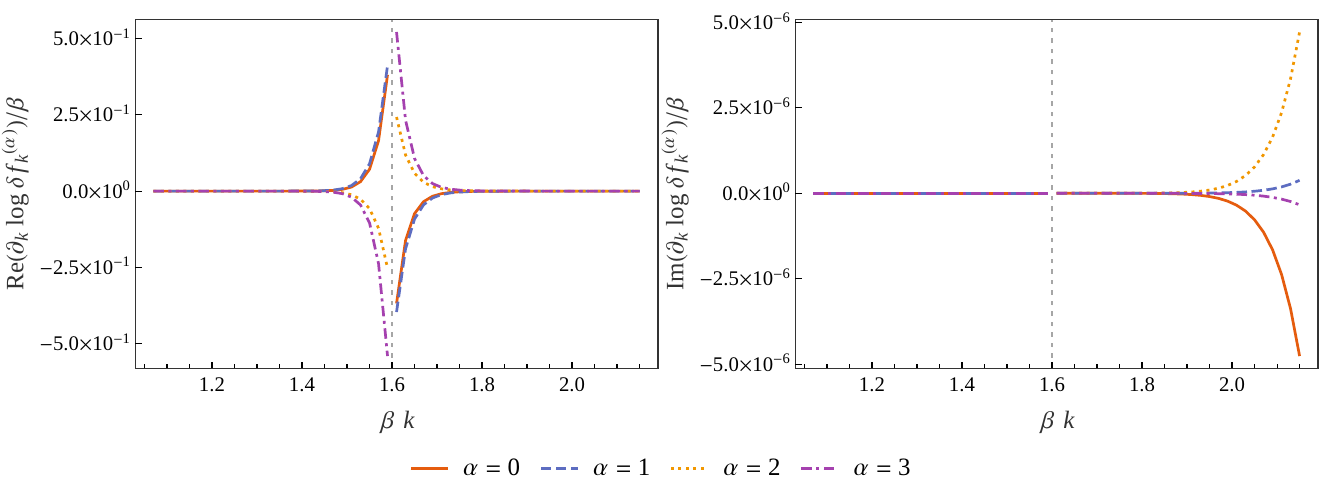}
  \caption{Real and imaginary parts of the discrete derivatives $\partial_k \log \delta f_k^{(\alpha)}/\beta$ near the source, as functions of $\beta k$. The parameters are given in figure \ref{fig:full-log-profile}.}
  \label{fig:source-deviation-deriv}
\end{figure}

\begin{figure}[p]
  \centering
  \includegraphics[width=0.92\textwidth]{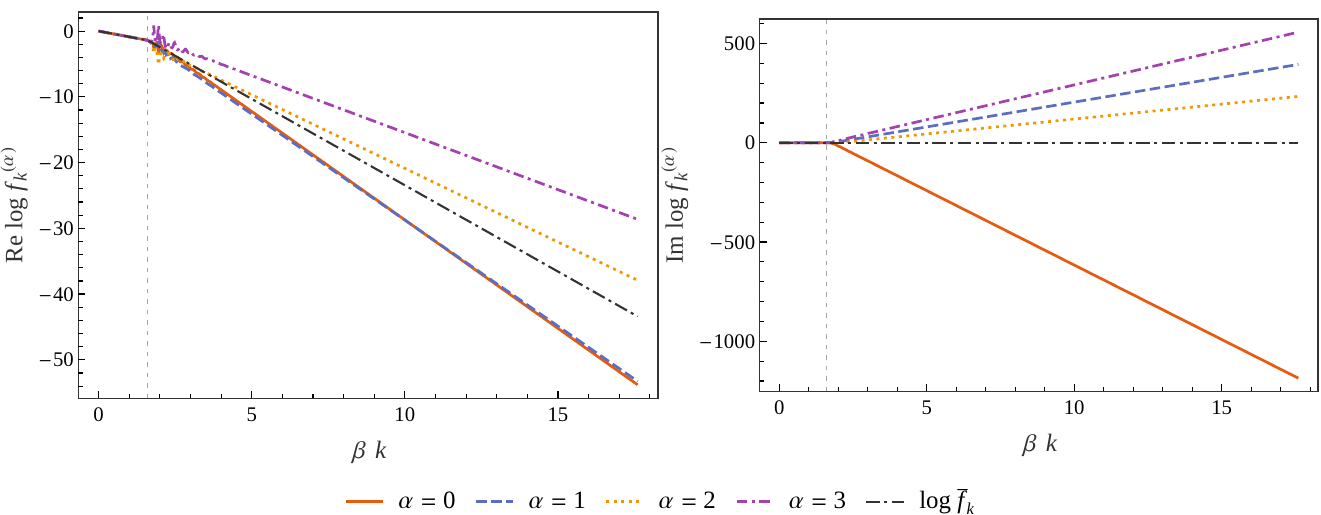}
  \caption{Real and imaginary parts of the numerical profiles $\log f_k^{(\alpha)}$ and the centre-of-mass profile $\log \bar{f}_k$ as functions of $\beta k$ for the second $K=4$ example.
  The parameters are identical to those in figure \ref{fig:full-log-profile}, except for the larger transverse imaginary part of $\v_{\alpha}$: $\v_{\alpha} = L (\Delta_1 - \Delta_2) / 2 K + 10^{-4} \cdot (10,-5,-2,-3) + 10^{-1}\cdot (15,-5,-3,-7) i$.}
  \label{fig:k4-v2-full-log-profile}

  \includegraphics[width=0.92\textwidth]{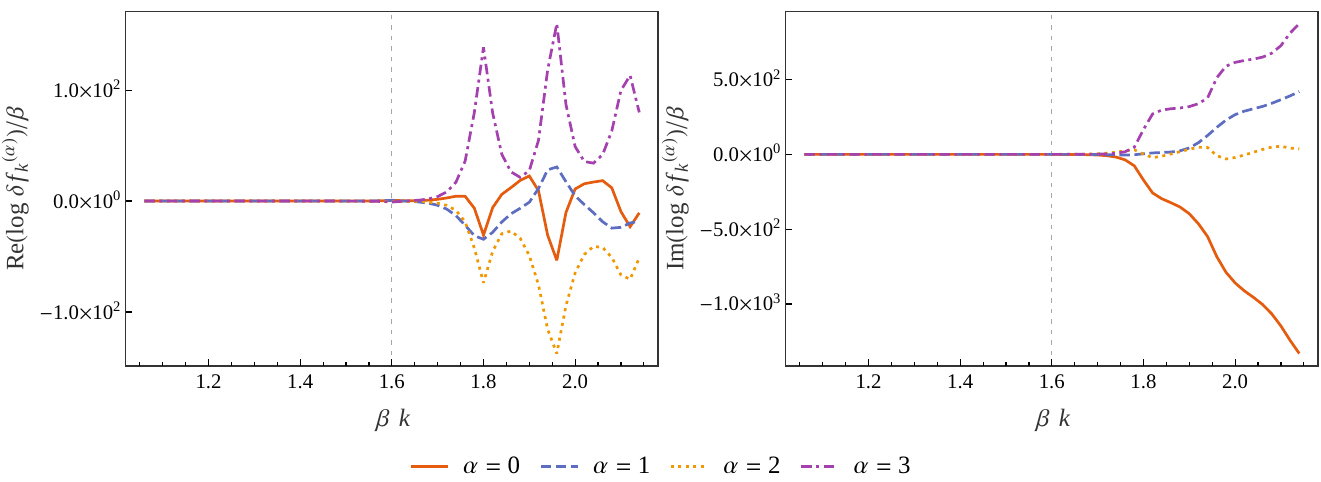}
  \caption{Real and imaginary parts of the transverse deviations $\log \delta f_k^{(\alpha)}/\beta$ near the source, as functions of $\beta k$, for the second $K=4$ example. The parameters are given in figure \ref{fig:k4-v2-full-log-profile}.}
  \label{fig:k4-v2-source-deviation}

  \includegraphics[width=0.92\textwidth]{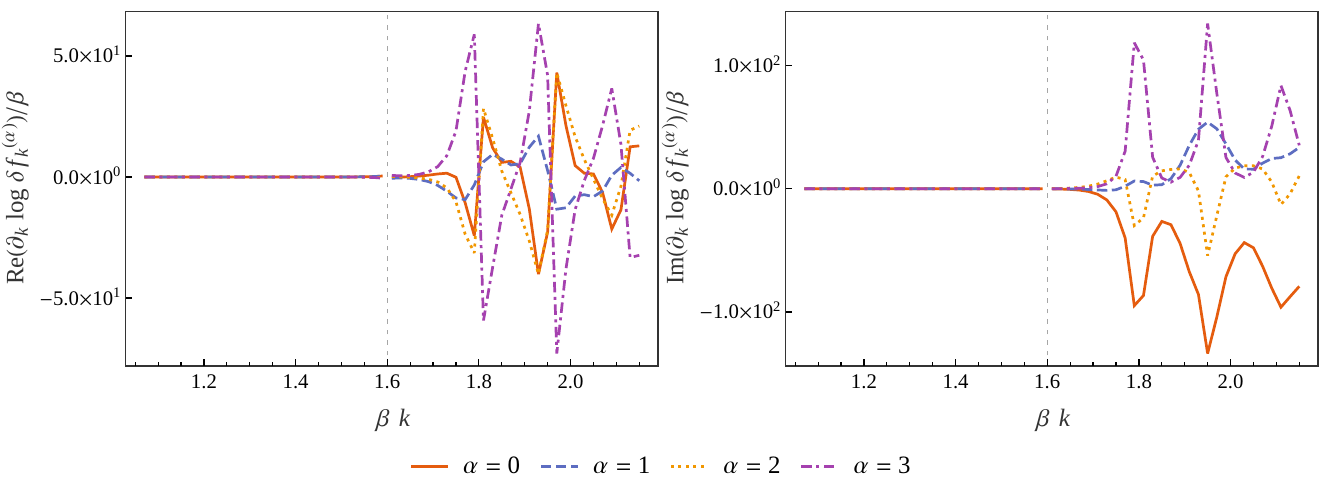}
  \caption{Real and imaginary parts of the discrete derivatives $\partial_k \log \delta f_k^{(\alpha)}/\beta$ near the source, as functions of $\beta k$, for the second $K=4$ example. The parameters are given in figure \ref{fig:k4-v2-full-log-profile}.}
  \label{fig:k4-v2-source-deviation-deriv}
\end{figure}

Example solutions for $K=3$ and $K=4$ with generic values satisfying \eqref{eq:eq-bounds-224} can be found in figures \ref{fig:k3-full-log-profile}, \ref{fig:full-log-profile} and \ref{fig:k4-v2-full-log-profile}.%
\footnote{
  These solutions were computed in \texttt{Mathematica}, using a script created with the assistance of \texttt{ChatGPT}.
  The solver starts from the known symmetric solution with equal sources and fugacities and gradually turns on the imbalances.
  The centre-of-mass mode $\log \bar{f}_k$ is used exactly, while the deviations $\log \delta f_k^{(\alpha)}$ are obtained numerically.
  The solver truncates the last sum in \cref{eq:saddle-eom-log} at a finite $l = N$ and imposes the boundary condition \eqref{eq:asymp-slope} there.
  It solves the saddle point equation for multiple values of $N$ to ensure that the differences decrease and the $N \to \infty$ limit converges.
}
The transverse deviations and their discrete derivatives near the source are shown in figures \ref{fig:k3-source-deviation} and \ref{fig:k3-source-deviation-deriv} for the $K=3$ example, in figures \ref{fig:source-deviation} and \ref{fig:source-deviation-deriv} for the first $K=4$ example, and in figures \ref{fig:k4-v2-source-deviation} and \ref{fig:k4-v2-source-deviation-deriv} for the second.
While the transverse real part of $\v_{\alpha}$ has to stay sufficiently small,%
\footnote{By the transverse part we mean the part $\v_{\alpha}^+$ of $\v_{\alpha}$ summing to $\sum_{\alpha=0}^{K-1} \v_{\alpha}^{+} = 0$.
This excludes the common shift of $L (\Delta_1 - \Delta_2) / 2K$.
In particular, as $\Delta_1, \Delta_2 \to 0$, the transverse real part of $\v_{\alpha}$ has to go to zero---else, there would exist an $\alpha$ violating $\operatorname{Re} \v_{\alpha} > 0$.
}
there are no restrictions on the transverse imaginary part.
As this becomes larger, the saddle point solutions tend to behave more chaotically (compare figures \ref{fig:full-log-profile} and \ref{fig:k4-v2-full-log-profile}).
However, this behaviour seems to start after $k = z$, leaving \eqref{eq:bound-dev-1} and \eqref{eq:bound-dev-2} intact.\footnote{The derivatives in figures \ref{fig:k3-source-deviation-deriv}, \ref{fig:source-deviation-deriv} and \ref{fig:k4-v2-source-deviation-deriv} appear to have a gap at $\beta k = \beta z$. This does not signal a divergence. Rather, it reflects the aesthetic choice of connecting the lattice points in the rest of the figure---the underlying lattice is always discrete. The values just before and after the jump correspond to the differences $\log \delta f_{z}^{(\alpha)} - \log \delta f_{z-1}^{(\alpha)}$ and $\log \delta f_{z+1}^{(\alpha)} - \log \delta f_{z}^{(\alpha)}$ respectively.}

\begin{figure}[p]
  \centering
  \includegraphics[width=0.92\textwidth]{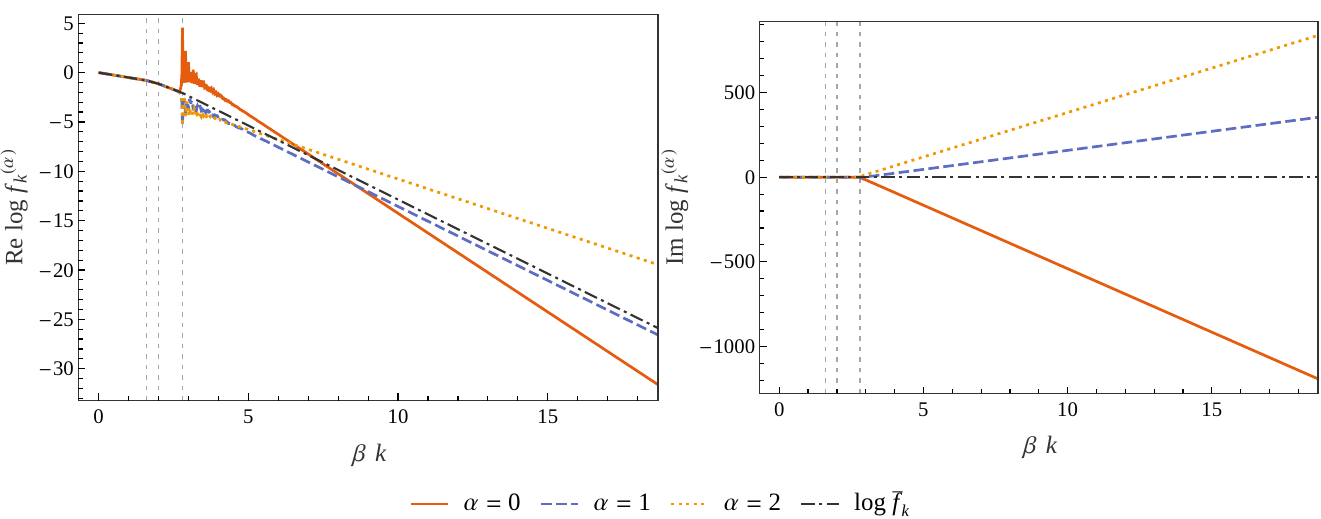}
  \caption{Real and imaginary parts of $\log f_k^{(\alpha)}$ and $\log \bar{f}_k$ as functions of $\beta k$ for $K=3$ with large transverse $\v_{\alpha}$.
  The parameters are $L_{\alpha}=(2,1,0)$, $\Delta_1=3/100$, $\Delta_2=-1/100$, $\v_{\alpha}=L(\Delta_1-\Delta_2)/(2K)+i(3/2,-9/20,-21/20)$, $\beta m_{0,1} = 8/5$, $\beta m_{0,2} = 14/5$ and $\beta m_{1,1} = 2$.}
  \label{fig:baryonic-large-vperp-full-log-profile}

  \includegraphics[width=0.92\textwidth]{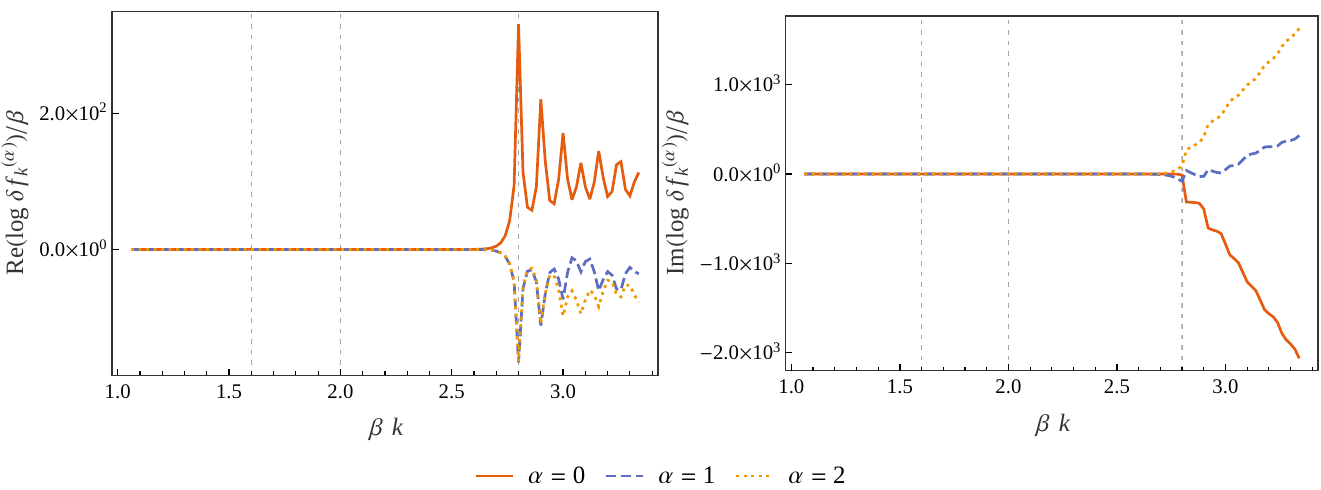}
  \caption{Real and imaginary parts of the transverse deviations $\log \delta f_k^{(\alpha)}/\beta$ near the sources, as functions of $\beta k$, for the example in figure \ref{fig:baryonic-large-vperp-full-log-profile}.}
  \label{fig:baryonic-large-vperp-source-deviation}

  \includegraphics[width=0.92\textwidth]{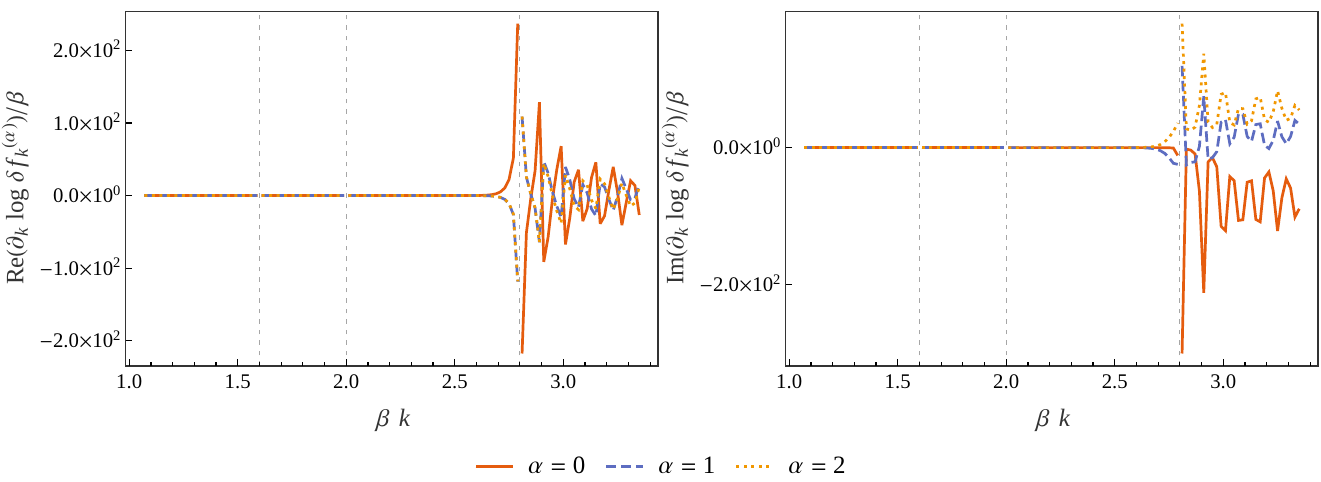}
  \caption{Real and imaginary parts of the discrete derivatives $\partial_k \log \delta f_k^{(\alpha)}/\beta$ near the sources, as functions of $\beta k$, for the example in figure \ref{fig:baryonic-large-vperp-full-log-profile}.}
  \label{fig:baryonic-large-vperp-source-deviation-deriv}
\end{figure}

For comparison, figures \ref{fig:baryonic-large-vperp-full-log-profile}, \ref{fig:baryonic-large-vperp-source-deviation} and \ref{fig:baryonic-large-vperp-source-deviation-deriv} show an example with unequal background fluxes $m_{\alpha,i}$ and a large transverse imaginary chemical potential. Unlike the cases above, the deviations $\log \delta f_k^{(\alpha)}$ and their derivatives are large at the last source, resulting in a breakdown of equations \eqref{eq:bound-dev-1} and \eqref{eq:bound-dev-2}.
These and further simulations suggest two regimes: a colour-locking regime for small $\beta k$, where $\log \delta f_k^{(\alpha)}$ are small%
\footnote{One can argue for $\mathcal{O}(\beta)$ deviations at the sources in this regime more rigorously by expanding the equations of motion around the centre-of-mass mode. In particular, numerical results suggest that the deviation is exactly linear in $\beta$. This can be explained by the form of the source term in the saddle point equation, which is proportional to $\beta$.}
and an asymptotic regime with slopes dictated by the $\v_{\alpha}$.
Larger sources push the transition regime to larger $\beta k$, while larger transverse chemical potentials $\v_{\alpha}$ push it towards smaller $\beta k$.
If both are sufficiently large, it can happen that one (or multiple) sources lie in the transition regime and deviate from the $\mathcal{O}(\beta)$ behaviour.%
\footnote{In the case of the simulation in figure~\ref{fig:baryonic-large-vperp-full-log-profile}, placing the last source at $\beta m_{0,2} = 11/5$ instead puts it back into the colour-locking $\mathcal{O}(\beta)$ regime.}
Consequently, investigating the behaviour at large sources, where the geometric interpretation of the giant graviton coefficients becomes relevant, may require sending $\v_{\alpha} \to 0$ as well.

\clearpage
\bibliographystyle{JHEP}
\bibliography{biblio.bib}

\end{document}